%
%
%
%
%
%
\newif\ifshowfigs
\showfigstrue
%
%
\font\Cc=cmcsc10

\def\Thsty{\it}
\def\Lesty{\it}
\def\Cl#1{\centerline{#1}}
\baselineskip=15pt
\hfuzz=2pt
\overfullrule=0pt

%
%
%
%
\def\endpage{\vfill\eject}

\def\someroom{\removelastskip\par\vskip 30 true pt 
plus 10 truept minus 10 truept} 
\def\hroom{\hglue 10pt }
\def\pni{\par\noindent}
\def\sni{\someroom\noindent}
\def\leftit#1{\par\noindent\hangindent=40pt\hangafter=1
           \hbox to 30pt{\hglue10pt#1\hss}\hglue 10pt\ignorespaces}
\def\hypit#1{\par\noindent\hangindent=60pt\hangafter=1
           \hbox to 40pt{\hglue5pt#1\hss}\hglue 20pt\ignorespaces}
\headline={\ifodd\pageno\rightheadline \else\leftheadline\fi}
\def\rightheadline{\hfil}
\def\leftheadline{\hfil}
\footline={\ifodd\pageno\rightfootline \else\leftfootline\fi}
\def\rightfootline{\hss\tenrm\folio\hss}
\def\leftfootline{\hss\tenrm\folio\hss}
%
%
%
%
%
%
\immediate\openout8=\jobname.txs
\newcount\sectno
\newcount\chapno
\newcount\equano
\newcount\theono
\newcount\refno
\sectno=0
\chapno=0
\equano=0
\theono=0
\refno=0
\def\eqhead{ }
\def\\{\backslash}
\newif\ifintrmk
\def\irmkanf{\vglue 8pt\noindent\hbox to\hsize{\hrulefill}\pni}
\def\irmkend{\pni\hbox to\hsize{\hrulefill}\vglue 8pt\noindent}

\font\bfe=cmbx12
\def\chapskip{\removelastskip\par\vglue 40pt}
\def\sectskip{\removelastskip\par\vskip 40pt}
\def\blwskip{\removelastskip\par\vskip 20pt}
\def\chap#1{\equano=0 \sectno=0 \theono=0 \global\advance\chapno by 1%
\def\eqhead{\number\chapno .}%
\chapskip\goodbreak\centerline{\bfe \eqhead \hglue 5pt #1}
\blwskip}%
\def\sect#1{\global\advance\sectno by 1
\sectskip\goodbreak%
\noindent{\bf \eqhead\number\sectno  \hglue 5pt #1}
\blwskip}%
\def\subsect#1{\removelastskip\vskip 25pt plus5pt minus10pt\noindent%
{\bf #1}\par\vskip 15pt\noindent}
\def\appendix#1#2{\par\chapskip\goodbreak\centerline{\bfe Appendix #1. #2}%
\blwskip%
\equano=0\sectno=0\theono=0\def\eqhead{#1.}}
\edef\Raum{ }
\def\eqn{{\hbox{\global\advance\equano by 1}}%
\eqno (\eqhead\number\equano )}%
\def\EQN#1{\eqn\edef\Zwi{(\eqhead\number\equano )}%
\immediate\write8{EQN  \Zwi\Raum = \noexpand#1}
\global\let #1=\Zwi
}
\def\queq#1{$#1$}
\def\The#1{{\global\advance \theono by 1 \someroom \noindent%
{\bf Theorem
\eqhead\number\theono }\hroom
\edef\Zwi{\eqhead\number\theono}
\immediate\write8{STM  \Zwi\Raum = \noexpand#1    Theorem }
\global\let#1=\Zwi
}}
\def\Pro#1{{\global\advance \theono by 1 \someroom \noindent%
{\bf Proposition
\eqhead\number\theono }\hroom
\edef\Zwi{\eqhead\number\theono}
\immediate\write8{STM  \Zwi\Raum = \noexpand#1    Proposition }
\global\let#1=\Zwi}}
\def\Rem#1{{\global\advance \theono by 1 \someroom \noindent%
{\bf Remark
\eqhead\number\theono }\hroom
\edef\Zwi{\eqhead\number\theono}
\immediate\write8{STM  \Zwi\Raum = \noexpand#1    Remark }
\global\let#1=\Zwi}}
\def\Cor#1{{\global\advance \theono by 1 \someroom \noindent%
{\bf Corollary
\eqhead\number\theono }\hroom
\edef\Zwi{\eqhead\number\theono}
\immediate\write8{STM  \Zwi\Raum = \noexpand#1    Corollary }
\global\let#1=\Zwi}}
\def\Def#1{{\global\advance \theono by 1 \someroom \noindent%
{\bf Definition
\eqhead\number\theono }\hroom
\edef\Zwi{\eqhead\number\theono}
\immediate\write8{STM  \Zwi\Raum = \noexpand#1    Definition }
\global\let#1=\Zwi}}
\def\Lem#1{{\global\advance \theono by 1 \someroom \noindent%
{\bf Lemma
\eqhead\number\theono }\hroom
\edef\Zwi{\eqhead\number\theono}
\immediate\write8{STM  \Zwi \Raum = \noexpand#1    Lemma }
\global\let#1=\Zwi
}}
\def\Proof{\someroom \noindent{\it Proof:}\hroom}
\def\endproof{\hfill
\hbox{\vrule width 7pt depth 0pt height 7pt} \vskip 25pt plus
5pt minus 10pt}
\def\eqndprf{\qquad
\hbox{\hfill\vrule width 7pt depth 0pt height 7pt\hglue 8pt}}
\def\refit#1{\par\noindent\hangindent=1.5cm\hangafter=1
           \hbox to 1cm{#1\hss}\hglue 0.5cm\ignorespaces}
\def\Ref#1#2{\refit{[#1]} #2}
\def\quref#1{[#1]}
\def\today{\ifcase\month\or
 January\or February\or March\or April\or May\or June\or
 July\or August\or September\or October\or November\or December\fi
 \space\number\day, \number\year}
\def\getArchdate{\openin7=\jobname.arx
                 \ifeof7 \def\Archdate{\today}
                 \else \read7 to \Archdate\fi
                 \closein7}
%
%
\newcount\figno
\figno=0
\def\nextfig#1{\global\advance\figno by 1%
\edef\Zwi{\number\figno}%
\immediate\write8{FIG  \Zwi\Raum = \noexpand#1   }%
\global\let#1=\Zwi $\Zwi$}
%
%
\def\senzafig#1{\global\advance\figno by 1%
\edef\Zwi{\number\figno}%
\immediate\write8{FIG  \Zwi\Raum = \noexpand#1   }%
\global\let#1=\Zwi}
%
%
%
%
%
\def\N{{\rm I\kern-.16em N}}
\font\twblk=cmss10
\font\tenblk=cmss8
\font\eiblk=cmss8
\def\Z{\mathchoice{{\hbox{\twblk Z\kern-.35emZ}}}
{{\hbox{\twblk Z\kern-.35emZ}}}
{{\hbox{\tenblk Z\kern-.30emZ}}}
{{\hbox{\eiblk Z\kern-.24emZ}}}}
\def\Q{{\rm \kern.25em\vrule height1.4ex
depth-.12ex width.06em\kern-.31em Q}}
\def\R{{\rm I\kern-.2emR}}
\def\C{{\rm \kern.25em\vrule height1.4ex
depth-.12ex width.06em\kern-.31em C}}
\def\K{{\rm I\kern-.2emK}}
\def\L{{\rm I\kern-.2emL}}
\def\bbbone{{\mathchoice {\rm 1\mskip-4mu l} {\rm 1\mskip-4mu l}    
{\rm 1\mskip-4.5mu l} {\rm 1\mskip-5mu l}}}
%
%
\def\al{\alpha}
\def\be{\beta}
\def\ga{\gamma}
\def\de{\delta}
\def\ep{\epsilon}
\def\veps{\varepsilon}
\def\et{\eta}
\def\ze{\zeta}
\def\th{\theta}
\def\vth{\vartheta}

\def\ka{\kappa}
\def\la{\lambda}
\def\rh{\rho}
\def\si{\sigma}
\def\ta{\tau}

\def\ph{\phi}
\def\vphi{\varphi}
\def\ch{\chi}
\def\ps{\psi}
\def\om{\omega}
\def\vphi{\varphi}
\def\veps{\varepsilon}
%
%
\def\Ga{\Gamma}
\def\De{\Delta}

\def\Si{\Sigma}

\def\Ph{\Phi}

\def\Om{\Omega}
%
%

\def\chq{{\bar\chi}}
\def\psq{{\bar\psi}}

%

%
%

%
%
\def\cA{{\cal A}}
\def\cB{{\cal B}}
\def\cC{{\cal C}}

\def\cF{{\cal F}}
\def\cG{{\cal G}}

\def\cJ{{\cal J}}

\def\cL{{\cal L}}
\def\cM{{\cal M}}
\def\cN{{\cal N}}
\def\cO{{\cal O}}
\def\cP{{\cal P}}
\def\cQ{{\cal Q}}
\def\cR{{\cal R}}
\def\cS{{\cal S}}
\def\cT{{\cal T}}
\def\cU{{\cal U}}
\def\cV{{\cal V}}
\def\cW{{\cal W}}

%
%
\def\zer{{\oldstyle 0}}
\def\one{{\oldstyle 1}}
\def\two{{\oldstyle 2}}
\def\thr{{\oldstyle 3}}
\def\prP{{\bf P}}
\def\db{{\mkern2mu\mathchar'26\mkern-2mu\mkern-9mud}}
\def\tr{\hbox{ tr }}
\def\I2{{$I_2$}}
\def\del{\partial}
\def\dth{{\del_\theta}}
\def\gtoas#1{\quad\mathop{\longrightarrow}\limits_{#1}\quad}
\def\ve#1{{\bf #1}}

\def\tst#1{{{\theta_{#1}}^*}}

\def\nat#1{\{ 1,\ldots,#1 \} }

\def\abs#1{{\left\vert #1 \right\vert}}

\def\openkrnl#1{\mathop{#1}\limits^{\;_\zer}}
%
%
\def\True#1{\; \bbbone\left( #1 \right) \;}
%
%
\def\deqalign#1{\vcenter{\openup1\jot \mathsurround=0pt \ialign{
                \strut\hfil$\displaystyle{##}$&$\displaystyle{{}##}$\hfil&
                $\displaystyle{{}##}$\hfil\crcr
                #1\crcr}}}         
%
%
\def\meqalign#1{\vcenter{\openup1\jot \mathsurround=0pt \ialign{
                \strut\hfil$\displaystyle{##}$&&$\displaystyle{{}##}$\hfil&
                \quad\hfil$\displaystyle{##}$\crcr
                #1\crcr}}}         
\def\dst{\displaystyle}

\def\tst{\textstyle}
\def\frac#1#2{\dst {#1\over#2}}     
\def\sfrac#1#2{{\tst{#1\over#2}}}   

\def\dotcup{\mathop{{\mathop{\cup}\limits^\cdot}}}
\def\Const{\hbox{ \rm const }}
\def\supp{\hbox{ \rm supp }}
\def\NOL{non--\-over\-lap\-ping}
\def\OL{{o\-ver\-lap\-ping}}

%


\def\suffix{ps}
\newcount\system
\global\system=3   

\def\ifundefined#1{\expandafter\ifx\csname#1\endcsname\relax}
\ifundefined{figdir}\def\figdir{}\fi
%
%
\newcount\firstline
\newdimen\pswidth  \newdimen\xleft
\newdimen\psheight \newdimen\ytop \newdimen\ybot
\newcount\justx \newcount\justy
\global\justx=0 \global\justy=0
\newdimen\vpos \newtoks\label 
\newread\labelfile \newdimen\xcoord \newdimen\ycoord
\newif\ifdoit 
\newbox\labox
\newdimen\xdvikwid 
\newdimen\xdvikht
\newdimen\pspoints
\newdimen\rwi
\pspoints=1bp
\newcount\temp
\def\readdim#1{\global\read\labelfile to \temp
\global #1=\temp pt}
%
%
%
%
\def\figcrop#1{\par
\openin\labelfile=\figdir#1.lbl                                              
\global\read\labelfile to\firstline\message{#1}               
\global\read\labelfile to\temp
\readdim{\ybot}
\readdim{\xleft}
\readdim{\ytop}
\global\read\labelfile to\justx
\global\read\labelfile to\justy
\global\read\labelfile to\label
\readdim{\pswidth}
\global\advance\pswidth by -\xleft
\readdim{\psheight}
\global\advance\ybot by -\psheight
\global\advance\psheight by -\ytop
\global\read\labelfile to\justx
\global\read\labelfile to\justy
\global\read\labelfile to\label
\vbox to\psheight{\vfill
\ifnum\system=1
\fi 
\ifnum\system=2
\fi 
\ifnum\system=3
\fi         
\ifnum\system=4
\fi         
\ifnum\system=1
\hbox to \pswidth{\kern-\xleft\special{postscriptfile \figdir#1.\suffix }\hfil}\fi
\ifnum\system=2
\hbox to \pswidth{\kern-\xleft\special{ps: plotfile \figdir#1.\suffix }\hfil}\fi
\ifnum\system=3
\hbox to \pswidth{\kern-\xleft\includegraphics{\figdir#1.\suffix}\hfil}\fi
\ifnum\system=4
\hbox to \pswidth{\kern-\xleft\includegraphics{\figdir#1.\suffix}\hfil}\fi
\ifnum\system=5
\hbox to \pswidth{\kern-\xleft\includegraphics{\figdir#1.\suffix}\hfil}\fi 
\ifnum\system=6
   \xdvikwid=\pswidth
   \xdvikht=\psheight
   {\global\divide\xdvikwid by \pspoints}
   {\global\divide\xdvikht by \pspoints}
   \rwi=\xdvikwid
    {\global\multiply\rwi by 10}
\hbox to \pswidth{\kern-\xleft\includegraphics{\figdir#1.\suffix\space}\hfil}\fi                   
\vskip -\baselineskip
\vskip -\ybot 
\vskip-\psheight %
\hbox to\pswidth  {\hss}%
\parindent=0pt\offinterlineskip                                       
\vpos=0 pt%
\loop\readdim{\xcoord}                                 
\ifdim \xcoord < -999pt \doitfalse\else\doittrue\fi                        
\ifdoit \advance \xcoord by -\xleft
\readdim{\ycoord}
\advance \ycoord by -\ytop                              
\global\read\labelfile to\justx                                       
\global\read\labelfile to\justy                                       
\global\read\labelfile to\label
\global\setbox\labox=\hbox{\label\hskip-0.3em}%
\advance\vpos by-\ycoord                                              
\vskip-\vpos \vpos=\ycoord                                         
\hbox to\pswidth{\hskip\xcoord %
\hbox to 0pt{\ifnum\justx>0\hss\fi%
\vbox to0pt{%
\ifnum\justy<2\vss\fi%
\copy\labox\kern0pt%
\ifnum\justy>0\vss\fi}%
\ifnum\justx<2\hss\fi}%
\hss}%
\repeat%
\advance\vpos by-\psheight%
\vskip-\vpos %
}\closein\labelfile}
%
%
%
\def\figplace#1#2#3{
\openin\labelfile=\figdir#1.lbl
\ifeof\labelfile\immediate\write16{Can't find \figdir#1.lbl; I quit!}\end\fi 
\closein\labelfile
\null\hskip#2\raise #3 \hbox{\figcrop{#1}}
}
%
%
\def\herefig#1{%
\ifshowfigs
\midinsert
\centerline{\figplace{#1}{0in}{0in}}
\endinsert
\fi
}

\def\figdir{}
\def\FST{{\tt I}} 
\def\k{\ve{k}}
\def\p{\ve{p}}
\def\q{\ve{q}}
\def\Q{\ve{Q}}
\def\r{\ve{r}}

\def\t{\ve{t}}
\def\t{\ve{t}}
\def\v{\ve{v}}
\def\={=}
\def\half{\sfrac{1}{2}}
\def\set#1#2{\{ #1 : #2\}}

\def\ulth{\underline{\Th}}
\def\ulph{\underline{\ph}}
\def\ulj{\underline{j}}
\def\ult{\underline{t}}
\def\AOne#1{{\sl (H1)$_{#1}$}} 
\def\ATwo#1{{\sl (H2)$_{#1}$}} 
\def\AThr{{\sl (H3)}} 
\def\AFou{{\sl (H4)}} 
\def\AFoup{{\sl (H4')}} 
\def\AFiv{{\sl (H5)}} 
\def\SYmm{{\sl (Sy)}} 
\def\ssni{\par\vskip 15pt \noindent}
\def\half{{\sfrac{1}{2}}}

\def\THREE{{\tt III}}
\def\SiRPA{\Sigma_{\rm RPA}}
\def\KRPA{K_{\rm RPA}}
\def\hxp{{h}} 
%
%
\def\qCP{B.3}
\def\compcont{B.4}
%
%
{\nopagenumbers
\rightline{December 1996}
\vglue 3 true cm

\Cl{\bfe Perturbation Theory around Non--Nested Fermi Surfaces}
\Cl{\bfe II. Regularity of the Moving Fermi Surface: RPA Contributions}
 \vglue 2 true cm
\Cl{\Cc Joel Feldman$^{a,}$\footnote{$^1$}{\rm 
feldman@math.ubc.ca, http://www.math.ubc.ca/$\sim$feldman/}, 
Manfred Salmhofer$^{b,}$\footnote{$^2$}{\rm 
manfred@math.ethz.ch, http://www.math.ethz.ch/$\sim$manfred/manfred.html},
and Eugene Trubowitz$^{b,}$\footnote{$^3$}{\rm trub@math.ethz.ch}}
\vglue 1 true cm
\Cl{$^a$\sl Mathematics Department, The University of British Columbia,
Vancouver, Canada V6T 1Z2}
\Cl{$^b$\sl Mathematik, ETH Z\" urich, CH--8092 Z\" urich, Switzerland}
\vglue 2 true cm
\Cl{\bf Abstract}

{\noindent
Regularity of the deformation of the Fermi surface under short-range
interactions is established for all contributions to the RPA 
self--energy (it is proven in an accompanying paper that 
the RPA graphs are the least regular
contributions to the self--energy). 
Roughly speaking, the graphs contributing to the RPA 
self--energy are those constructed by contracting
two external legs of a four--legged graph that consists of a
string of bubbles.
This regularity is a necessary 
ingredient in the proof that renormalization does not change the model.
It turns out that the self--energy is more regular when derivatives are
taken tangentially to the Fermi surface than when they are taken 
normal to the Fermi surface. 
The proofs require a very detailed analysis of the singularities that
occur at those momenta $\p$ where the 
Fermi surface $S$ is tangent to $S+\p$.
Models in which $S$ is not symmetric under the 
reflection $\p \to -\p$ are included.
}
\endpage}
%
%
\chap{Introduction}
\pni
This paper is a continuation of \quref{FST}, which we refer to as \FST\ in 
what follows. Together with an accompanying paper, 
hereafter referred to as \THREE, it contains the proof that the 
counterterm function of \FST\ is regular enough 
for the solution of the equations giving the renormalization
of the Fermi surface (discussed in detail below) to exist.
We shall recall briefly the motivation and the setting of the problem, 
as well as 
some results of \FST, to make this paper as self--contained as possible. 
For a detailed motivation, definitions of the scale flow and 
renormalization we refer the reader to \FST\ and to [S].  
All of this paper can be understood if one knows 
only the most elementary properties of scaled propagators, as
stated in Lemma 2.3 of \FST. In particular, no familiarity with the 
formalism of the Gallavotti--Nicol\` o \quref{GN} 
tree expansion is required. 
\sect{The Problem}
\noindent
We consider a many--fermion system in a crystal background 
or on a lattice in $d \ge 2$ spatial dimensions, 
defined by a band structure and an interaction
with a small coupling constant $\la $. The action for the model is 
$$
\cA= \int\frac{d^{d+1}p}{(2\pi)^{d+1}} 
\psq(p) (ip_\zer - E(\p))\ps(p)
\; + \; \la V(\ps,\psq)
\eqn $$
The basic assumptions we make 
are that the free band structure $E$ and the Fourier transform of the potential
are both at least $C^2$ (\AOne\ and \ATwo{}), 
that the curvature of the Fermi surface $S$ is 
everywhere positive (\AThr), and, for $d=2$,
 that the filling factor is so small that 
certain umklapp processes do not happen in second order (\AFiv). 
For instance, in the two--dimensional Hubbard model, this restriction means that
the filling factor $n$ has to obey $n < 0.369$, where $n=1$ is half--filling.
For $d \ge 3$, we need no filling restriction.
We do not assume that $E(-\p) = E(\p)$
for all $\p$. If this symmetry does not hold, we call $E$ asymmetric. 
If $E$ is asymmetric, we also assume that the curvature at a point 
$\p \in S$ and at its antipode $\ve{a}(\p)$ do not differ by
too much (\AFou), but that they coincide at only finitely many
points (\AFoup).
We believe that all these assumptions are necessary and sufficient, that is, 
if one of them does not hold, not only the proofs, but also at
least one conclusion of our theorems break down. The filling restriction 
\AFiv\ is needed only in two dimensions, and only for special parts
of our proofs.

The assumptions \AOne{}--\AFiv\ are spelled out in
detail in Chapter 2. They are stronger 
than those of \FST, so all results of \FST\ apply. The dynamics is 
given by the limit of the grand canonical ensemble as the volume tends
to infinity and the temperature to zero. The interaction produces a
self--energy of the electrons which one 
wants to calculate for small $\la$ by perturbation theory. 
The interaction may have drastic nonperturbative effects such as
superconductivity, but in these weakly coupled systems, the 
correct way to begin the analysis, and to decide if this happens, 
is to study perturbation theory \quref{FT1,FST}.

It is well--known that the naive (unrenormalized) perturbation 
theory is infrared divergent because of the slow decay of the 
fermion two--point function in position space. 
The inherent slow decay 
is responsible for most of the physical 
(e.g.\ conductance) properties of these systems. In momentum space,
this behaviour manifests itself as the singularity of the propagator
on the Fermi surface, which, at zero temperature, is the boundary 
of the set of occupied states. As discussed in \FST, 
the infrared divergence of perturbation theory 
is not a problem of the model but of the way the expansion is done.
Because of the self--energy effects, 
the Fermi surface moves when the interaction is turned on, 
and this motion of the singularity of the propagator 
causes the infrared divergences of a naive expansion. 
In \FST, we showed that, by making the band structure a function
of $\la$, one can keep the Fermi surface fixed as the 
interaction is turned on, and thereby gave an infrared finite 
renormalized perturbation expansion. More precisely, 
for a given band structure $e$ and its 
associated covariance $(ip_\zer -e(\p))^{-1}$, 
one can identify those parts $K(e,\la V,\p)$ of the self--energy
$\Si (p_\zer,\p)$ that move the 
surface. Then the model with the modified band structure 
$e(\p)+K(e,\la V,\p)$ and interaction $\la V$ has a well--defined 
and locally Borel summable perturbation series. However, since 
$$
K(e,\la V,\p) =\sum\limits_{r=1}^\infty K_r (e,V,\p) \la^r
\EQN\Kfps $$ 
is a 
functional of $e$ and a function of the coupling $\la $ and the
spatial part $\p$ of the momentum,
simply replacing $e(\p)$ by $e(\p) + K(e,\la V,\p)$ 
changes the model in a rather complicated way. 
To construct the original  model with the given fixed free band 
relation $E$ and a given 
interaction one must solve $E=e+K$ for $e$. The free Fermi surface is the 
zero set of $E$, while the interacting Fermi surface is the zero set 
of $e$. The central problem is thus:
fix a suitable $E$, and determine $e$ from 
$$
E=e+K(e, \la V )
.\EQN\Wltfrml $$
That is, invert the map from the renormalized to the bare band structure.
Once this is done, the counterterm function $K$ acquires a new role. 
It describes the deformation of the Fermi surface under the interaction.
Thus proving regularity properties of $K$ as a function of $\p$, 
as we do here, is proving regularity of the moving Fermi surface.

In \FST, we proved uniqueness of the solution to \queq{\Wltfrml}
for a very large class of Fermi surfaces and interactions. 
This paper and its sequels are devoted to proving existence of the solution 
for the class of $E\in C^2$ which 
give rise to a strictly convex Fermi surface. We now discuss the main 
reasons why this is a rather nontrivial problem and give an outline
how we solve it.

\sect{Main Results}

\noindent
It was explained in detail in \quref{FST} why many terms in the unrenormalized 
perturbation expansion are divergent.  In a nutshell, the coefficients 
in the series for e.g.\ the self--energy contain integrals over arbitrary
powers of the free propagator $(ip_\zer - e(\p))^{-1}$, which is not in $L^2$.
For the same reason, it is far from trivial to prove any regularity of 
the self--energy and the counterterm function $K$ even after renormalization:
every derivative increases the power of the denominator by one, 
and thus potentially introduces new divergences. Naive power
counting suffices only to show H\" older continuity of degree
$\al < 1$ of the self--energy, which is not even sufficient to
renormalize the theory. Moreover, 
it is evident that one will be able to solve \queq{\Wltfrml} 
for generic $E\in C^2$ only if 
one can prove that $K$ is also a $C^2$ function of momentum. 
In \FST, we proved volume improvement
estimates implying sharper power counting 
bounds that allow us to show that the theory can be renormalized
and that the self--energy and the 
counterterm function are finite and $C^{1+\al}$ for some $\al
>0$, determined by the Fermi surface
geometry. We introduced the notion of overlapping loops 
(see Definition 2.19 of \FST) and showed that 
whenever a graph has overlapping loops, the integral for 
the phase space volume contains a subintegral bounded by
$$
\cW(\veps) = \sup\limits_{\q \in \cB}
\max\limits_{v_i \in \{ \pm 1\}} 
\int\limits_{S^{d-1}\times S^{d-1}} d\th_\one \, d\th_\two \,
\True{\abs{e(v_\one \p(0,\th_\one) + v_\two \p (0,\th_\two ) + \q )} \leq \veps}
\EQN\cWdef $$
(see \FST, (A.6)). Here $\cB$ is the Brillouin zone, 
$\True{X}=1$ if $X$ is true, and zero otherwise.
$\p(\rh,\th)$ denotes a parametrization of a neighbourhood of the Fermi 
surface $S$ with $\rh=0$ corresponding to the Fermi surface itself.
(This parametrization will be given in detail in Section 2.2.
$\th$ runs over $S^{d-1}$ because, under \ATwo{}--\AThr, 
the Fermi surface is diffeomorphic to $S^{d-1}$).
We showed in \FST\ that under a very general non--nesting 
condition, there is $0<\ep <1$ such that
$$
\cW(\et ) \leq \Const \et^\ep
,\EQN\ONEvol $$
and that this implies volume--improved power counting bounds for all 
two--legged graphs and for all four--legged graphs except for the ladder
graphs. Any $\al<\ep$ can be used in the
above statement that the self--energy is $C^{1+\al}$.
Under the strict convexity condition imposed here, the bounds in \FST\ imply that
$$
\cW( \veps) \leq \Const \veps^{\sfrac{d-1}{d}}
\EQN\FSTvol $$
(see Proposition 1.1 and Lemma A.2 in \FST). In this paper, we  
prove 

\The{\bestvol}{\Thsty Assume \ATwo{2,0}, \AThr, and \AFou. Then 
there is a constant $Q_V\ge 1$ such that for all $\veps > 0$
$$
\cW(\veps) \leq Q_V \; \veps \; \cases{|\log \veps | & if $d=2$\cr
1 & if $d \ge 3$.}
\eqn $$ 
The constant $Q_V$ depends only on $|e|_\two=\sup\limits_{\p \in \cB}
\sum\limits_{|\al| \le 2} \abs{D^\al e(\p)}$, and the numbers
$r_\zer$, $g_\zer$, and $w_\zer$ (defined in Chapter 2).} 

\ssni
As mentioned, the precise assumptions \AOne\ -- \AFiv\ are stated 
at the beginning of Chapter 2. Assumption \AFiv\ is not needed
in the proof of Theorem \bestvol, which is given in Appendix B.
In the application, $\veps = M^j$ (with $M>1$ and $j<0$) 
is a small energy scale.
The self--energy $\Si$ and the counterterm $K$ are given as
scale sums, e.g.\
$$
\Si(\p) = \sum\limits_{j < 0} \Si^{(j)}(\p)
.\EQN\Sise $$
The standard
power counting bound for this function is $\abs{\Si^{(j)}} \le
M^j$, and every derivative multiplies this bound by a factor
$M^{-j}$. In \FST\ we showed that actually  $\abs{\Si^{(j)}} \le
M^{j(1+\ep)}$, because of the extra volume improvement factor $M^{\ep j}$
coming from \queq{\ONEvol}. This allows one to take almost $1+\ep$ derivatives
and still have a convergent series in \queq{\Sise}.

Theorem \bestvol\ 
implies that the volume improvement factor $M^{\ep j}$ of \FST\ can 
be replaced by $|j|M^j$ wherever it appears. Thus, the bound
for the $\Si^{(j)}$ (and the similar bound for $K^{(j)}$)
improves from the standard behaviour $M^j$ to 
$|j|M^{2j}$ just by this volume estimate and Theorem 2.46 $(i)$ in \FST. 
Counting the effect of derivatives in the way described,
we see that the scale sum over $I< j < 0$ of the second 
derivative of $K$ or $\Si$ diverges at most as $|I|^2$, not as $M^{|I|}$,
as $I \to -\infty$, as unimproved power counting would suggest. 
In other words, Theorem \bestvol{} ensures that the divergence of the second derivative
can be, at worst, marginal and one may hope for convergence with a little more
care in doing the bounds.

The volume bound stated above is best possible, however, and so one has
to go beyond volume estimates even to solve the problem in second order. 
It turns out that this requires a much more detailed analysis.
Because the Fermi surface is a $d-1$--dimensional submanifold
of $d$--dimensional space, there are angular variables parametrizing
it, and which also occur as integration variables.
The idea is 
now to use these angular integration variables to prevent derivatives
from degrading the scale behaviour, by moving the dependence on the 
external momentum from propagators ${1\over ip_\zer-e(\p)}$ into a Jacobian. However, this Jacobian has singularities
whose location depends on the external momentum. 
For the second order contribution, we do a careful analysis of
these singularities and show that under our hypotheses, in particular
because the Fermi surface has strictly positive curvature,
the second order counterterm  
has exactly the same degree of differentiability as $e$. 

\The{\zweireg}{\Thsty 
\leftit{$(i)$} Let $d=2$. There exists $\hxp > 0$ such that
if $0 \le \hxp' \le \hxp$, $k \ge 2$, and
\AOne{k,\hxp'},\ATwo{k,\hxp'},\AThr--\AFiv\ hold, 
then $K_\two(e,V,\p)$, given in \queq{\Kfps}, 
is $C^{k,\hxp'}$ in $\p$.
Moreover, the second order self--energy $\Si_\two$ is $C^{1,\ga}$ in $\p$
for any $\ga \in (0,1)$.
\leftit{$(ii)$} Let $d\ge 3$. There exists $\hxp > 0$ such that
if $0 \le \hxp' \le \hxp$, and 
\AOne{2,\hxp'},\ATwo{2,\hxp'},\AThr, and \AFou\ hold,
then $K_\two$ and $\Sigma_\two$ are $C^{2,\hxp'}$ in $\p$.
\pni
Here  $C^{k,\hxp}$ is the set of functions all of whose $k^{\rm th}$
order derivatives are H\"older continuous of index $\hxp$. $\hxp'=0$ is allowed.
}
\ssni 
Note that Theorem \zweireg\ states for $d=2$ that if $e$ and
$\hat v$ are $C^2$, then the counterterm $K_\two$ is also $C^2$, 
whereas the self--energy $\Si_\two$ is only shown to be 
$C^{1,\ga}$ for any $\ga <1$,
(that is, more loosely speaking, $\Si_\two \in C^{2-\veps}$ for
any $\veps >0$), even if $e\in C^k$ and $\hat v \in C^k$ for
some $k\ge 2$. 
We can prove convergence of the second derivative of the 
second--order self--energy 
$\Sigma_\two$ in two dimensions {\it only if the 
derivative is taken tangential to $S$}. For derivatives taken 
in the direction $\rh$ transversal to $S$, or with respect to $p_\zer$, 
we show that in $d=2$, the second derivative is at most logarithmically 
divergent. 
The calculations in \quref{F} indicate that this logarithm is indeed there
in two dimensions, i.e.\ that
$$
\Sigma_\two (p_\zer,\p) \sim {p_\zer}^2 \log |p_\zer|
.\EQN\Siglog $$
Stated differently, at positive temperature $T>0$ (which
provides a natural infrared cutoff), the behaviour of $\Si_\two$
is $T^2 \log T$. 
The extra logarithm played a role in the discussion about the 
existence of two--dimensional Fermi liquids \quref{F}.
Our results imply that this logarithmic singularity is (if it exists) 
not an obstruction to the  perturbative solution of \queq{\Wltfrml}, 
{\it because we are doing renormalization using the more regular
function $K$ instead of $\Si$ itself}. 
At present we do not know
a way of making the skeleton expansion, where one subtracts 
$\Si$, not only $K$, rigorous in $d=2$. 
We explain these problems further in the next section.  

The statement that $\Si \in C^{1,\ga}$ for any $\ga <1$ also
holds for the full perturbative self--energy, by 
Theorem \bestvol. We did not state it in Theorem \zweireg\ because the proof,
although an easy combination of Theorem 2.46 of \FST\ and 
the methods developed in Section 3.4, requires familiarity
with the tree expansion. We shall give it in \THREE. 

As stated in Theorem \zweireg, in three dimensions, the self--energy
is $C^2$ in $\p$, so there is no logarithm in the second derivative
with respect to $\rh$. Thus for $d\ge 3$, the skeleton expansion
can be made rigorous by our results.
One can also combine the methods of Chapter
3 of \FST\ and the ones developed here to show that in the $p_\zer$--%
dependence of $\Sigma$, the logarithm of \queq{\Siglog} is absent
in three dimensions, i.e., that $\Si$ is $C^2$ also in $p_\zer$. 

Although elementary, our detailed analysis of the second order counterterm is rather
subtle, and a generalization to all two--legged 1PI graphs (those 
contributing to $\Sigma$ and $K$) 
gives a system of equations for the singularities
that looks rather hopeless. Fortunately, a generalization to all two--legged 1PI graphs
is not required.
In \THREE, we give a new graph classification which isolates 
the only graphs that require the detailed analysis done here,
and that can exhibit behaviour as in \queq{\Siglog}
for $d=2$. These graphs constitute, in a sense to be specified
later, the (generalized) random--phase approximation 
contributions $\SiRPA$, $\KRPA$ 
to the self--energy and the counterterm.
See Chapter 4.
For these graphs, a detailed analysis of singularities is possible
and we show 

\The{\RPAreg}{\Thsty Let $\SiRPA$ be the RPA self--energy and $\KRPA$
the RPA counterterm.
\leftit{$(i)$}
If $d=2$ and $e(-\p)=e(\p)$ for all $\p$, 
there exists $\hxp > 0$ such that
if $0 \le \hxp' \le \hxp$, $k\ge 2$
and if \AOne{k,\hxp'}, \ATwo{k,\hxp'}, \AThr, and \AFiv\ hold, 
then $\KRPA$ is a $C^{k,\hxp'}$ function of $\p$. 
$\hxp'=0$ is allowed.
\leftit{$(ii)$}
If $d=2$, there exists $\hxp > 0$ such that
if $0 < \hxp' \le \hxp$
and if \AOne{2,\hxp'},\ATwo{2,\hxp'},\AThr--\AFiv\ and \AFoup\ hold, then 
$\KRPA$ is $C^{2,\hxp'}$ in $\p$. In this case, the condition
$e(-\p)=e(\p)$ for all $\p$ would force a violation of \AFoup.
$\hxp'=0$ is not allowed.
\leftit{$(iii)$}
If $d \ge 3$, there exists $\hxp > 0$ such that
if $0 \le \hxp' \le \hxp$ and \AOne{2,\hxp'}, \ATwo{2,\hxp'}, and \AThr\
(and, for asymmetric $e$, \AFou) hold,
then $\KRPA$ and $\SiRPA$ are $C^{2,\hxp'}$ in $\p$.
$\hxp'=0$ is allowed.
 
}

\ssni
Note that, as in Theorem \zweireg, the self--energy is shown to be
$C^2$ in $d \ge 3$. In two dimensions, one expects (both from the bounds
we derive and calculations in the literature) that RPA graphs also 
produce a behaviour as in \queq{\Siglog}. Note also that for asymmetric
$e$, we only prove $K \in C^2$, not $C^k$ with $k \ge 3$. The detailed 
analysis done in Chapter 4 suggests that the third derivative of $\KRPA$ 
may actually not converge if $e$ is asymmetric.

\ssni
The extension of Theorems \zweireg\ and \RPAreg\ to 
the exact self--energy and the exact $K$ is proven in \THREE.
\ssni
Thus, in any fixed order $r$ of perturbation theory, we can start 
to look for solutions of \queq{\Wltfrml} by iteration. 
The uniqueness theorem proven in \FST\ then guarantees that if 
the iterative sequence has an accumulation point, 
it converges. But our bounds are good enough to prove convergence 
directly, so that one has a solution of \queq{\Wltfrml}.

\sect{Consequences}
\noindent
To put these results into context, we now discuss how one gets from
the solution of \queq{\Wltfrml} to a rigorous
version of what is usually called 
`self--consistent renormalization'. 
Given a free model with band structure $E$ (including the chemical 
potential $\mu$), we want to do a  formal power series expansion
expansion in $\la $ for the interacting model given by the generating 
functional for connected amputated Green functions
$$
e^{\cG (\ps,\psq)} =\frac{1}{Z_E} \int d\mu_{C_E} (\ch,\chq ) 
e^{-\la V(\ps+\ch,\psq+\chq)}
\eqn $$
with $d\mu_{C_E}$ the Grassmann Gaussian measure with propagator 
$1/(ip_\zer-E(\p))$. The constant $Z_E$ normalizes $\cG (0,0) = 0$. 
This may be defined as the 
limit $\be \to \infty$ of the grand canonical ensemble with 
partition function $\tr e^{-\be(H-\mu N)}$. 
Denote the solution to equation \queq{\Wltfrml} by $e= F(E,\la )$.
Let $\ka (E, \la ) = K(F(E,\la ), \la )$. Then $E=e+ \ka= e + K(e,\la )$.
We now move $K$ from the Gaussian measure into the interaction 
by the standard Gaussian shift formula, so that $K$ now acts as a counterterm.
Since $E=e+K$, this leaves $e$ in the propagator, and
$$\eqalign{
e^{\cG (\ps,\psq)}  & = 
\frac{1}{Z_e} \int d\mu_{C_e}(\ch,\chq)e^{-(\chq,K\ch)}\; 
e^{-\la V(\ps+\ch,\psq+\chq)} \cr
& = \frac{1}{Z_e}  e^{-(\psq,K\ps)}
\int d\mu_{C_e}(\ch,\chq)\; 
e^{-(\psq+\chq,K(\ps+\ch))}e^{-\la V(\ps+\ch,\psq+\chq)} 
e^{(\psq,K\ch)+(\chq,K\ps)} 
.\cr}\EQN\rencG $$  
The change in normalization factor from $Z_E$ to $Z_e$ is irrelevant
for any correlation function, and the extra source terms in the integrand 
just modify the external legs in a trivial way. 
Effectively, external vertices are not renormalized.
{\it Because $e$ is given by the solution to \queq{\Wltfrml}, the model 
has not been changed in any way}. What has changed is our way of looking at it.
Splitting $E=e+\ka $ means going from the bare to the interacting Fermi surface.
After that, the other interaction effects can be calculated at fixed surface.
In fact, everything is already arranged such that the expansion 
of \queq{\rencG} in $\la$ is precisely the renormalized 
expansion of \FST. All theorems of \FST\ apply. In particular, 
there are no infrared divergences.
 
In other words, the interaction effects are calculated 
in two steps: first, we determine the interacting Fermi surface, then the
self--energy and the $n$--point functions.

It is often stated that the renormalization problem can be dealt
with by doing a skeleton expansion in which on all lines, the
free propagator is replaced by the interacting propagator
(sometimes, this is also called `self--consistent renormalization').
That is, one calculates the values of skeleton diagrams, using
for the $r^{\rm th}$ order in $\la$
the interacting propagator $(ip_\zer-e(\p) -\Si_{r-1}(p))^{-1}$,
where $\Si_{r-1}$ is the self--energy up to order $r-1$ in $\la$, 
instead of the free propagator $(ip_\zer-e(\p))^{-1}$.
However, a regularity problem similar to \queq{\Wltfrml}
also arises in this procedure: to show that the values of 
skeleton diagrams with 
propagators containing $e+\Si $ are well--defined, one has to assume that 
$\Si $ has the same regularity
properties as $e$. However, $\Si $ is not a function
one is free to choose. One has to show regularity of the self--energy. 
This regularity 
problem is harder than the one necessary to invert \queq{\Wltfrml},
because $\Si$ is even less regular than $K$, and in fact,
it may not have a solution in $d=2$ dimensions. 
Let us be more specific about why regularity of $\Si$ (or $K$) is needed. 
To show that the most 
elementary power counting estimates hold, one needs that the volume
of a shell of thickness $\veps$ around the Fermi surface is bounded by a 
constant times $\veps$. This can be shown for the surface
$S(e) = \set{\p}{e(\p)=0}$ if $e\in C^1$, and if the gradient of $e$ does not
vanish on $S(e)$ (see \ATwo{}).
To have the same statement for the surface $e(\p)+\Si(0,\p)=0$, 
one also has to show that $\Si(0,\p)$ is $C^1$ in $\p$.
But to show that $\Si$ is $C^1$, 
one needs upper and lower bounds on the curvature
of $S(e)$ -- which requires $e\in C^2$ -- already in second order
(if the Fermi surface has flat sides, or if the system is 
one--dimensional, $\Si$ is typically not $C^1$). Since the second order 
$\Si$ appears on `interacting propagators' of higher order graphs,
one needs $\Si \in C^2$. It was  mentioned above that 
most likely, $\Si \not\in C^2$, because of the logarithm in
\queq{\Siglog}, so proceeding this way may be impossible. 
Instead we take a counterterm function $K$ which agrees with
$\Si$ only on the Fermi surface. Tangential derivatives of $K$
agree with tangential derivatives of $\Si$, but normal derivatives
do not agree. A priori, proving 
$K\in C^2$ could require $e \in C^3$, and, hence in the next order
$K \in C^3$, and so on. However, this sequence
stops already at $k=2$: we show that $e \in C^2$ implies $K \in C^2$.
This is equivalent to saying that the iteration for the solution of \queq{\Wltfrml}
stays in a fixed set of $C^2$ band structures.

In renormalization group studies, a wave function renormalization
is often introduced. For instance, in one--dimensional systems,
it is crucial for taking into account the anomalous dimension 
correctly \quref{BGPS,BM}.
In dimension  $d\ge 2$, our results show that a wave function
renormalization is {\it not} necessary for renormalizing formal 
perturbation theory, 
although it may  be convenient.
One can easily retrieve the perturbative wave function renormalization,
and show that it is finite to all orders, 
in the usual way from the self--energy:
we proved in \FST\ that $\Si \in C^{1}$. 
Moreover, our assumption that $\hat v(-p_\zer,\p) = 
\overline{\hat v (p_\zer,\p)}$ (see \AOne{}) implies that 
$\Si (-p_\zer,\p) = \overline{\Si(p_\zer,\p)}$, so that 
$\Si (0,\p) \in \R$ for all $\p$. Since $\Si$ is $C^1$ in $p$,
this implies $(\del_\zer \Si )(0,\p) \in i\R$ and 
$\nabla \Si(0,\p) \in \R$ for all $\p$. Recalling that 
by Theorem 1.2 $(iii)$ of \FST, $\Si(0,{\bf P}(\p))=0$
(here ${\bf P}$ denotes the projection onto the Fermi surface),
we get by Taylor expansion 
$$
\Si(p) = p_\zer (\del_\zer\Si)(0,{\bf P}(\p)) + 
(\p - {\bf P}(\p)) \cdot \nabla \Si(0,{\bf P}(\p))
+ \tilde\Si(p)
\eqn $$
where $\tilde \Si$ vanishes faster than linear as $p_\zer$ approaches
$0$ and $\p$ the Fermi surface $S$. By the result that$\Si \in
C^{1,\ga}$ for all $\ga < 1$, we know that $\tilde \Si$ vanishes almost 
quadratically in that limit. Combining the linear term 
in $p_\zer$ with the $ip_\zer$ of the free
propagator, we get the usual formula
$$
Z(\p) = 1+i (\del_\zer\Si)(0,{\bf P}(\p)) \in \R
\eqn $$
for the prefactor of $ip_\zer$ (whether one calls $Z$ or $1/Z$
the wave function renormalization is, of course, a matter of
convention).
The $\nabla \Si$ term gives the usual correction to the 
Fermi velocity. As mentioned above, the shift from the free to
the interacting Fermi surface is given by $K(\p)$. 

What does this perturbative analysis imply for the full, nonperturbative 
model? This depends on $e$. If $e(-\p)=e(\p)$, the perturbation series
is not convergent, and thus defines only formal power series,
because of the factorial growth 
of the contribution of the ladder diagrams. 
It is well-known that the particle--particle ladder graphs really
produce such factorials if $e(-\p)=e(\p)$.
It is proven very generally in \FST\ that for many-fermion models,
only the ladders can produce those factorials. Thus there are no other 
graphs whose contribution could cancel them, and they prevent 
convergence of the renormalized perturbation series. 
(Do not confuse this with the infrared divergences discussed above.
In the renormalized expansion $\Si = \sum\la^r \Si_r$, all $\Si_r$ are finite
$C^1$ functions, but the convergence radius of the power series 
in $\lambda$ is zero). 
An improvement of power counting (called loop improvement), 
and the corresponding statement
that the ladders give the only singular contribution to the four-point
function was also stated in \quref{BG} for the case where the Fermi surface
is a sphere. 
It is proven in \quref{FMRT} by implementing the Pauli principle that
at least for spatial dimension $d=2$ 
there are no other obstacles to the convergence of the renormalized 
perturbation series. 
It is proven in \quref{FT2} that for an attractive
interaction, the perturbative RG flow, which is dominated 
by the ladder diagrams,
leads to a symmetry--breaking fixed point given by Cooper pairing.

All this changes very much for the class of $e$ which violates
the symmetry $e(-\p) = e(\p)$ (the precise class is specified 
in \AFou\ and \AFoup),
to which a big part of our analysis is devoted. 
This asymmetry suppresses the Cooper instability, 
i.e.\ it removes the factorial growth of the ladders.
By the result proven in \FST\ that the contribution from all other graphs
to the four--point function is nonsingular, one may suspect 
that perturbation theory converges.  
It is proven nonperturbatively in \quref{FKLT} in $d=2$ that the renormalized
perturbation expansion converges in these models. Thus models with 
 $e(-\p) \neq e(\p)$ can be proven to be Fermi liquids. 
The results of \FST\ and of the present paper provide the perturbative 
part of this proof: in \FST\ we have shown that the renormalized Green 
functions are all finite, and that only ladder subdiagrams produce factorial
growth in the value of individual diagrams. In the present paper and \THREE,
we prove that renormalization does not change the model. 

While differentiability of the self--energy may look like a rather
technical problem at first sight, we should like
to remark that a self--energy that is not $C^1$ is a common feature
of various proposals for non--Fermi--liquid behaviour in two
dimensions (see \quref{S} for further discussion), so that
the differentiability issue does have a physical significance.

We end this introduction with some more detailed remarks.
As anybody who actually reads the proofs will see, being in a
$C^2$ (or $C^{2,\hxp}$) class of functions poses some rather severe
technical restrictions, which show up in various technical details.
For instance, the proof of Theorem \bestvol\ requires a version
of the Morse lemma for $C^2$ functions. Since this
lemma is not completely standard, we prove it in Appendix A.
Moreover, the critical point analysis 
involves the antipode $\ve{a}(\p)$ of $\p \in S$, 
which is the point on $S$ where $\nabla
e(\ve{a}(\p))$ is antiparallel to $\nabla e (\p)$. For a $C^k$ surface,
$\ve{a}$ is in general only a $C^{k-1}$ function. It requires careful
arguments to show that this does not cause a deterioration of  the differentiability
properties of $K$. 

To do renormalization without changing the model, we may
use the renormalized expansion only after inverting
\queq{\Wltfrml}. This restricts us to a class
of $C^{2,\hxp}$ functions even if the starting $E$ had a higher
degree of differentiability, because it is the differentiability
of $e$ that enters the bounds. 
One consequence of this is that the higher--tangency--argument 
used in \quref{FKLT} to show the absence of the Cooper instability
(i.e., the boundedness of the particle--particle ladders)
does not apply. 
We define a notion of minimal change
of the curvature (similar to the definitions in {\bf A3} of \FST)
for our class of $C^{2,\hxp}$ functions in Assumption \AFoup\
and show in Appendix C 
that the particle--particle ladders are bounded under
this weaker condition. We also need this 
for the regularity proofs. It is at this point that we actually
need the extra H\" older continuity stated in the above Theorems.
In particular, it is the reason why $h'=0$ is not allowed in
Theorem \RPAreg\ $(ii)$. In a $C^2$ class of functions, 
a natural definition of a minimal change in curvature would be
impossible. 
  
Finally, we note that although the filling restriction \AFiv\
that we imposed
for $d=2$ may seem peculiar, numerical results indicate that in its absence, 
the behaviour of the self--energy is indeed different. These effects 
may also be of physical interest.

We give the definition of our class of models in Chapter 2 
and prove Theorem \zweireg\ in Chapter 3, and Theorem \RPAreg\
in Chapter 4.  
Appendix A contains the $C^2$ Morse Lemma, 
and Appendix B contains the proof of Theorem \bestvol.  
Appendix C contains the proof of the one--loop volume bound needed 
to prove Theorem \RPAreg\ in case $e$ is asymmetric. 
In Appendix D, we prove regularity properties of the scale zero
effective action, which relates the theory without a cutoff in
$p_\zer$ to the one with a cutoff.

\vfill\eject

\chap{Definitions and Assumptions}
\pni
We denote the $\veps$--neighbourhood of a set $A$ 
$$
U_\veps (A) = \{ y: \exists x \in A \hbox{ with } \abs{y-x} < \veps\}
\eqn $$
and, as in \FST, we denote the norm 
$$
\abs{f}_k = \sup\limits_{p \in \R\times \cB} \; \sum_{\abs{\al } \leq k}
\abs{D^\al f(p)}
\eqn $$
where  $\al =(\al_\zer, \ldots , \al_d) \in \Z^{d+1}$,
$\al_i \geq 0$ for all $i$, 
is a multiindex, 
$\abs{\al} = \sum\limits_{i=0}^{d} \al_i$, and 
$\del^\al = \left({\del \over \del p_\zer}\right)^{\al_\zer} \ldots 
\left({\del \over \del p_d}\right)^{\al_d}$. 

Let $0 < \hxp \le 1$.
We denote the space of $C^k$ functions on a set $\Om$
whose $k^{\rm th}$ derivative is $\hxp$--H\"older continuous 
by $C^{k,\hxp}(\Om)$, and use the norm
$$
\abs{f}_{k,\hxp} = \sup\limits_{p \in \R\times \cB} \; \sum_{\abs{\al } \leq k}
\abs{D^\al f(p)}
+\max\limits_{\al: |\al| =k}
\sup\limits_{x,y\in \Om\atop x\ne y} \frac{\abs{D^\al f(x)-D^\al f(y)}}{\abs{x-y}^\hxp} 
.\eqn $$
For $h=0$, we define $C^{k,0}(\Om)=C^k(\Om)$.

\sect{Band structure and interaction}
\pni
Our analysis takes place in momentum space, 
given by $\R \times \cB$, where $\cB$ is the 
torus $\R^d/\Ga^\#$, with $\Ga^\#$ the dual lattice to the position 
space lattice $\Ga$, e.g.\ for $\Ga = \Z^d$, $\cB = \R^d/2\pi\Z^d$.
Let $\cF\subset \R^d$ be a fundamental domain for the action of the
translation group $\Ga^\#$, and $\openkrnl{\cF}\subset \R^d$ its interior, 
e.g.\ for $\Ga^\# = 2\pi \Z^d$, and $\cB = \R^d/2\pi\Z^d$,
$\cF = [-\pi,\pi)^d$ and $\openkrnl{\cF} = (-\pi,\pi)^d$.
We could also take $\cB$ to be a bounded subset of $\R^d$.

The one--electron problem provides a band structure $e(\p)$ which 
enters the propagator associated to lines \ifshowfigs\hskip 4pt\hbox to 0pt{\hss\includegraphics{arrow.ps}\hss}
\hskip 36pt plus 5pt\fi. We include the chemical potential $\mu$ in $e$.
For example, for the free electron gas, $e(\p)=\frac{\p^2}{2m}-\mu$.
The interaction is given by the vertex 
\herefig{vertex}
where $\de^\#$ is a $\de $ function on $\R$ for $p_\zer$ and 
a $\de$ function on $\cB$ (i.e.\ one on $\R^d$ modulo $\Ga^\#$) 
for the spatial part $\p$. The vertex function has the symmetry
$$
\langle p_2\; p_4\mid V \mid p_3 \; p_1\rangle =
\langle p_4 \; p_2 \mid V \mid p_1 \; p_3 \rangle
.\eqn $$
For simplicity, we assume that 
$$
\langle p_2\; p_4\mid V \mid p_1 \; p_3\rangle = \hat v(p_\two-p_\one)
.\EQN\twobodyint $$
We make the following assumptions on the interaction $\hat v$, 
the band structure $e(\p)$ and the Fermi surface $\ S=\set{\p\in\cB}{e(\p)=0}$.
For some $k \geq 2$,
\ssni
\hypit{\AOne{k,\hxp}} {\Lesty $\hat v \in C^{k,\hxp}(\R \times \cB,\C)$ 
with all derivatives of 
order at most $k$ uniformly bounded on $\R\times\cB$, and
$\hat v$ satisfies
$$
\hat v(-p_\zer , \p) = \overline{\hat v(p_\zer,\p)}
.\EQN\evenv $$
There is a bounded real--valued $C^{k,\hxp}$ function $\tilde v(\p) $ such that
$$
\lim\limits_{p_\zer \to \infty} \hat v(p_\zer,\p) = 
\tilde v(\p)
\EQN\contildev $$
and there are $\al > 0$, $K_\zer>0$,  and  $\pi_\zer>0$ such that 
$$
\forall \abs{p_\zer} \ge \pi_\zer \forall \p \in \cB : \qquad 
\abs{\hat v (p_\zer,\p) - \tilde v(\p)} \le K_\zer \; \abs{p_\zer}^{-\al}
\EQN\UVdecay $$

}
\ssni
\queq{\evenv} implies that the counterterms $K$, defined below,
are real--valued, $K(\p) \in \R$ for all $\p\in \cB$. \queq{\evenv}
and \queq{\contildev } also imply that
$\lim\limits_{p_\zer \to -\infty} \hat v(p_\zer,\p) = \tilde
v(\p)$.

The decay condition
\queq{\UVdecay} assures us that the scale zero effective action,
defined below, gives rise to $C^{k,\hxp}$ vertex functions, and that
therefore the ultraviolet (large $p_\zer$) part of the problem
can be separated
from the infrared (small $p_\zer$) part by the semigroup structure
of the flow of effective actions. If the interaction is instantaneous,
i.e., if $\hat v$ is independent of $p_\zer$, 
\queq{\UVdecay} holds trivially.
\ssni
\hypit{\ATwo{k,\hxp}} {\Lesty  $e\in C^{k,\hxp}(\cB,\R)$ and 
$\nabla e(\p)\ne 0$ for all $\p\in S$. }
\ssni
Assumption \ATwo\ implies that $S$ is a $C^k$ submanifold of $\cB$. 
Since $k\geq 2$, its curvature $\ka$ is therefore well--defined.
Since $\cB$ is compact, $e$ is continuous, and $S=e^{-1}(\{ 0\})$ 
is a preimage of zero, $S$ is compact.
\ssni
\hypit{\AThr}  {\Lesty $\ka$ is strictly positive everywhere. }
\ssni
In $d>2$ this is meant in the matrix sense. \AThr\ implies that
$S$ bounds a strictly convex set. In two dimensions, $S$ is a simple
closed curve. 
Let $\ve{n} = \frac{\nabla e}{|\nabla e|}$ be the unit normal to $S$. 
Strict convexity implies that the equation 
$$
\ve{n} (\ve{a}(\p)) = - \ve{n}(\p)
\eqn $$
has, for any $\p \in S$, a unique solution $\ve{a}(\p)\in S$. 
Necessarily, $\ve{a}(\p) \neq \p$. We call $\ve{a}(\p)$ the antipode of $\p$. 
Since $\ve{n}$ is $C^{k-1}$, $\ve{a}\in C^{k-1}(S,S)$. 

We shall not assume in general that $e(-\p ) = e(\p)$. 
We call a band structure $e$ symmetric if
\ssni
\hypit{\SYmm} {\Lesty For all $\p \in \cB$, $e(-\ve{p} ) = e(\ve{p} )$.}
\ssni
and asymmetric otherwise. If \SYmm\ holds, $\ve{a}(\p)=-\p$.
In the asymmetric case, we assume
\ssni
\hypit{\AFou} {\Lesty For all curves $t \mapsto \p(t)$ in $S$, 
$$
\abs{1-\abs{\frac{\del\p(t)}{\del t}}^{-1}\;
\abs{\frac{\del \ve{a}(\p(t))}{\del t}}}
\le \frac{1}{8}
\EQN\upplo $$

}
\ssni
If $e$ is symmetric, \AFou\ holds trivially by \SYmm\ (the left
hand side of \queq{\upplo} is zero). We shall give a restatement
of \AFou\ in terms of the curvature below. Essentially, the 
curvature at the points $\p$ and $\ve{a}(\p)$ must not differ
by too much. 
\ssni
For asymmetric $e$ and $d=2$, we need another assumption, \AFoup, which
we state after we have introduced coordinates, because it is
more easy to state, and understand, after some preparations.
This assumption forces the curvature at almost all points $\p$
on the Fermi surface to differ from the curvature at the antipode
$\ve{a} (\p)$.  
\ssni 
The last assumption is
\ssni
\hypit{\AFiv} {\Lesty 
$\set{u\p+v\q}{\p,\q\in S, u,v=\pm 1}\subset \openkrnl{\cF}$.}
\sni
Assumptions \AOne\ and \ATwo\ are identical to the assumptions
{\bf A1} and {\bf A2} of \FST, save for the condition \queq{\UVdecay}
on $\hat v$ which we use for the ultraviolet part of the problem
(which was discarded in \FST), and the extra (optional) H\" older
continuity. They are satisfied for interactions
that decay fast enough, and under natural conditions on the one--particle
problem. For a detailed discussion of their meaning, see Section 1.5 in \FST.
Assumption \AThr\ implies {\bf A3} of \FST\ (the class of Fermi surfaces 
considered there was much bigger than that of positive curvature). 
Thus all theorems of \FST\ apply to this situation. 

Assumption \AFou\ is a technical assumption
that ensures that the locations of the singularities of 
certain Jacobians depend
continuously on the external momenta. The specific number
$\sfrac{1}{8}$ in \queq{\upplo} is certainly not optimal.

Assumption \AFoup\ is needed in 
the $d=2$ asymmetric case (where typically $\ve{a}(\p)\ne -\p $).
That the curvatures
differ at almost every $\p \in S$ and its antipode 
is needed in a volume bound for the particle--particle
bubble (see Appendix C); this is the bound that also implies
that the Cooper channel is turned off in the asymmetric case.

Assumption \AFiv\ restricts the density $n$ to be small enough,
to avoid certain umklapp processes. E.g. in the 
Hubbard model, where $e(\p) = -2(\cos p_\one + \cos p_\two) - \mu$,
this is fulfilled for densities $n < 0.369$
(here $n=1$ is half-filling).
We shall need Assumption \AFiv\ only to prove statements about
the second--order graph and the RPA graphs. 
In particular, the optimal volume improvement
bound (and the bounds for values of large graphs in \THREE\
do not require \AFiv).
Thus a detailed investigation of the role of \AFiv\ can be done
essentially in second order. 
The explicit role of \AFiv\ in our proofs will be discussed 
in Section 3.3.

\sect{Some elementary consequences}
\pni
\ATwo\ and \AThr\ 
imply that there are $\veps_\one >0$ and 
$g_\zer > 0$ such that 
$$
\forall \ve{p} \in U_{\veps_\one} (S) : \quad 
\abs{\nabla e (\ve{p} )} \geq g_\zer 
.\eqn $$
The filling restriction \AFiv\ has the following consequence.

\Lem{\byAfive} {\Lesty Let $S$ fulfil \AThr\ and \AFiv. Then
\leftit{$(i)$} If $\p,\q,\r \in S$ and $2\q=\p+\r$, then $\p=\q=\r$
\leftit{$(ii)$} For a given $\p_\two \in S$, the equation 
$$
\p_\one - \p = \ve{a}(\p_\two ) - \p_\two, \quad \p_\one,\p \in S 
\EQN\lngstchrd $$
has only the solution $\p_\one= \ve{a}(\p_\two)$, $\p=\p_\two$.}
\Proof 
By \AFiv, both sides of the equations in $(i)$ and $(ii)$ are vectors
in $\openkrnl{\cF}$, which is an open convex subset of $\R^d$. 
Therefore, to determine their solutions, 
we may consider $S$ as a subset of $\R^d$ instead of $\cB$.
$(i)$ If $2\q=\p+\r$ then the three points
$\ \q,\ \r=2\q-\p=\q+(\q-\p)\ $ and $\ \p=\q-(\q-\p)\ $ are collinear. 
But $\p,\q,\r\in S$, so by strict convexity collinearity can hold only
if $\p=\q=\r$.
$(ii)$ It is obvious that $\p_\one= \ve{a}(\p_\two)$, $\p=\p_\two$ is a solution
of \queq{\lngstchrd}. 
We only have to show that there is no other solution.
Let $\r=\sfrac{\p_\two - \ve{a}(\p_\two)}{|\p_\two - \ve{a}(\p_\two)|}$.
We show uniqueness by proving that the chord from 
$\p_\two$ to its antipode $\ve{a}(\p_\two)$ is strictly longer than any other 
chord of $S$ in direction $\r$. This is sufficient because for $\p$ and 
$\p_\one $ to be a solution of \queq{\lngstchrd}, the chord given by
$\p_\one-\p$ must point in direction $\r$, 
i.e.\ $\p_\one-\p=|\p_\two - \ve{a}(\p_\two)|\; \r$. 

Consider the family of lines $L_\p=\set{\p-t\r}{t\in \R}$, 
parametrized by $\p \in S$. As $\p$ is varied, the line
$L_\p$ slides over $S$, and thus defines a chord in direction $\r$.
All chords in direction $\r$ are produced by an $L_\p$ for some $\p$.
Consider the tangent planes to $S$ at $\p_\two $ and $\ve{a}(\p_\two)$.
They are parallel. Hence any line segment parallel to $\r$ whose 
endpoints lie strictly between the two tangent planes has length
strictly less than $|\p_\two - \ve{a}(\p_\two)|$.
Because $S$ is strictly convex, it remains on one side of each of the 
tangent planes, and it intersects the tangent planes only at 
$\p_\two $ and $\ve{a}(\p_\two)$. Thus every other chord in direction 
$\r$ is shorter than that from $\p_\two $ to $\ve{a}(\p_\two)$. 
\endproof 

\subsect{Radial and angular coordinates}
For $r_\zer > 0$, let $\cT = (-2r_\zer ,2r_\zer ) \times S$.
Then there is an $r_\zer>0$ and a $C^k$--diffeomorphism
$\phi :\cT \to \phi(\cT )=\cU (S) \subset \cB$,
$(\rho , \si ) \mapsto\ve{p} = \phi (\rho, \si )$
such that $e( \phi (\rho, \si ))=\rho$, and such that 
$$
|\nabla e (\ph(\rh,\si))| \ge g_\zer > 0
\EQN\gzerinit $$ 
for all $|\rh|<r_\zer$ and all $\si$.
$\ph$ is constructed explicitly in Lemma 2.1 in \FST, 
using the integral curves of a $C^\infty$ vector field $u$ that 
is transversal to $S$ in the sense that
$$
u(\p) \cdot \nabla e(\p) \ge u_\zer \ge \frac{g_\zer}{2}
.\EQN\uzerinit $$
We shall assume that $\ph$ is given by this
specific construction and use its properties, and therefore 
ask you to recall it from \FST. 
For ease of notation, we shall also write 
$\ve{p} (\rho, \si ) $ for $\phi (\rho , \si )$, 
and $\p (\si ) $ for $\p (0,\si )$.
Since $\p $ is $C^{k}$ in $\rh$ and $\si $, the Jacobian
$$
J(\rh,\si) =\abs{ \det \ve{p}^\prime (\rho, \si )}
\EQN\JacobianJ $$ 
of the corresponding
change of variables $\ve{p} \to (\rho, \si ) $ is $C^{k-1}$.

We also denote $S_\rh = \set{\p \in \cB }{e(\p ) =\rh }$. For $\rh $ small 
enough, $S_\rh $ is also strictly convex, and we denote the antipodal map 
on $S_\rh $ by $\ve{a}_\rh$. It is defined for $\p \in S_\rh$ by 
$$
\ve{n}(\ve{a}_\rh(\p))=-\ve{n}(\p) \qquad \hbox{ and } 
\qquad \ve{a}_\rh (\p) \in S_\rh
\eqn $$

We assume that $r_\zer $ is chosen such that
strict convexity of $S_\rh $ holds for all $\abs{\rh}\leq r_\zer $.
\sni
$r_\zer$ will be chosen smaller in what follows, it depends, however,
only on $\abs{e}_2$ and the geometry of the Fermi surface,
i.e., on $g_\zer$ and the constant $w_\zer$ defined in this section
(which is related to the minimal curvature of $S$).
The antipodal map $\ve{a}_\rh$ induces a map $a_\rh(\si)$ by
$$
\p(0,a_\rh(\si))=\ve{a}_\rh(\p(0,\si))
\eqn $$
For $\rh=0$, we denote $a(\si)=a_\zer(\si)$.

There is a $C^k$--diffeomorphism that maps $S$ to the unit sphere
$S^{d-1}$, $\si \mapsto \th$. We use this variable $\th$ in the
following and keep the same notation for the map $(\rh,\th) \to
\p(\rh,\th)$, the Jacobian $J$ and
the antipode $a$. The diffeomorphism from $S$ to $S^{d-1}$
depends on $e$, but it is
a $C^k$--diffeomorphism because $S$ is a $C^k$ surface, and 
the Jacobian and its derivatives are bounded uniformly for all
$e$ satisfying \ATwo{k,0}\ and \AThr. 

\subsect{Two Dimensions}
For $d=2$, the variable $\theta \in S^1$ is simply an 
angular variable. The map $\ph $ is
$(\rho, \theta ) \mapsto \ve{p} (\rho, \theta )$. It is 
$C^{k}$ in both variables.
The vector $\dth\ve{p}$ is nonzero at all points in $\cT$ 
because $\ph$ is a diffeomorphism.
By definition of the coordinates $\rh$ and $\th$ (see Lemma 2.1 in \FST),
the variable $\th$ is constant on the integral curves of $u$, which 
are fixed independently of $\rh$. 
Therefore, if we now regard $\th$ as a variable in $\R$, 
the period of the map $\th \mapsto \p(\rh,\th)$ is the 
same for all $\rh$ with $|\rh|<r_\zer $. 

Define the matrix $e''(\p)$ by  
$$
\left(e''(\p)\right)_{ij} = \frac{\del^2 e}{\del p_i\del p_j}(\p)
,\EQN\edpdef $$
let $(a,b)=\sum_i a_i b_i$ and let 
$$
w(\p) = (\del_\th \p , e''(\p) \del_\th\p )
.\EQN\wpdef $$
At fixed $\rh$, $\p(\rh,\th)$ is a parametrization of the curve $S_\rh$. 
Denote the unit tangent by 
$$
\ve{t} (\p) = \frac{\del_\th\p}{|\del_\th\p|}
\eqn $$
and define the arclength as $s(\rh,\th) = \int\limits_0^\th d\vth
|\del_\th \p(\rh,\vth)|$. Then
$$
\del_\th^2 \p (\rh,\th) =\frac{\del^2 s}{\del\th^2}\; \ve{t}(\p(\rh,\th))
\pm \ka(\rh,\th) \left(\frac{\del s}{\del\th}\right)^2\; \ve{n}(\p(\rh,\th))
.\EQN\frene $$
The sign is $-1$ if $e(\p)< 0$ inside $S$ and $+1$ otherwise. 

\Lem{\wBound}{\Lesty By \ATwo, for all $|\rh|<r_\zer$ and all $\th$,
$$
\ka(\rh,\th) |\nabla e(\p(\rh,\th))| = 
-\frac{w(\p(\rh,\th))}{|\del_\th\p(\rh,\th)|^2}
.\eqn $$
By \AThr, there is $\ka_\zer > 0$ such that $\ka(0,\th) \geq 2\ka_\zer >0$
for all $\th$. Thus $r_\zer > 0$ can be chosen so small that for all 
$|\rh|<r_\zer$ and all $\th$, $\ka(\rh,\th) \ge \ka_\zer$, and there is 
$w_{min}> 0$ such that for all $|\rh|<r_\zer$ and all $\th$
$$
\abs{w(\p(\rh,\th))} \ge w_{min} |\del_\th\p(\rh,\th)|^2
.\eqn $$
Moreover, $|\del_\th^2 \p(\rh,\th)|\ge\sfrac{w_{min}}{|e|_\one}
|\del_\th\p(\rh,\th)|^2$.
}
\Proof $e(\p(\rh,\th))=\rh$ for all $\th$, so 
$$\eqalign{
\nabla e (\p(\rh,\th))\cdot\del_\th\p(\rh,\th) & = 0 \cr
\hbox{ and }\quad w(\p(\rh,\th))+\nabla e(\p(\rh,\th))\cdot
\del_\th^2\p(\rh,\th) &= 0
.\cr}\EQN\Trick $$
We insert \queq{\frene} and use $\ve{t}\cdot\ve{n}=0$ and 
$\sfrac{\del s}{\del \th}=|\del_\th \p|$, to get
$$
w(\p(\rh,\th)) \pm \ka(\rh,\th) |\del_\th \p|^2 \; |\nabla e(\p(\rh,\th))|
=0
.\eqn $$
Thus $\abs{w(\p(\rh,\th))} \ge w_{min}|\del_\th\p(\rh,\th)|^2$ holds with 
$$
w_{min}=\ka_\zer g_\zer 
,\eqn $$
and, by \queq{\Trick}, 
$$
|e|_\one\; |\del_\th^2 \p(\rh,\th)|\ge
|\nabla e(\p(\rh,\th)| \; 
|\del_\th^2 \p(\rh,\th)\cdot \ve{n}(\p(\rh,\th))|\ge
w_{min} |\del_\th \p(\rh,\th)|^2 
.\eqndprf\eqn $$
\sni
We transform to a new angular variable, which is a multiple of 
the arclength of the curve $\p(0,\th)$, so that in the new variables,
$\del_\th\p$ is of constant length for $\rh=0$, and  
normalized such that the 
period is $2\pi$. We give the argument in detail to show 
that no differentiability is lost by this change of variables.  
Let $|\del_\th \p (\rh, \th)|=v(\rh,\th)$. Since $v(\rh, \th) > 0$
for all $|\rh|\leq r_\zer $ and all $\th$, we can define a new variable 
$\tilde \th$ as follows. Let $\th_\zer $ be fixed and $P>0$. Define
$$
\tilde \th(\th) = P \int\limits_{\th_\zer}^\th d\vth\; v(0,\vth)
\eqn $$
$v(0, \th) $ is $C^{k-1}$, so the map 
$\th \mapsto \tilde \th $ is $C^k$. It is a diffeomorphism
because $v(0, \th) >0$, and $\tilde \th(\th_\zer )=0$. Denoting its
inverse by $\th (\tilde \th)$ and $\tilde \p (\rh, \tilde \th ) = 
\p (\rh, \th(\tilde \th))$, we have
$$
\frac{\del}{\del \tilde \th} \tilde \p (\rh, \tilde \th ) =
\frac{\del \th}{\del\tilde\th} \; \del_\th \p (\rh, \th(\tilde\th)) =
\frac{1}{P v(0, \th)} \del_\th \p (\rh, \th)
\eqn $$
so 
$$
\abs{\del_{\tilde \th} \tilde \p (0, \tilde \th )} = \frac{1}{P} 
\quad \hbox { for all } \tilde \th
.\eqn $$
Unless stated otherwise, we take the variable $\tilde \th$ and choose
$P$ such that the period is $2 \pi$, i.e.\ $\tilde \th \in \R/2\pi \Z$.
We also drop the tilde, so that in summary 
$$
\th \in \R / 2\pi \Z, \qquad \abs{\del_\th \p (0,\th)} = \frac{1}{P} 
.\eqn $$
Calling 
$$
w_\zer=\frac{w_{min}}{P^2}=\frac{\ka_\zer g_\zer}{P^2}
,\eqn $$
the statement of Lemma \wBound\ is: for all $|\rh|<r_\zer $ and all $\th$,
$$
\quad\abs{w(\ve{p} )} \geq w_\zer 
\EQN\wz$$
In terms of the new coordinate $\th$, 
the antipodal map $a$ on the Fermi surface satisfies
$$
\dth \ve{p} (0, a(\theta) ) = - \dth \ve{p} (0, \theta )
.\EQN\antidef $$  
$a$ is $C^{k-1}$ in $\th $.
If \SYmm\ holds, $a$ is $C^\infty$ because $a(\th) = \th+ \pi$ for all $\th$.
If $e$ is asymmetric, \AFou\ implies that for all $\th$,
$$
\frac{7}{8}\leq \frac{\del a}{\del \th } \leq \frac{9}{8}
.\EQN\antidel $$
In the symmetric case this is trivially true because $a(\th) = \th+ \pi$.
By \queq{\antidef},
$$
\del_\th^2\p(0,a(\th)) \; \frac{\del a(\th)}{\del \th} =
- \del_\th^2\p(0,\th)
.\EQN\deltwop $$
By choice of the coordinate $\th$, 
$\del_\th^2\p(0,\th) \propto \ve{n}(\p(0,\th))$, so 
$$
\frac{\del a}{\del\th}= 
\frac{|\del_\th^2\p(0,\th)|}{|\del_\th^2\p(0,a(\th))|}=
\frac{\ka(0,\th)}{\ka(0,a(\th))}
.\EQN\curvratio $$
Thus \AFou\ is simply a condition on the ratio of the curvatures 
at $\p$ and its antipode. 

\subsect{The assumption \AFoup}
For $d=2$, and $e$ not obeying \SYmm, we impose
\ssni
\hypit{\AFoup} {\Lesty
The function $\sfrac{\del a}{\del \th}$ 
obeys $\sfrac{\del a}{\del \th}=1$ only at finitely 
many points $\th^{(1)}, \ldots, \th^{(N)}$. There is $\de_\zer>0$
such that for all $k,l\in \nat{N}$: if $k \ne l$, 
$U_{2\de_\zer} (\th^{(k)}) \cap U_{2\de_\zer}(\th^{(l)}) = \emptyset$,
$\sfrac{\del a}{\del \th}$ is monotonic on $U_{2\de_\zer} (\th^{(k)}) $,
and there is a constant $K_a>0$ such that for all $k \in \nat{N}$  
and all $\th,\th' \in U_{\de_\zer}(\th^{(k)})$, 
$$
\abs{\sfrac{\del a}{\del \th}(\th) - \sfrac{\del a}{\del \th}(\th')}
\ge K_a \abs{\th -\th'}
.\eqn $$

}
\sni
It should be clear that for any asymmetric surface, points
$\th^{(1)}, \ldots, \th^{(N)}$, where $\sfrac{\del a}{\del \th}=1$,
must exist: $\sfrac{\del a}{\del \th}$
is a $2\pi$--periodic continuous function that satisfies
$$
\frac{\del a}{\del \th}(a(0))= 
\frac{\ka(0)}{\ka(a(0))} =
\left(\frac{\ka(0)}{\ka(a(0))}\right)^{-1} =
\left(\frac{\del a}{\del \th}(0)\right)^{-1} 
\eqn $$
by \queq{\curvratio}. By the intermediate value theorem, it must
either be constant (then $e$ is symmetric), or it must take the
value $1$ at least twice. 
If we suppose for a moment that $\frac{\del a}{\del \th}$
is differentiable, a similar argument shows that $\frac{\del^2 a}{\del \th^2}$
must also have zeros. \AFoup\ states (without referring to such
additional derivatives), that at those $\th$ where the curvature
ratio is one, this ratio varies at least linearly. 
The curvature ratio turns up, and hence \AFoup\ will be used,
only in the proof of Theorem \RPAreg\ $(ii)$, more specifically
in the one--loop volume estimate that shows boundedness of the
particle--particle ladder if \SYmm\ does not hold (if
\AFoup\ holds instead). 

\subsect{Higher Dimensions}
Let $\del^2 e$ denote the $d\ge 3$ analogue of the matrix defined
in \queq{\edpdef}. By \AFou, there is, for each $\th\in S^{d-1}$,
a basis $\v_1,\cdots,\v_{d-1}$ for the orthogonal complement to
$\nabla e(\p(\th))$ such that the matrices
$$\eqalign{
C_\th&=\left[\left(\v_i,\partial^2 e\big(\p(0,\th)\big)\v_j\right)\right]_{1\le i,j\le d-1}\cr
A_\th&=\left[\left(\v_i,\partial^2 e\big(\p(0,a(\th))\big)\v_j\right)\right]_{1\le i,j\le d-1}\cr
}\eqn $$
obey
$$
\abs{\sfrac{|\nabla e(\p(\th))|}{|\nabla e(\p(a(\th)))|} 
A_\th C^{-1}_\th-v\bbbone}\le 1/5
\eqn $$
for some $v\in\{\pm1\}$ (in the symmetric case, $C_\th=A_\th$
and $|\nabla e(\p(\th))|= |\nabla e(\p(a(\th)))|$).
The matrix $C_\th$ is invertible because $S$ has positive curvature 
everywhere.

\sect{Covariance}
\noindent
The propagator for the independent ($\la =0$) electrons is
$$
\left(G_\zer\right)_{\al\al'}(p) = \de_{\al\al'} C(p_\zer,e(\p))
\eqn $$
with 
$$
C(\om,E)= \frac{e^{i\om 0^+}}{i\om-E}
\eqn $$
This notation means the usual boundary value for distributions, e.g.\
$$
\int d^{d+1} p \frac{e^{ip_\zer 0^+}}{ip_\zer-e(\p)}= 
\lim\limits_{\ta \downarrow 0} 
\int d^{d+1} p \frac{e^{ip_\zer \ta}}{ip_\zer-e(\p)}
.\EQN\prescri $$
The reason for this limiting prescription is that by Fourier transformation,
at temperature $T=\sfrac{1}{\be}$, the Fermi distribution function, 
which determines the thermodynamics of independent electrons, is 
given by
$$
\frac{1}{1+e^{\be E}} = \lim\limits_{\ta \downarrow 0} 
\frac{1}{\be} \sum\limits_{n \in \Z} \frac{e^{i\om_n \ta}}{i\om_n-E}
\eqn $$
where $\om_n = (2n+1)\pi T$. In the limit $T \to 0$, the frequency sum 
becomes an integral.

The behaviour of perturbation theory is determined by the 
properties of $C$ and the interaction potential $\hat v$. It is obvious that
$\int_1^\infty |C(p)| dp_\zer$ is infinite so that one must not take 
the absolute values inside loop integrals containing only one propagator. 
As will be shown, this is harmless, however, 
and the mathematically difficult, and physically 
relevant, singularity is in the infrared, at $p_\zer =0$ and 
$e(\p)=0$. This singularity implies that 
$$
\int\limits_{p_\zer^2+E^2\le 1} |C(p_\zer,E)|^n = \infty
\eqn $$
for all $n\ge 2$, which causes the divergences of 
unrenormalized perturbation theory \quref{FT1,FT2,FST}. 
The singularity of $C$ at $p_\zer^2+E^2=0$ is the important one because it 
determines the long--distance behaviour of the fermion 
two--point function. 

\subsect{The scale zero effective action}
To separate the harmless, short--distance part of the propagator
from the long--distance part, we first integrate over
all fields whose momenta are not in a neighbourhood of the 
Fermi surface. This includes in particular the fields with large $|p_\zer|$.
This integration produces an effective interaction, which 
we call the scale zero effective action, for the remaining fields,
which are then subject to a fixed ultraviolet cutoff. 

To integrate over the fields with momenta away from the Fermi surface, 
we split the propagator $C$ into a scale zero part $C_\zer$ where 
$p_\zer^2+E^2 \ge \Const > 0$ and an infrared part $C_{<0}$ where 
$p_\zer^2+E^2$ can get arbitrarily close to zero, as follows. 
Let $M\ge \max\{4^3,\sfrac{1}{r_\zer}\}$ (then $\abs{e(\p)} < M^{-1}$
implies $|\rh|<r_\zer$), and let $a \in C^\infty(\R_\zer^+,[0,1])$ be
such that $a(x)=0$ for $0 \le x \le M^{-4}$, $a(x) =1$ for $x \ge M^{-2}$, 
and $a'(x) > 0$ for all $x \in (M^{-4},M^{-2})$. Define
$$\eqalign{
C_\zer (p_\zer,E) &= a(p_\zer^2+E^2) C(p_\zer,E)\cr
C_{<0} (p_\zer,E) &= \left(1-a(p_\zer^2+E^2)\right) C(p_\zer,E)
\cr}\EQN\sczerC $$
Since $C=C_\zer+C_{<0}$,  the effective action 
$$
e^{\cG(\ch,\chq)} = \int d\mu_C(\ps,\psq) e^{\la V^{(0)} (\ps+\ch,\psq+\chq)}
\eqn $$
can be written as 
$$
e^{\cG(\ch,\chq)} = \int d\mu_{C_{<0}}(\ps,\psq) e^{\cV_{\rm eff}^{(0)} (\la,\ps+\ch,\psq+\chq)}
\eqn $$
where
$$
 e^{\cV_{\rm eff}^{(0)} (\la,\ch,\chq)} = 
\int d\mu_{C_{\zer}}(\ps,\psq) e^{\la V^{(0)} (\ps+\ch,\psq+\chq)}
.\eqn $$
$\cV_{\rm eff}^{(0)} (\la,\ps,\psq)$ is a formal power series in $\la$:
$$
\cV_{\rm eff}^{(0)} (\la,\ps,\psq) = 
\sum\limits_{r=1}^\infty \la^r
\sum\limits_{m=0}^{\bar m(r)} \int  
dp_\one \ldots dp_{2m}\;
\de\left(\sum\limits_{i=1}^m (p_i-p_{m+i})\right)
V_{m,r}^{(0)}(p_\one,\ldots,p_{2m}) \prod\limits_{i=1}^m 
\psq(p_i) \ps(p_{m+i}) 
\eqn $$
The propagator that appears in $\cV_{\rm eff}$ is $C_\zer$, 
which has no infrared singularity.

\Lem{\UVpart}{\Lesty Assume \AOne{k,\hxp}\ and \ATwo{k,\hxp},
with $k\ge 2$ and $\hxp \ge 0$.
The $V_{m,r}^{(0)}$ are all bounded and $C^{k,\hxp}$ in the external momenta, i.e.
$$
\abs{(V^{(0)})_{m,r} }_{k,\hxp} \le K(m)^r r!
\EQN\SSS $$
}

\Proof See Appendix D. \endproof

\Rem{\dummy}{
In fact, the scale zero effective action is analytic in $\la$ 
(so the $r!$ is not really there on the right side of \queq{\SSS}).
Since $C_\zer$ has no singularity, one can show that $\cU_v$ is 
analytic in $\la $ in a disk independent of the volume and the temperature, 
by adapting the determinant bound given in \quref{FMRT} suitably. 
Recall that momentum space is $\R \times \cB$ with $\cB = \R^d/\Ga^\#$
compact. So the spatial components of momenta are subject to a fixed
ultraviolet cutoff. Consequently, there is no `stability of matter'
problem. }

\ssni
The term $(V^{(0)})_{1,r}$ is a bilinear term in the fermions. 
Since it is $C^k$ and $O(\la)$, and since $\la$ is small, it does not change
the properties of $e$. We simply absorb it into $e$ without changing 
notation. 

Thus the only difference between the original model and 
the one with interaction $V_{\rm eff}^{(0)}$ is that the latter contains 
vertices with $2m \ge 6 $ external legs, whose vertex functions are 
at least of order $\la^2$, and which are $C^{k,\hxp}$ in the external
momenta with bounds that are scale independent. This is a minor
complication which is easily taken into account (it will mainly
concern us in \THREE). 

\subsect{The scale decomposition in the infrared}
We now turn to the essential part of the problem. 
We decompose the infrared part $C_{<0}$ of the propagator into slices  
as in \FST: Let $f(x) = a(x)-a(x/M^2)$, then $1-a(x)=\sum\limits_{j<0}
f(M^{-2j} x)$, and (omitting the $e^{ip_\zer 0^+}$ since the limits can
be taken inside the integrals because large $p_\zer $ do not occur
in the infrared part of the propagator)
$$\eqalign{
C_{<0}(p_\zer,E) & = \frac{1-a(p_\zer^2+E^2)}{ip_\zer -E}
= \sum\limits_{j<0} C_j(p_\zer,E) \cr
C_j(p_\zer,E) & = \frac{f(M^{-2j}(p_\zer^2+E^2))}{ip_\zer -E}
.\cr}\EQN\CpzE $$
The point of this decomposition is that the propagator on scale $j$, $C_j$,
has simple properties: it is easy to prove (see \FST, Lemmas 2.1 and 2.3)
that 
$$
\max\limits_{|\al | =s} 
|D^\al C_j (p_\zer,e(\p))|\leq W_s M^{-(s+1)j} 
\; \True{|ip_\zer- e(\p)| \leq M^j}
\EQN\shellj $$
where $D^\al$ is a derivative with respect to $p$ ($\al$ is a multiindex 
with $|\al|=s$, $0\leq s\le k$) of order $s$. 
The indicator functions take the value
$\True{X}=1$ if $X$ is true and $\True{X}=0$ otherwise. 
In words: on `slice' number $j$, the propagator is for all 
$(p_\zer, \p)$ of absolute value at most $M^{2-j}$,
($s=0$ in \queq{\shellj}; $W_\zer = M^2$), and every derivative produces
another large factor $M^{-j}$. The constant $W_s$ depends on $|e|_s$
and on $g_\zer$.
Moreover, the support of $C_j$ is contained in the product of 
an interval of length $2M^j$ in $p_\zer$
and a thin shell of thickness $\Const M^j$ around the Fermi surface $S$.
By our choice of $M$, $M^{-1} < r_\zer$, so 
for all $j < 0$ this shell is contained 
in the region where the variables $\rh$ and $\th$ can be used. 

In infinite volume, we introduce an infrared cutoff by restricting 
the sum in \queq{\CpzE} to $j \geq I$, where $I \in \Z$, $I < 0$
(see \FST\ for details). In infinite volume the limit 
$I \to -\infty$ is the definition of the model. 
We shall show in another paper that the so defined infinite volume and
zero temperature Green functions are perturbatively identical to those obtained 
by taking a finite volume, positive temperature model and letting
the volume tend to infinity and afterwards the temperature go to zero.

The heart of the analysis of this paper concerns regularity properties
of the self--energy in the spatial part $\p$ of the momentum.
It will be obvious from studying the proofs that this analysis
works uniformly for positive temperature as well as for temperature zero.  

As explained, $\int dp_\zer \; dE |C_{<0}(p_\zer,E)|< \infty$, but 
$\int dp_\zer \; dE |C_{<0}(p_\zer,E)|^2 = \infty$ 
because of the singularity at $p_\zer = E = 0$ 
(see \FST, Section 1.4). To get finite perturbative Green functions, 
we renormalize. To this end, we use the projection $\prP$ introduced in 
Section 2.2 of \FST, i.e.
$$
\prP(\p(\rh,\si))=\p(0,\si)
,\eqn $$
and for continuous functions $F(p)$, 
$$
(\ell F) (p_\zer,\p(\rh,\si)) = \cases{ F\big(0,\p(0,\si)\big) \ch(\rh) 
& $|\rh| < 2 r_\zer$ \cr
0 & otherwise\cr}
\eqn $$
where $\ch\in C^\infty(\R,\R)$, $\ch(x)=1$ for $|x|<r_\zer$ and
$\ch(x)=0$ for $|x|>2r_\zer$, and $\ch$ decreases in $|x|$.
The counterterm and the self--energy are then defined 
recursively in $r$, the order in $\la$, as follows. In first order,
let $\Si_\one^{unsub} (p)$ be the sum of the values of the first order
1PI two--legged graphs, evaluated according to the Feynman rules of 
the model. Then $K_\one (\p) = (\ell \Si_\one^{unsub}) (p) = 
\Si_\one^{unsub} (0,\prP(\p))$, and 
$\Si_\one (p) = \Si_\one^{unsub} (p)- K_\one (\p)$. Assuming that
$\Si_r$ and $K_r$ have been defined, $\Si_{r+1}^{unsub}$ is the 
sum of values of all 1PI two--legged graphs with all two--legged 
insertions 1PI $T(p)$ being of the form $\Si_s$, $s<r$, 
i.e.\ subtracted on the Fermi surface
($T(p) - T(0,\prP(\p))$ appears). Then $K_{r+1}=\ell \Si_{r+1} $
and $\Si_{r+1}= \Si_{r+1}^{unsub}- K_{r+1}$.
Note that $K_r^I$ actually only depends on $\prP (\ve{p} ) \in S$; 
this will be very important in the following. 
In the scale decomposition, with an infrared cutoff $I$,
$K_r^I$ appears as 
a sum over trees and compatible labelled graphs
(for details, not needed here, see \FST, (2.76)). This scale decomposition
is a way of dealing with the functions that allows us to do estimates
that are even hard to get in the few cases where one can use 
exact calculations. It is proven in \FST, Theorem 1.2, that the sequence of 
functions $K^I_r$ and their derivatives with respect to $\p$ 
converge uniformly as $I \to -\infty $ and that 
$$
K_r = \lim\limits_{I\to-\infty} K_r^I
\eqn $$
is $C^1$ in $\p$. In the next chapter, we show that $K_\two$ is 
$C^2$ if $e$ and $\hat V$ are $C^2$.

\goodbreak

\chap{Regularity in Second Order}
\noindent
In this chapter, we give a detailed explanation of the problem 
and its solution for the example of the second--order counterterm 
$K^{(2)}$ (the problem begins to be nontrivial in second order).
After motivating the problem, we prove Theorem \zweireg.
We also explain the basics of the scale decomposition
when we deal with the second order graph. 

\sect{Preparations}
\noindent
The graphs that contribute to first order are shown in Figure
\nextfig{\erschte}.
\herefig{firstord}
They have the values 
$$
Y_\one(q)=\int\limits_{\R\times\cB}\db p\; P(p,q) C(p_\zer, e(\p))
\eqn $$
with 
$$
P(p,q)=\cases{ \langle q\; p \mid V\mid p \; q\rangle 
& rainbow \cr
\langle q\; p \mid V\mid q \; p\rangle  
& tadpole. \cr}
\EQN\raintad $$
(we use matrix notation for the spin sums; the tadpole involves a spin
trace which we omitted in the notation because it is inessential for 
the regularity problem. Also, $\db p = \sfrac{d^{d+1}p}{(2\pi)^{d+1}}$).
Since the only factor that depends on $q$ is the interaction function, 
and since $C$ is integrable, it is trivial to take the limit $I\to -\infty$
of $Y_\one$, and it is $C^{k,\hxp}$ in the external momentum by \AOne{k,\hxp}.
Graphically speaking, the external momentum either
enters no line of the graph at all (for the tadpole term), or it
can be routed through the interaction line (for the rainbow term).

The second order gets contributions from the graphs shown in 
Figure \nextfig{\zwoate}.
(recall that only 1PI two--legged diagrams contribute to the 
self--energy and the counterterm). 
\herefig{secondord}
The shaded disk in the second order rainbow and  the tadpole
indicates an insertion of a first order diagram.
In our renormalized expansion, every two--legged insertion
is subtracted at the Fermi surface, for instance, 
the value of the second order rainbow is 
$$
Y_\two (q) = \int \db p \;  
\langle q\; p \mid V\mid p \; q\rangle C(p_\zer,e(\p))^2 
\left( Y_\one (p_\zer,\p)-Y_\one (0,\prP(\p))\right)
\eqn $$
where $\prP$ is the projection onto the Fermi surface defined in Lemma 2.1 
of \FST. As in first order, the contribution of the tadpole and rainbow graphs 
is $C^{k\hxp}$ by \AOne{k,\hxp}. 
The contribution from the scale zero six--legged
vertex is $C^{k,\hxp}$ because the external momentum does not enter
the fermion line and because the vertex function is $C^{k,\hxp}$ by
Lemma \UVpart.

We note in passing that although $\int |C(p)|^2 dp = \infty$, 
the value of the second order rainbow and the
second order tadpole would still be finite without
renormalization because of the special structure of 
these graphs, by  Lemma 2.42 of \FST. 
So divergences first appear in third order. However, the argument
of Lemma 2.42 in \FST\ has an additional derivative act on $\hat v$, 
and thus one can only show that the unrenormalized value of the
rainbow is $C^{k-1}$ if $\hat v$ is $C^k$. 

Moreover, being conditionally convergent, 
the unrenormalized value depends on how the limit is taken. 
In particular, this graph is one of the anomalous ones of Kohn and 
Luttinger that prevent convergence of the unrenormalized positive temperature
Green functions to their zero temperature counterparts. For the 
renormalized Green functions, there is no such problem. Their values 
at positive temperature converge to those at zero temperature.
This will be shown in another paper. Here we mention this only as a further 
motivation why the renormalized expansion is the correct one.
 
The `polarization' and `vertex' diagram (so called because a polarization 
bubble, resp. a vertex correction appear as pieces of the graph) have 
the value (denoting $p_k=(z_k,\p_k) \in \R \times \cB$)
$$
Y(p) = \int \left( \prod\limits_{k=1}^3 \db p_k\; 
C(z_k, e(\p_k)) \right) \de^\#(p_\one+p_\two-p_\thr -p)
P(p_\one,p_\two,p_\thr,p)
\EQN\socalled $$
with
$$
P(p_\one,p_\two,p_\thr,p)
=\cases{ \langle p\; p_\thr \mid V\mid p_\one \; p_\two \rangle^2
& `polarization' graph \cr
\langle p_\one\; p_\two \mid V\mid p \; p_\thr\rangle\;
\langle p_\thr \; p \mid V\mid p_\one \; p_\two\rangle
& `vertex' graph \cr}
\EQN\voilaP $$

By assumption \AOne{k,\hxp}, the function $P$ is $C^{k,\hxp}$ 
in the external momentum $p$ and in all $p_i$. But,
whichever way one uses the delta function to express one of the $p_i$
in terms of $p$ and the others, that momentum will depend on $p$, 
so one of the propagators depends on $p$. A derivative with respect to 
$p$ squares the denominator and thus makes the singularity stronger, 
and there is no cancellation in the frequency integration as in the 
rainbow graph. Naive power counting suggests that
already the first derivative is divergent. However, 
these graphs are \OL, and therefore the bounds in \FST\ imply an improvement
that makes the first derivative finite. 

We repeat this proof for the example at hand to motivate the 
more difficult bounds that follow. First, 
we briefly recall the tree structure associated to the scale decomposition.
If you are familiar with that, you can skip the remainder of this section.
Calling 
$$
\cJ = \set{(j_\one,j_\two,j_\thr)}{ I \leq j_k < 0}
\eqn $$
and inserting the scale decomposition \queq{\CpzE}, we get $Y(p) = 
\sum\limits_{(j_\one,j_\two,j_\thr)\in \cJ}  
Y_{j_\one,j_\two,j_\thr}(p)$ with
$$
Y_{j_\one,j_\two,j_\thr}(p) = 
\int \left(\prod\limits_{k=1}^3 \db p_k\; 
C_{j_k}(z_k, e(\p_k)) \right) \de^\#(p_\one+p_\two-p_\thr -p)
P(p_\one,p_\two,p_\thr,p)
.\eqn $$
At finite $I$, the integrand is bounded and $C^k$, so the integral for 
$Y$ is $C^k$ in $p$. All singularities arise in the limit $I \to -\infty$.
We now rearrange the scale sums such that $j_\one \le j_\two\le j_\thr$. 
$\cJ$ is the disjoint union
$
\cJ = D \dotcup U \dotcup T
$ 
where
$$\eqalign{
D&=\set{(j,j,j)}{j \in \{I, \ldots, -1\}}\cr
T&=\set{(j_\one,j_\two,j_\thr)\in \cJ}{
j_\one=j_\two\neq j_\thr \hbox{ or }
j_\one\neq j_\two = j_\thr \hbox{ or }
j_\one=j_\thr\neq j_\two } \cr
U&=\set{(j_\one,j_\two,j_\thr)\in \cJ}{\hbox{ if }
k\neq l, \hbox{ then } j_k \neq j_l }
,\cr}\EQN\allthosesets$$
so $Y$ is a sum $Y=Y_D+Y_U+Y_T$ where
$$
Y_{\cM}(p) = 
\sum\limits_{(j_\one,j_\two,j_\thr)\in \cM}  
Y_{j_\one,j_\two,j_\thr}(p)
.\eqn $$
For every triple $(j_\one,j_\two,j_\thr)\in U$, there is a unique 
permutation $\pi \in \cS_3$ such that 
$j_{\pi(1)} < j_{\pi(2)} < j_{\pi(3)}$, therefore 
$$
U= \set{(j_{\pi(1)}, j_{\pi(2)}, j_{\pi(3)})}{\pi \in \cS_3,
(j_\one,j_\two,j_\thr)\in \Om}
\eqn $$
with 
$$
\Om = \set{(j_\one,j_\two,j_\thr)\in \Z^3}{
I\le j_\one < j_\two < j_\thr < 0}
.\eqn $$
Thus, calling $\ulj=(j_\one,j_\two,j_\thr)$,
$$\eqalign{
Y_U(p)&=\sum\limits_{\ulj\in\Om}\sum\limits_{\pi\in \cS_3}
\int \prod\limits_{k=1}^3 \db p_k\; 
C_{j_{\pi(k)}}(z_k, e(\p_k)) \de^\#(p_\one+p_\two-p_\thr -p)
P(p_\one,p_\two,p_\thr,p) \cr
&=\sum\limits_{\ulj\in\Om}
\int \prod\limits_{k=1}^3 \db p_k\; 
C_{j_{k}}(z_k, e(\p_k)) 
\sum\limits_{\pi\in \cS_3}\de^\#(p_{\pi(1)}+p_{\pi(2)}-p_{\pi(3)} -p)
P(p_{\pi(1)},p_{\pi(2)},p_{\pi(3)},p) 
\cr}\eqn$$
We use the $\de^\#$ to `fix' $p_\thr$, i.e.\ to remove the 
$p_\thr$--integration, and get
$$
Y_U(p) = \sum\limits_{\pi \in \cS_\thr} Y_\Om^\pi (p) =
\sum\limits_{\pi \in \cS_\thr}
\sum\limits_{\ulj \in \Om} Y_{\ulj}^\pi (p)
\EQN\YUOm $$
with
$$
Y_{\ulj}^\pi (p) =  \int \db p_\one \db p_\two\;
C_{j_\one}(z_\one, e(\p_\one))\;
C_{j_\two}(z_\two, e(\p_\two))\;
C_{j_\thr}(\ze_{\pi(3)},e_{\pi(3)})\;
P_{\pi} (p_\one,p_\two,p)
\EQN\Yuljpi $$
where
$$
\ze_a=L_a(z_\one,z_\two,z), \qquad
e_a=e(L_a(\p_\one,\p_\two,\p))
,\EQN\eazea $$
$L_a$ is given by 
$$
L_a (p_\one,p_\two,p) = \cases{
 p+p_\one-p_\two & if $a=1$\cr
 p-p_\one+p_\two & if $a=2$\cr
-p+p_\one+p_\two & if $a=3$\cr}
,\EQN\Ladef $$
and $P_\pi(p_\one,p_\two,p) = P(p_{\pi(1)},p_{\pi(2)},p_{\pi(3)},p)$
with $p_\thr$ given by $L_{\pi(3)} (p_\one,p_\two,p)$. For the spatial part
$\p_\thr$, recall that the integral is over the torus $\cB$ and
that therefore the $\de $ function is the one on the torus $\cB$.
We shall bound the $Y_\Om^\pi$ separately for all $\pi$. The essential 
point is the dependence of $C_{j_\thr}$ on $p$, since $P_\pi$ is 
$C^{k,\hxp}$ by \AOne{k,\hxp}.
We therefore redraw the graph by collapsing the interaction lines
to vertices; it then looks like one of the graphs in Figure \nextfig{\figuone}.
The left graph corresponds to the case $a=3$ in \queq{\Ladef}, 
the right one to $a=2$.

\herefig{secord}
Similarly, $Y_T$ is a sum of terms of the form
$$\eqalign{
\sum\limits_{I\leq j_\one < j_\thr< 0} \int \db p_\one\db p_\two 
C_{j_\one}(z_\one,e(\p_\one))
& P(p_\one,p_\two,q,p) 
\Big( C_{j_\one}(z_\two,e(\p_\two))\; 
C_{j_\thr}(\ze,e(\q))+ \cr
+ & C_{j_\thr}(z_\two,e(\p_\two))\;
C_{j_\thr}(\ze,e(\q))\Big) \cr
\cr}\eqn$$
with $\ze$ and $\q$ again given one of the linear combinations 
in \queq{\Ladef}, and 
$$\eqalign{
Y_D(p)&=\sum\limits_{I\leq j < 0} \int \db p_\one\db p_\two 
\Big(C_{j}(z_\one,e(\p_\one))C_{j}(z_\two,e(\p_\two))\cr 
& C_{j}(z_\one+z_\two-z,e(\p_\one+\p_\two-\p)) 
P(p_\one,p_\two,p_\one+p_\two-p,p) 
.\cr}\eqn$$
Note that we have arranged that the external momentum $p$ is 
routed through the highest scale line $C_{j_\thr}$.
The arrangement of the scale sums into sums over these sets is none other
than the decomposition into Gallavotti--Nicol\` o trees \quref{GN}. The 
trees $t$ associated to $T,D$ and $\Om$ are drawn in Figure 
\nextfig{\figutwo}; 
the sum over all $\ulj$ and permutations $\pi \in \cS_\thr$ in \queq{\YUOm} 
corresponds to the sum over all labellings of $G$ that are consistent
with $t$. Also, fixing the momentum of the line with the highest scale
$j_\thr $ in terms of the others (i.e.\ putting that line into the 
spanning tree of the graph) is the natural choice for a spanning tree
that is consistent with the scale structure (see Section 2.6 in \FST).
This will be important in the estimates.

\herefig{gntrees}

\sect{Convergence Properties of Derivatives}

\noindent
After these preparations, we can start to discuss convergence questions.
Only $C_{j_\thr}$ depends on the external momentum $p$. Thus
$$
\abs{Y_{\ulj}^\pi}_s \leq \sup\limits_{p}
\int dp_\one \ dp_\two \; \abs{C_{j_\one}((p_\one)_\zer,e(\p_\one)) 
C_{j_\two}((p_\two)_\zer,e(\p_\two))} \; \abs{\sum\limits_{|\al| \le s}
\big(\sfrac{\del}{\del p}\big)^\al\ 
\left( C_{j_\thr} (\ze_{\pi(3)}, e_{\pi(3)}) P_\pi (p_\one,p_\two,p)
\right) }
\eqn $$
By the Leibniz rule and \queq{\shellj},
$$
\abs{Y_{\ulj}^\pi}_s \le {W_\zer}^2 M^{-j_\one-j_\two}
\sum\limits_{r=0}^s
\sum\limits_{t=0}^r m_{rt}\abs{P}_{r-t} W_{t} M^{-j_\thr(1+t)}
\sup\limits_p
\int dp_\one \ dp_\two \;\bbbone_{j_\one}(p_\one)\ \bbbone_{j_\two} (p_\two)\
\bbbone_{j_\thr}(p_\thr)
,\eqn $$
with $m_{rt}$ a combinatorial factor, and $\bbbone_j(p) = 
\True{\abs{ip_\zer-e(\p)} \le M^j}$.
We use the coordinates $(\rh,\th)$ and the notations $\p_i=\p(\rh_i,\th_i)$,
$\p=\p(\rh,\th)$, $p_i = (z_i, \p_i)$, $p=(z,\p)$, to get
$$
\abs{Y_{\ulj}^\pi}_s  \le K \;
M^{j_\one+j_\two-j_\thr (1+s)} \;
\sup\limits_\p  F(\p)
\eqn $$
with 
$$
F(\p) = 
\int\limits_{{\rh_\one}^2+{z_\one}^2 \le M^{2j_\one}} d\rh_\one\ dz_\one 
\int\limits_{{\rh_\two}^2+{z_\two}^2 \le M^{2j_\two}} d\rh_\two\ dz_\two
\int d\th_\one \ d\th_\two \; 
1\Big(|e(L_b(\p(\rh_\one,\th_\one),\p(\rh_\two,\th_\two),\p))|
\leq M^{j_\thr} \Big) 
\eqn $$
and 
$$
K=K(s,|e|_s,|\hat v|_s,|u|_{s+1})= {W_\zer}^2 |J|_\zer^2
\sum_{r\leq s} \sum_{t=0}^r m_{rt} |P_\pi|_{r-t} W_{t}
.\eqn $$
We have kept only the support function of the 
spatial part of the momentum in the integral for $F$. 
As in \FST, (A.2)--(A.6), we use
$\max\{ j_\one,j_\two\} \leq j_\thr$ and Taylor expansion to bound
$$
1\Big(|e(L_b(\p(\rh_\one,\th_\one),\p(\rh_\two,\th_\two),\p))|
\leq M^{j_\thr} \Big) \leq 
1\Big(|e\big(L_b(\p(0,\th_\one),\p(0,\th_\two),\p)\big)| 
\le (1+2\sfrac{|e|_1}{u_\zer})M^{j_\thr}\Big) 
\EQN\famous $$
on the support of the integrand, and obtain $F(\p) \le \cW 
\Big(\big(1+2\sfrac{|e|_1}{u_\zer}\big)M^{j_\thr}\Big)$, where
$\cW$ is the function defined in \queq{\cWdef}. Thus
$$
\abs{Y_{\ulj}^\pi}_s \le \pi^2 K\;
M^{j_\one+j_\two-j_\thr (1+s)} \; 
\cW\left(\big(1+2\frac{|e|_1}{u_\zer}\big)M^{j_\thr}\right)
.\eqn $$ 
\ssni
The crude bound $\cW \leq \left( \int d\si\right)^2$ gives the ordinary 
power counting bound
$$
\abs{Y_{\ulj}^\pi}_s \le \Const \; M^{j_\one+j_\two-j_\thr (1+s)}
.\EQN\ordbou $$
We now do the scale sum. In case $\Om$, the sum is over
$I \leq j_\one < j_\two < j_\thr < 0$. 
$$
\sum_{j_\thr>j_\two} M^{-j_\thr(1+s)} \le 
M^{-j_\two (1+s)} \sum_{k\geq 0} M^{-k(1+s)} =
M^{-j_\two (1+s)} \frac{1}{M^{1+s}-1}
\eqn $$
and similarly,
$$
\sum_{j_\two>j_\one} M^{-j_\two (1+s)} M^{j_\two} \leq
\cases{ |j_\one| & $s=0$\cr
\Const M^{-j_\one s} & $s>0$.\cr}
\eqn $$
Since 
$$
\sum_{j_\one < 0} M^{j_\one} |j_\one| < \infty
,\eqn $$
the scale sum over the value of the undifferentiated graph 
($s=0$) is majorized by a convergent sum.

However, if $s \ge 1$, 
$$
\sum_{j_\one=I}^{-1} M^{j_\one (1-s)} \geq 
\cases{|I| & $s=1$\cr
M^{|I| (s-1)} & $s\ge 2$\cr}
\eqn $$
diverges as $I \to -\infty$. The scale sums for $T$ and $D$ behave 
similarly (the scale sum in $D$ is particularly simple, because 
there is only a sum over one scale $j$; the right side of \queq{\ordbou}
simply reads $M^{(1-s)j}$). 
 
Using the optimal volume improvement bound, Theorem \bestvol, 
we obtain the improved power counting bound
$$
\abs{Y_{\ulj}^\pi}_s \le \Const \; M^{j_\one+j_\two - j_\thr s} |j_\thr|
,\EQN\thrthty $$
e.g.\ for $\ulj \in D$, the bound is now $|j|M^{(2-s)j}$.
It implies that convergence holds for $s<2$. This proves that 
the first derivative of $\Si$ exists and is a continous function. 
For $s=2$, the sum for the bound diverges as $|I|^2$. 

Above, we have not given the argument why being majorized by 
a convergent sequence implies that the function defined
by the scale sum converges as  $I \to -\infty$. 
This is a consequence of the dominated convergence theorem, 
by the argument that we used in Theorem 2.46 $(iv)$ in \FST: 
the sequence $(y^I)_{I<0}$ 
in $\cL=\ell^1 (\Z_-,\cC^\zer(\R\times \cB, \C))$, given by 
$y^I=(y^I_j)_{j<0}$, where
$$
y^I_{j_\one} = \cases{\sum\limits_{j_\two,j_\thr \atop j_\thr>j_\two>j_\one}
Y_{\ulj}^\pi (p) & $j_\one \geq I$\cr
0 & otherwise, \cr}
\eqn $$
converges pointwise in the space of sequences of continuous functions. 
Every element $y^I$ is pointwise bounded by $(|j| M^j)_{j<0}\in \cL$. 
Thus the sequence converges to an element $(y_j)_{j<0}$ of $\cL$, and 
$\Big( \sum\limits_{j\ge I} y_j^I\Big)_{I < 0}$
converges to $\sum\limits_{j<0} y_j$ as $I \to -\infty$. 

We have now demonstrated, for the example of the second order graph,
some of the statements that hold for all two--legged 1PI graphs by
the results of \FST\ and the optimal volume bound, Theorem \bestvol.
While too weak to make the second derivative convergent, the 
volume improvement given by Theorem \bestvol\ implies that the scale 
sum is only `marginally divergent' with $|I|$, i.e.\ it grows only 
as a polynomial of $I$, not as an exponential $M^{|I|}$. 
Thus one may hope that with a little more care in doing the bound,
the second derivative may be shown to converge.
However, for the self--energy, we cannot show this. But the 
counterterm function $K$ depends on fewer variables than $\Si$,
and therefore the regularity conditions on $K$ are weaker than
those on $\Si$. More precisely, 
in the coordinates $(\rh,\th)$ defined by the integral curves of $u$,
derivatives with respect to $\rh$ act transversally to $S$, while 
derivatives with respect to $\th$ act tangentially to $S$. From the 
general intuition that problems arise when the singularity is 
moved one may expect that the tangential derivative is better behaved
than the transversal one. The counterterms are defined as
$$
K_r = \left( \ell \Si_r^{unsub}\right) (p) =
\Si^{unsub}_r (0,\p(0,\th))
\eqn $$
so they depend only on the tangential variable $\th$, but not
on $\rh$. In other words, $K$ is constant
along the integral curves of $u$. It is therefore easier to check 
the differentiability of $K$ than that of $\Si$ because $p_\zer$
and $\rh$ are fixed to zero in $K$ and we only have to control tangential 
derivatives. 
In the spherically symmetric case
the dependence on $\th $ drops out as well by rotational invariance, 
and the function is a constant. 
In the nonspherical case in $d=2$, the proof of regularity of $Y$ in $\th$
is surprisingly tricky. We have

\Lem{\jaythree} {\Lesty 
Let $d=2$. For every $s \leq k$, there is a constant
$\De_s$, depending on $|e|_s$, $g_\zer$, and $w_\zer$, such that for 
all $\ulj $ with $\max\{j_\one,j_\two\}\le j_\thr<0$,
$$
\sup\limits_{\th \in S} \abs{\frac{\del^k Y_{\ulj}^\pi}{\del\th^k}
(0,\p(0,\si))} \le \De_s |P|_s \ M^{j_\one+j_\two} \ 
\cases{ M^{-j_\thr} |j_\thr| &  if $s\le k-1$\cr & \cr
M^{-\sfrac{3}{2}j_\thr}  |j_\thr| &  if $s=k$\cr}
\EQN\jthree $$
}
\sni
This Lemma immediately implies Theorem \zweireg\ for $d=2$ because 
the scale sum over $M^{j_\one+j_\two-\sfrac{3}{2}j_\thr}\ |j_\thr|$
converges by the above analysis.
The next section contains the proof of Lemma \jaythree.

\sect{The Singularities of the Jacobian}
\noindent
In this section, we will use \AFiv. We choose 
$\th$ as in Section 2.1, i.e.\ $\th \in \R/2\pi\Z$ for all $\rh$, and 
$|\del_\th \p (0,\th)| = \Const$ if $d=2$, and $\th \in S^{d-1}$
for $d \ge 3$. The $\th$--dependent part of
the integral is
$$\eqalign{
\om_{j,b}(\rh_\one,\rh_\two,P,p ) = 
\int\limits_0^{2\pi}d \th_\one 
\int\limits_0^{2\pi}d \th_\two\; 
J(\rh_\one,\th_\one) J(\rh_\two,\th_\two)
P(p_\one,p_\two,p) C_j (\ze_b,e_b) 
\cr}\EQN\omjdef$$
where $\p_k=\p(\rh_k,\th_k)$ and the external momentum is 
$p=(z,\p(\rh,\th))$. $e_b$ and $\ze_b$ given by \queq{\eazea}. 
In this expression,
the derivative with respect to $\th$ can act on the propagator and 
degrade the scale behaviour of the integral, which leads to the 
bounds derived above. However, 
one may try to use a further change of variables from 
$\th_\one,\th_\two $ to new, $\th$--dependent, variables to make 
$C_{j}$ independent of $\th $. The integration variables $\rh_\one$ and 
$\rh_\two$ are not useful for that purpose because 
making $\rh_\one$ or $\rh_\two$ $\th$--dependent only transfers the 
dependence on $\th$ from one of the propagators into another one in 
the integral \queq{\Yuljpi} for $Y$. 
Then the problem that the derivative can degrade 
the scale behaviour arises in another part of the integrand, 
so nothing has been gained. A change of variables in $\th_\one$, $\th_\two$
puts the $\th$--dependence only into the factors $J$ and $P$,
as well as the new Jacobian. Such a change of variables  
is possible in most, but not all, of the integration region,
because the resulting Jacobian has singularities that depend on 
$\th$. The main problem in showing regularity is to control these
singularities, that is, to show that the derivative does not make
them nonintegrable. This requires a detailed analysis of the dependence 
of the singularities on $\th$. We shall split the integral into 
contributions from a regular region, where the Jacobian and its
derivatives are bounded, and a singular region, which contains the 
singularities. The regular region will be easy to treat by a change 
of variables. We shall show that the singular region can be chosen 
fixed, i.e.\ independent of $\th$, if $\th$ varies in a small 
neighbourhood of a given, fixed, $\th^{(0)}$. 
The singularities of the Jacobian as a function of $\th_\one,\th_\two$
lie in neighbourhoods of the critical points of 
$$\et_b (\th_\one, \th_\two, \th ) = \pm
e(L_b(\ve{p} (0 , \th_\one ), \ve{p} ( 0 , \th_\two ),
\ve{p} ( 0 , \th )))
.\EQN\etabdef $$
The map $\et_b$ is $C^k$ in all variables. After the analysis
of the critical points of $\et_b$, the dependence of 
$e(L_b(\p(\rh_\one,\th_\one),\p(\rh_\two, \th_\two),\p(0,\th)))$
on the $\rh_i$ will be controlled by a Taylor expansion.
At the critical points, 
$$
{\del \et_b \over \del \th_i } (\th_\one, \th_\two, \th) = 
\nabla e (L_b(\p(0,\th_\one),\p(0,\th_\two),\p(0,\th)) 
\cdot \del_\th \ve{p} (0, \th_i ) = 0 \hbox{  for $i = 1,2$} 
.\EQN\etbcrit $$

\Lem{\findCP}{\Lesty Let $d\ge 2$. Assume \ATwo{2,0}, \AThr\ and \AFiv. Then all
solutions $(\th_\one , \th_\two )$ of the critical point equation 
\queq{\etbcrit} for which 
$L_b(\p_\one,\p_\two,\p(0,\th))\in S$ holds 
are given by $(\th_\one^* , \th_\two^* )= c_{b,l} (\th)$, with
$$\eqalign{
c_{b,\one} (\th ) &= (\th,\th)\cr
c_{b,\two} (\th ) &= \cases{(a(\th),\th)& $b=1,3$\cr
                            (\th,a(\th))& $b=2$\cr}\cr
c_{b,\thr} (\th ) &= \cases{(a(\th),a(\th)) & $b=1,2$\cr
                        (\th,a(\th))    & $b=3$ \cr}
\cr} \EQN\thatsthem $$
Here $a$ denotes the antipodal map on the 
Fermi surface $S$, see \queq{\antidef}.}
\Proof
Denote $\p_i=\p(0,\th_i)$ and $\Q = L_b(\p_\one,\p_\two,\p)=\p(0,\th')$. Then 
$$\eqalign{
\nabla e(\Q)\cdot\del_\th\p(0,\th_\one )&\=0\hskip.25in \Longrightarrow
\hskip.25in \p_\one\in\{\Q,\ve{a}(\Q)\}\cr
\nabla e(\Q)\cdot\del_\th\p(0,\th_\two)&\=0\hskip.25in \Longrightarrow
\hskip.25in \p_\two\in\{\Q,\ve{a}(\Q)\}\cr
}\EQN\therego $$
so in particular, $\th_\two = \th_\one $ or $\th_\two=a(\th_\one)$. 
Let $b=1$. Then $\p=\Q-\p_\one+\p_\two$. Thus, given $\th_\one$, 
we have the table 
\vskip.1in\noindent
\null\hfil\vbox{\offinterlineskip
\hrule
\halign{\vrule#&
         \strut\hskip.1in$#$\hfil&
         \hskip.1in\vrule#\hskip.1in&
          \hfil$#$\hfil&
           \hskip.1in\vrule#\hskip.1in&
          \hfil$#$\hfil&
           \hskip.1in\vrule#\cr
height2pt&\omit&&\omit&&\omit&\cr
&\th_\two&&\th'&&\p&\cr
\noalign{\hrule}
&\th_\one&&\th_\one&&\p_\one&\cr
height2pt&\omit&&\omit&&\omit&\cr
&\th_\one&&a(\th_\one)&&\ve{a}(\p_\one)&\cr
height2pt&\omit&&\omit&&\omit&\cr
&a(\th_\one)&&\th_\one&&\ve{a}(\p_\one)&\cr
height2pt&\omit&&\omit&&\omit&\cr
&a(\th_\one)&&a(\th_\one)&&2 \ve{a}(\p_\one)-\p_\one&\cr
height2pt&\omit&&\omit&&\omit&\cr
}\hrule}\hfil
\vskip.1in
\noindent 
Given $\th$, the last column fixes $\th_\one$, and the first one then 
fixes $\th_\two$. The first three rows produce the critical points 
$c_{\one,\one}, c_{\one,\thr}$ and $c_{\one,\two}$. 
The last row drops out because, 
calling $\q=\ve{a}(\p_\one)$ and $\r=\p_\one$, Lemma \byAfive\ implies
that $\q=\r$, which is a contradiction.
The case $b=2$ is gotten from $b=1$ by exchanging $\th_\one$ and $\th_\two$.
$b=3$ has $\p= -\Q+\p_\one+\p_\two $. 
The solution for $c_\thr (\th)$ comes from the case
$\Q=\ve{a}(\p)$ (and $\th_\two = a(\th_\one)$). \endproof

\noindent
\subsect{The significance of the filling condition}
The significance of this Lemma and of \AFiv\ in its proof is: 
given $\p(0,\th)$, at the critical points, $\th_\one $ and $\th_\two$ 
take the values $\th$ or $a(\th)$. 
The derivative $\sfrac{\del \et}{\del \th_i}=0$, but by Lemma \findCP,
$\sfrac{\del\et}{\del \th} =0$ as well at these points. Therefore, 
the equation to make $\et$ independent of $\th$ by a change in 
$\th_\one $ and $\th_\two$, 
$$
\sfrac{\del}{\del \th}\et(\th_\one (\th),\th_\two (\th),\th) = 0
,\EQN\Consis $$
remains consistent at these points, so a solution may still 
exist (and will be shown to exist below). This is not the case for the 
radial derivative, which is nonzero at the critical points.

Note that the collinearity argument could not have been applied 
without \AFiv\ because if $2\ve{a}(\p_\one) - \p_\one $ is on a copy of $S$
obtained by translation by some $\ga \in \Ga^\#$, the three vectors 
could differ (see Figure \nextfig{\figuthr}) 
for particular values of $\p$. In general, at these points,
\queq{\Consis} has no nonsingular solution for $\sfrac{\del \th_k}{\del \th}$.
This means that the derivative with respect to $\th$ really acts on the 
denominator and degrades the scale behaviour. The existence of the 
second derivative would then require another cancellation. 
A sketch of this situation for the case where \SYmm\ holds, 
so that $2\ve{a}(\p_\one)-\p_\one =-3\p_\one$ is given in Figure $\figuthr$. 
To visualize the situation, we have
drawn a periodic picture on $\R^2$ instead of a torus. The copies of 
$S$ obtained by translation by vectors in $\Ga^\#$ are to be
identified with $S$.  $\sfrac{\del\et}{\del \th} \neq 0$ because 
$\del_\th\p$ is not parallel to $\del_\th\p(0,\th')=-\del_\th
\p (0,\th_\one) $.
\herefig{nonA5}
In the Hubbard model, \AFiv\ gives the filling restriction stated
in the Introduction. As mentioned, such a filling restriction
is not as unnatural as it may look because numerical results
indicate a value of the filling below half--filling,
where the Fermi surface stays fixed. This filling corresponds
to $\mu =1/2$, which is the first value of $\mu$ where $2S$ touches
a translate of $S$.  
\Lem{\Unif}{\Lesty $(i)$ \ There is $\de=\de(w_\zer)>0$ such that for all 
$l,b\in \{ 1,2,3\}$ and for all $\th^{(0)}\in \R/2\pi\Z$, there 
is a $C^{k-1}$--diffeomorphism 
$$\eqalign{
\Ga_{bl}:U_{3\de}(0,0)\times U_\de(\th^{(0)})&\to 
\Ga_{bl}\big(U_{3\de}(0,0)\times U_\de(\th^{(0)})\big)
\subset\R^2\times U_\de(\th^{(0)})\cr 
(t_\one,t_\two,\th) &\mapsto (x,y,\th)=\Ga_{bl}(t_\one,t_\two,\th)
,\cr}\eqn $$ 
and a map $m_{bl}: \Ga_{bl}(U_{3\de}(0,0))\times U_\de(\th^{(0)})\to \R$,
with the following properties:
$$
\et_b(c_{bl}(\th)+(t_\one,t_\two),\th)=m_{bl}(x,y,\th)\; xy
,\EQN\noconsterm $$
$m_{bl}$ is $C^{k-2}$, and the maps $(x,y,\th)\mapsto x\ m_{bl}(x,y,\th)$ and
$(x,y,\th)\mapsto y\ m_{bl}(x,y,\th)$ are $C^{k-1}$ in all variables.
Moreover, 
for all $(x,y,\th)\in \Ga_{bl}\big(U_{3\de}(0,0))\times U_\de(\th^{(0)})\big)$,
$$
|m_{bl}(x,y,\th)| \ge \frac{w_\zer}{2}
\EQN\msnonzero $$
and 
$$
\det \Ga'_{bl}= \left|\matrix{ \sfrac{\del x}{\del t_\one} &
\sfrac{\del x}{\del t_\two}\cr
\sfrac{\del y}{\del t_\one} & \sfrac{\del y}{\del t_\two}\cr}
\right| \; = \; 1.
\EQN\noJac $$
$(ii)$ \ For $\veps > 0$ and $(t_\one,t_\two) \in U_{3\de}(0,0)$, 
$$
\abs{\et_b(c_{bl}(\th)+(t_\one,t_\two),\th)} < \veps
\EQN\epscond $$
implies
$$
\abs{xy} < \frac{2\veps}{w_\zer}
\EQN\xyeps $$
and there are constants $Q_s$, depending only on 
$w_\zer$, $r_\zer$ and $g_\zer$, and $\abs{e}_s$, such that
$$\eqalign{
\abs{\lbrack\frac{\del^s}{\del\th^s}\et_b\rbrack_{x,y}} &\le
Q_s\abs{e}_{s+2} 
\ \veps 
\qquad\hbox{ for all } s \le k-2\cr
\abs{\lbrack\frac{\del^{k-1}}{\del\th^{k-1}}\et_b\rbrack_{x,y}} &\le
Q_k \abs{e}_{k} 
\ \veps^{\sfrac{1}{2}}
\cr}\EQN\deriveps $$
In \queq{\deriveps}, $\lbrack\sfrac{\del}{\del\th}\rbrack_{x,y}$ 
is the partial de\-ri\-va\-tive with re\-spect to $\th$ with $x,y$ held fixed,
i.e.\ 
$$
\lbrack\frac{\del^{s}}{\del\th^{s}}\et_b\rbrack_{x,y}=
\frac{\del^{s}}{\del\th^{s}} \et_b\big((c_{bl}(\th),0)+ {\Ga_{bl}}^{-1}
(x,y,\th)\big)
\eqn $$
}

\Proof $(i)$ \ 
{\it Symmetric Case:} We assume \SYmm\ to hold. Then the antipodal
map is $a(\th)=\th+\pi$, hence $C^\infty$, and by \SYmm\ and a transformation
$\p_\one \to - \p_\one$ or $\p_\two \to -\p_\two$, we can always choose the
signs such as to make $e_b=e(\p_\one+\p_\two-\p)$. The critical points are 
then given by $b=3$ in \queq{\thatsthem}. We define 
$$
f_l(u_\one,u_\two,\th)=\cases{
e(\p(\th+u_\one)+\p(\th+u_\two)-\p(\th)) & $l=1$\cr
e(\p(a(\th)+u_\one+u_\two)+\p(\th+u_\two)-\p(\th)) & $l=2$\cr
e(\p(\th+u_\one)+\p(a(\th)+u_\one+u_\two)-\p(\th)) & $l=3$\cr
}
\eqn $$
so
$$\eqalign{
f_l(0,u_\two,\th)& =0 \quad\hbox{ for all }u_\two, \th\cr
f_l(u_\one,0,\th)& =0 \quad\hbox{ for all }u_\one, \th
\cr}\eqn $$
In the cases $l=2$ and $l=3$, we use that by \SYmm\
$$
\p(a(\th)+u)= \p(a(\th+u)) = - \p(\th+u).
\eqn $$
Since $a\in C^\infty$, $f_l$ is $C^k$ in all variables. By Taylor expansion,
$$
f_l(u_\one,u_\two,\th)= u_\one u_\two \ m_l(u_\one,u_\two,\th)
\eqn $$
with the $C^{k-2}$ function 
$$
m_l(u_\one,u_\two,\th)=\int\limits_0^1 d\al
\int\limits_0^1 d\be \; \del_\one\del_\two f(\al u_\one,\be u_\two, \th)
.\eqn $$
At $(u_\one,u_\two)=(0,0)$, 
$$
m_l (0,0,\th) = \cases{
w\big(\p(\th)\big) & l=1\cr
-w\big(\p(a(\th))\big) & l=2,3 \cr}
\eqn $$
so $|m_l(0,0,\th)| \ge w_\zer $ for all $\th$. Since $k \ge 2$, $m_l$ is 
continuous, so there is $\de >0$ such that for all $\th$ and for all 
$(u_\one,u_\two)\in U_{3\de}(0,0)$, $|m_l(u_\one,u_\two,\th)| 
\ge \sfrac{w_\zer}{2}$. Also,
$$
u_\one\ m_l(u_\one,u_\two,\th)=\int\limits_0^1 d\be \; 
\del_\two f_l(u_\one, \be u_\two, \th)
\eqn $$
is $C^{k-1}$, and similarly $u_\two m_l $ is $C^{k-1}$. Let 
$m_{bl}=m_l$, $x=u_\one$ and $y=u_\two$. For $l=1$, 
$\Ga_{bl}$ is the identity. For $l=2$, $\Ga_{bl}(t_\one,t_\two,\th)
=(t_\one-t_\two,t_\two,\th)$, and for $l=3$, 
$\Ga_{bl}(t_\one,t_\two,\th) = (t_\one, t_\two-t_\one,\th)$.
In all cases, \queq{\noJac} is obvious.
\ssni
{\it Nonsymmetric case:} The change here is that the antipodal map 
$a\in C^{k-1}$, and that the map $\Ga$ will now be nontrivial if the 
antipode is involved. We consider the $c_{bl}$ separately. The case
$l=1$ is identical to the symmetric case, and all other cases are 
similar either to $c_{22}$, where
$$
f(t_\one,t_\two,\th)=e(-\p(\th+t_\one)+\p(a(\th)+t_\two)+\p(\th))
,\eqn $$
or to $c_{1,3}$, where
$$
\tilde f(t_\one,t_\two,\th)=e(\p(a(\th)+t_\one)-\p(a(\th)+t_\two)+\p(\th))
.\eqn $$
We deal with $f$ first. Fix $\th^{(0)}\in \R/2\pi\Z$. Obviously, 
$f(0,t_\two,\th)=0$ for all $t_\two$ and $\th$, but $f(t_\one,0,\th)=0$
will in general hold only for $t_\one=0$ because of the asymmetry. 
We construct a $C^{k-1}$ function $t_\two = Y(t_\one,\th)$ such that
$$
f(t_\one, Y(t_\one,\th),\th)=0
\eqn $$
as follows. By Taylor expansion, 
$$
f(t_\one,t_\two,\th)=t_\one \ g(t_\one,t_\two,\th)
\EQN\ftoneg $$
with 
$$
g(t_\one,t_\two,\th)=\int\limits_0^1 d\al \; 
\del_\one f (\al t_\one,t_\two,\th)
,\eqn $$
in particular, 
$$
g(0,0,\th)=  \del_\one f(0,0,\th) = 
- \nabla e(\p(a(\th)) \cdot \del_\th\p(\th) = 0
.\eqn $$
Thus, by a further Taylor expansion in $t_\two $ and in $t_\one$,
$$
g(t_\one,t_\two,\th)=g(t_\one,0,\th) + h(t_\one,t_\two,\th)\ t_\two =
t_\one\ \ell(t_\one,\th)+ h(t_\one,t_\two,\th)\ t_\two
\EQN\geqn $$
with the $C^{k-2}$ functions 
$$
h(t_\one,t_\two,\th)= \int\limits_0^1d\be\; 
\del_\two g(t_\one,\be t_\two,\th)=
\int\limits_0^1 d\al
\int\limits_0^1 d\be\; \del_\two\del_\one f(\al t_\one,\be t_\two,\th)
\eqn $$
and 
$$
\ell(t_\one,\th) = \int\limits_0^1 d\al
(1-\al) \del_\one^2 f(\al t_\one,0,\th)
.\eqn $$
At $(t_\one,t_\two)=(0,0)$, 
$$
h(0,0,\th)=\del_\two\del_\one f(0,0,\th) = 
-\left( \del_\th\p(a(\th)), e''(\p(a(\th))\del_\th\p(\th)\right) 
= w(\p(a(\th)) \neq 0
.\eqn $$
The function $Y$ is constructed as the solution to the equation
$$
g(t_\one,Y,\th)=t_\one\ \ell(t_\one) + Y\ h(t_\one, Y, \th)=0
.\EQN\Yeq $$
A point where the equation holds is $(t_\one,Y)=(0,0)$, and
by \queq{\geqn}, and because $h$ is $C^{k-2}$, hence at least 
continuous since $k \ge 2$,
$$
\del_\two g(0,0,\th) = \lim\limits_{t_\two \to 0} 
\frac{g(0,t_\two,\th)-g(0,0,\th)}{t_\two}
= \lim\limits_{t_\two \to 0} 
h(0,t_\two,\th)=h(0,0,\th)\neq 0
.\eqn $$
So the solution to \queq{\Yeq} exists by the implicit function theorem,
and $Y$ inherits its differentiability properties from $g$.
The crucial point is now that $g$ is $C^{k-1}$ in all variables because
$$
g(t_\one,t_\two,\th)=\int\limits_0^1 d\al\; 
\nabla e(\p(a(\th)+\al t_\one)-\p(a(\th)+t_\two)+\p(\th))
\cdot \del_\th \p (a(\th)+\al t_\one)
,\eqn $$
and because $a$, $\nabla e$, and $\del_\th \p$ are all $C^{k-1}$. Thus 
$a \in C^{k-1}$ does not cause any loss in differentiability of $Y$.
$|h|$ is uniformly continuous in $\th$, so the size of the 
neighbourhood of $(0,0)$ where $Y$ is defined can be chosen 
uniformly in $\th$ by compactness of $\R/2\pi\Z$. 
Thus there is $\de_\one>0$ and a function 
$Y\in C^{k-1}((-\de_\one,\de_\one)\times U_{\de_\one}(\th^{(0)}),\R)$
such that  for all $\th\in U_{\de_\one}(\th^{(0)})$:
$Y(0,\th)=0$, and $g(t_\one,Y(t_\one,\th),\th)=0$ for all $|t_\one|<\de_\one$.
This implies 
$$
f(t_\one,t_\two,\th)=t_\one\ (t_\two-Y(t_\one,\th)) \  
\tilde h(t_\one,t_\two,\th)
.\EQN\Factors $$

Choose $\de$ such that $3\de \le \de_\one$ and such that for all 
$(t_\one,t_\two) \in U_{3\de}(0,0)$, $|\tilde h(t_\one,t_\two,\th)|\ge 
\sfrac{w_\zer}{2}$. Define $\Ga$ by 
$$
x=t_\one, \quad y=t_\two-Y(t_\one,\th)
\eqn $$
and $m_l(x,y,\th)=\tilde h(x,Y(x,\th)+y,\th)$, then 
$$
\left|\matrix{ \sfrac{\del x}{\del t_\one} &
\sfrac{\del x}{\del t_\two}\cr
\sfrac{\del y}{\del t_\one} & \sfrac{\del y}{\del t_\two}\cr}
\right| \; = \;\left|\matrix{ 1 & 0 \cr 
-\sfrac{\del Y}{\del t_\one} & 1\cr}
\right| \; = \; 1
\eqn $$
so $\Ga(\cdot, \cdot, \th)$ is a $C^{k-1}$--diffeomorphism at fixed $\th$, and 
\queq{\noJac} holds. It follows from \queq{\ftoneg} and the properties of $Y$
that 
$$
y\ m_l (x,y,\th) = (t_\two - Y(t_\one,\th))\ h (t_\one,t_\two,\th) 
= g(x,Y(x,\th)+y,\th)
\eqn $$
is $C^{k-1}$. Moreover, $f(t_\one,Y(t_\one,\th,\th))=0$ 
implies by Taylor expansion
$$
f(t_\one,t_\two,\th)=(t_\two-Y(t_\one,\th)) 
\int\limits_0^1 d\al\; (\del_\two f)
(t_\one, \al t_\two + (1-\al) Y(t_\one,\th),\th)
.\eqn $$
Comparing that to \queq{\Factors}, we see that 
$$
x \ m_l (x,y,\th) = \int\limits_0^1 d\al\; (\del_\two f)
(x, Y(x,\th) + \al y,\th)
\eqn $$
is also $C^{k-1}$. This proves all statements of the Lemma for 
the case $f$. The case $\tilde f$ is similar: defining
$$
\bar f(\bar t_\one,t_\two,\th)= 
\tilde f (t_\two+\bar t_\one,t_\two,\th)
,\eqn $$
we have again that $\bar f(0,t_\two,\th)=0$ for all $t_\two$ and all
$\th$, and we have thus reduced the case to the previous one by the 
change of variables $(t_\one,t_\two) \to (t_\one-t_\two,t_\two)$. 
The Jacobian of this change of variables is one, so \queq{\noJac} 
still holds. 
\ssni
$(ii)$ \queq{\xyeps}
is obvious from \queq{\epscond} and \queq{\msnonzero}. The first
bound in \queq{\deriveps} follows because $m_{bl} $ is $C^{k-2}$ and
$\abs{\sfrac{\del^s}{\del\th^s} m_{bl}}\leq \Const \abs{e}_{s+2}$.
The constant depends on $w_\zer$, $g_\zer$, and $r_\zer$.
For the second bound, we use that $x \ m_{bl}$ and $y\ m_{bl}$ are
$C^{k-1}$. This implies 
$$
\abs{\frac{\del^{k-1} \et_b}{\del\th^{k-1}}} = 
\abs{x\frac{\del^{k-1}}{\del\th^{k-1}} y\ m_{bl}(x,y,\th)} \leq
\Const \abs{e}_k \ \abs{x}
\eqn $$
and, exchanging $x$ and $y$,
$$
\abs{\frac{\del^{k-1} \et_b}{\del\th^{k-1}}} \le 
\Const \abs{e}_k \ \abs{y}
.\eqn $$
Combining these two equations, we get
$$
\abs{\frac{\del^{k-1} \et_b}{\del\th^{k-1}}} \le 
\Const \abs{e}_k \min\{|x|,|y|\} \le
\Const \abs{e}_k \sqrt{|xy|}
\eqn $$
which implies \queq{\deriveps}. \endproof

\noindent
To take the $\rh$--dependence into account, we  use the following Lemma.
\Lem{\contagain}{\Lesty
For any $\veps >0$ 
there is $\de > 0$ such that for all $\th^{(0)}$ and 
for all $\rh_\one $ and $\rh_\two $ with 
$\abs{\rho_i} < \de$  and all $\th$ with 
$\abs{\th-\th^{(0)}} < \de$,  
all the solutions to the critical point equations
$$
\nabla  e\left(L_b(\p(\rh_\one,\th_\one),\p(\rh_\two,\th_\two),
\p(0,\th))\right)
\cdot \del_\th \p(\rh_i,\th_i) = 0 
 $$ 
for $i = 1,2$ are in $\veps$--neighbourhoods of the
critical points $c_{b,l}(\th^{(0)})$.
}
\Proof  See Lemma \qCP\ and Remark \compcont. \endproof   

\goodbreak
\def\SP#1#2{S_{#1,j_{#1}}^{(#2)}}
\def\YY{Z}
\def\ulnu{\underline{\nu}}
We now prove a more general statement that implies Lemma \jaythree,
and which we shall use later to bound more general graphs and
also derivatives with respect to $e$. The generalization is about
the $\th$--behaviour of the propagators associated to lines.
The free propagator $C_j$ is independent of the angular variable,
that was the whole point of the above strategy of removing the
dependence on the external angle from $C_{j_\thr}$. However, one
can prove the same bounds for more general propagators if they depend on $\th$
in a way that taking $\th$--derivatives does not deteriorate
their scale behaviour. The strings of two--legged
subdiagrams that occur in general graphs and their 
derivatives with respect to $e$ have this property, 
i.e.\ they satisfy the hypotheses
of the following theorem.  
\The{\dreipro}{\Thsty 
Let $d=2$, and \ATwo{k,0}
and \AThr--\AFiv hold. 
Let $A \in C^{k-1}((\R\times \cN)^3, \C)$ and for $\nu \in \{1,2,3\}$
let $\SP{1}{\nu}$, $\SP{2}{\nu}$, $\SP{3}{\nu}$ be in 
$C^{k-1} (\R \times \cN, \C)$ and satisfy for $l \in \{1,2,3\}$
$$
\abs{D^\al \SP{l}{\nu} (p)} \le 
\Ga_{l,\nu,|\al|} M^{-j_l (\nu+|\al|)}
\True{|ip_\zer-e(\p)| \le M^{j_l} } \qquad
\forall \; 0 \le |\al| \le k-1
\EQN\onlyyou $$
and for all $s \le k-1$
$$
\abs{\sfrac{\del^s}{\del\th^s} \SP{l}{\nu} (p_\zer,\p(\rh,\th))}
\le \Ga_{l,\nu,s} M^{-\nu j_l}\; \True{|ip_\zer-\rh| \le M^{j_l} } 
\EQN\thunif $$
(where the constants $\Ga_{l,\nu,s}$ are increasing in $s$).
Let $\ulj = (j_\one,j_\two,j_\thr)$ with $j_\one \le j_\two \le
j_\thr$, 
$\ulnu=(\nu_\one,\nu_\two,\nu_\thr)$, and
$$
X_{\ulj}^{b,\ulnu} (p) = 
\int dp_\one \; dp_\two \; \SP{1}{\nu_\one}(p_\one) 
\SP{2}{\nu_\two} (p_\two) \SP{3}{\nu_\thr} \left( L_b(p_\one,p_\two,p)\right)
A(p_\one,p_\two,p)
\EQN\Xuljbulnudef $$
where $L_b$ is defined in \queq{\Ladef}.
Then, for all $s \le k-1$, there is $Q_s > 0$ such that for all
$\nu_\one,\nu_\two,\nu_\thr \in \{1,2,3\}$ and all $b \in\{1,2,3\}$
$$
\sup\limits_{\th}\sup\limits_{|p_\zer-i\rh| \le M^{j_\one}} \;
\abs{\sfrac{\del^s}{\del\th^s }X_{\ulj}^{b,\ulnu} (p_\zer, \p(\rh,\th))}
\le Q_s \abs{A}_s \; M^{j_\one(2-\nu_\one)+j_\two(2-\nu_\two)}\;
\cases{
|j_\thr|\ M^{(1-\nu_\thr) j_\thr }  & if $s \le k-2$
\cr
|j_\thr|\ M^{({1 \over 2}-\nu_\thr) j_\thr } & if $s=k-1$.}
\EQN\Xuljb $$
The constant $Q_s$ depends only on $s$, the constants $\Ga_{l,\nu,s}$,
$r_\zer$, $w_\zer$, and $|e|_s$.
}

\Proof We introduce the coordinates $(\rh,\th)$ such that 
$e(\p(\rh,\th))=\rh$, write
$p_i=(\om_i,\p(\rh_i,\th_i))$, and denote $p_\thr^{(b)} = 
L_b(p_\one,p_\two,p)$. Then
$$
X_{\ulj}^{b,\ulnu} (p) = \int d\om_\one d\rh_\one 
\int d\om_\two d\rh_\two \YY_{\ulj}^{b,\ulnu} 
(p,\om_\one,\rh_\one,\om_\two,\rh_\two) 
\eqn $$
with
$$
\YY_{\ulj}^{b,\ulnu} = \int
d\th_\one J(\rh_\one,\th_\one)\; 
\int d\th_\two J(\rh_\two,\th_\two)
\; \SP{1}{\nu_\one} (p_\one)\; \SP{2}{\nu_\two} (p_\two)\; 
\SP{3}{\nu_\thr}(p_\thr^{(b)}) \; A(p_\one,p_\two,p)
\eqn $$
The momentum 
$p_\thr^{(b)}=L_b(p_\one,p_\two,p)$ defines $\om_\thr^{(b)}$,
$\rh_\thr$, and $\th_\thr$ by 
by $p_\thr^{(b)}=\big(\om_\thr^{(b)},\p(\rh_\thr,\th_\thr)\big)$ where
$$
\rh_\thr=e\left(L_b(\p_\one,\p_\two,\p)\right)
\eqn $$
and where $\th_\thr$ is also a function of $p_\one,p_\two$ and
$p$. In particular, both $\rh_\thr $ and $\th_\thr$ depend on
$\th$. In view of \queq{\thunif}, it is desirable to change variables
in this integral, to make $\rh_\thr$ independent of $\th$ because
then the $j_\thr$--behaviour of the integral is not deteriorated
by derivatives with respect to $\th$. 

Fix $b\in \{ 1,2,3\}$. 
Fix $\de>0$ so small that it satisfies the hypothesis of Lemma \Unif, 
and such that the $3\de$--neighbourhoods of the critical points are
mutually disjoint.
Fix $r_\zer > 0$ so small that for all $|\rh_i| <r_\zer$, all $\th$ with
$|\th-\th^{(0)}| < r_\zer$, and all $l\in\{ 1,2,3\}$, 
$$
\abs{c_{bl}(\th)-c_{bl}(\th^{(0)})} < \frac{\de}{2} 
\eqn $$
(this is possible by Lemma \contagain). Fix $\th^{(0)}\in \R/2\pi\Z$.
Denote $\ulth=(\th_\one,\th_\two)$. 
Let 
$$
R=\left(\R/2\pi\Z\right)^2 \setminus \bigcup\limits_{l=1}^3
\{ \ulth : \abs{\ulth - c_{bl}(\th^{(0)})} < \de\}
,\eqn $$
then every $\ulth \in R$ has distance 
at least $\sfrac{\de}{2}$ from the critical points. 
$R$ is compact, so there is $\xi=\xi(\de)>0$ such that 
for all $|\rh_i| <r_\zer$, all $\th$ with
$|\th-\th^{(0)}| < r_\zer$, and all $\ulth\in R$
$$
\abs{\nabla e(L_b(\p(\rh_\one,\th_\one),\p(\rh_\two,\th_\two),\p(0,\th)))}
\ge 2 \xi
.\eqn $$
$\nabla e$ is uniformly continuous on that set, so $r_\zer$ can be chosen 
so small that there is $N=N(\de)\in \N$ 
and a partition of unity $1=\sum\limits_{l=1}^N \ch_l$
with 
$$
\ch_l(\ulth) = \cases{
1 &$\abs{\ulth-c_{bl}(\th^{(0)})} <  \de$ \cr
0 &$\abs{\ulth-c_{bl}(\th^{(0)})} > 2\de$ \cr}
\qquad \hbox{ for } l \in \{1,2,3\}
,\EQN\cinftyhat $$
and such that for all $l \ge 4$, 
$\ulth\in \supp \ch_l$ implies either 
$$
\inf\limits_{|\rh_\one| < \de, |\rh_\two| < \de}\;
\inf\limits_{|\th'-\th^{(0)}| \le r_\zer} \;
\abs{\frac{\del }{\del \th_\one} e\big( L_b(\p(\rh_\one,\th_\one),
\p(\rh_\two,\th_\two),\p(0,\th')\big)}\ge \xi
\eqn $$
or
$$
\inf\limits_{|\rh_\one| < \de, |\rh_\two| < \de}\;
\inf\limits_{|\th'-\th^{(0)}| \le r_\zer} \;
\abs{\frac{\del }{\del \th_\two} e\big( L_b(\p(\rh_\one,\th_\one),
\p(\rh_\two,\th_\two),\p(0,\th')\big)}\ge \xi
.\EQN\letssay $$
Note that the just constructed functions $\ch_l$ depend only on $\th^{(0)}$
and $\de$; they are independent of $\th$. Note also that $\de$, $\xi$, 
and the $\ch_l$ are independent of the scales $j_\one,j_\two,j_\thr$.
Let $\th$ be such that $|\th-\th^{(0)}| < r_\zer$.   
We insert the partition of unity in the integral for $\YY$ and get
$$
\YY_{\ulj}^{b,\ulnu} =\sum\limits_{l=1}^N \YY_{\ulj,l}^{b,\ulnu}
\eqn $$
where the integrand for $ \YY_{\ulj,l}^{b,\ulnu} $ is that of 
$\YY_{\ulj}^{b,\ulnu}$ times $\ch_l(\ulth)$. We do not write all
the arguments $(p,\om_\one,\rh_\one,\om_\two,\rh_\two)$ of 
$\YY_{\ulj}^{b,\ulnu}$ for brevity. Note that $\YY_{\ulj}^{b,\ulnu}$
vanishes if $|i\om_\one - \rh_\one | > M^{j_\one}$ or if
 $|i\om_\two - \rh_\two | > M^{j_\two}$  because of the 
support properties of the propagators $\SP{k}{\nu_k}$, so that
in deriving bounds below we may always assume that 
$|\om_k|$ and $|\rh_k|$ are at most $M^{j_k}$. 

For $l\ge 4$, an ordinary change of variables from $\th_\one $ or 
$\th_\two $ to $e$ is possible. If, for instance, for the given $l$
\queq{\letssay} holds, then we 
can write $\th_\two$ as a function of $e$ and the other variables,i.e.,
$$
\th_\two=\th_\two(\th,e,\th_\one,\rh_\one,\rh_\two)
.\eqn $$
By the implicit function theorem, this function exists, and 
$\th_\two$ is a $C^k$ function of $(\th,e,\th_\one,\rh_\one,\rh_\two)$.
The Jacobian  
$$
\tilde J(\th,e,\th_\one,\rh_\one,\rh_\two) = 
\left\lbrack\left( \big(\frac{\del e_b }{\del \th_\two}\big)\big( 
L_b(\p(\rh_\one,\th_\one), \p(\rh_\two,\th_\two),\p(0,\th))\big)
\right)^{-1}
\right\rbrack_{\th_\two=\th_\two(\th,e,\th_\one,\rh_\one,\rh_\two)}
\EQN\tilJac $$
of this change of variables is $C^{k-1}$ in all its arguments. 
By \queq{\letssay}, it satisfies, for all $l \leq k-1$,
$$
\abs{\frac{\del^l \tilde J}{\del \th^l}} \leq 
2^l \; l!
\ \xi^{-l-1}\ \left(1+\abs{\ph}_{l}\right)^l
\EQN\tiJnu $$
where $\ph(\th) = \sfrac{\del e_b }{\del \th_\two} (\th)$. 
By construction, $\abs{\ph}_{l}$ is also bounded by a function of
$\abs{e}_{l+1}$ and $\xi$. 
Changing variables to $e$ in the integral for $\YY$, we get
$$\eqalign{
\YY_{\ulj,l}^{b,\ulnu} &= \int d\th_\one J(\rh_\one,\th_\one)
\int de\; J(\rh_\two, \th_\two(\th,e)) \; \tilde J(\th,e)\;
\ch_l(\th_\one,\th_\two(\th,e)) A(p_\one,p_\two,p)\cr
&\SP{1}{\nu_\one} (p_\one) \;
\SP{2}{\nu_\two} \left(\om_\two,\p(\rh_\two,\th_\two(\th,e))\right)
\;
\SP{3}{\nu_\thr}\left(\om_\thr^{(b)},\p(e,\th_\thr^{(b)} (\th,e))\right)
.\cr}\EQN\YYyy $$
Although $\th_\two (\th,e)$ and $\th_\thr^{(b)}$ also depend 
on $\th_\one$, $\rh_\one$,
and $\rh_\two$, we suppressed this in the notation since this
dependence is harmless in the following estimates by \queq{\letssay}
and \queq{\tiJnu}. 
By the assumptions on $\SP{i}{\nu_i}$ and $A$, and since $J \in C^{k-1}$ because
$e\in C^k$, the integrand is $C^{k-1}$ in $\th$. 
We take $s \le k-1$ derivatives with respect to $\th$, to get 
$$\eqalign{
\abs{\sfrac{\del^s}{\del\th^s}\YY_{\ulj,l}^{b,\ulnu}} & \le 
\ga M^{-j_\one\nu_\one-j_\two\nu_\two-\nu_\thr j_\thr}\; 
\True{|i\om_\one-\rh_\one| \le M^{j_\one}} \;
\True{|i\om_\two-\rh_\two| \le M^{j_\two}}
\int de\; \True{|i\om^{(b)}_\thr -e| \le M^{j_\thr}}
\cr}\eqn $$
with 
$$\eqalign{
\ga &=2\pi|J|_\zer \Ga_{1,\nu_\one,0} 
\sum\limits_{s_1+s_2+s_3+s_4=s} \frac{s!}{s_1!s_2!s_3!s_4!}
\Ga_{2,\nu_\two,s_2} \Ga_{3,\nu_\thr,s_3}
\sup \abs{\sfrac{\del^{s_\one}}{\del \th^{s_\one}} (J\tilde J \ch_l)}
\sup\abs{\sfrac{\del^{s_4}}{\del \th^{s_4}} A} \cr
& \le 2 \pi |J|_\zer 4^s |A|_s \Ga_{1,\nu_\one,s}
\Ga_{2,\nu_\two,s} \Ga_{3,\nu_\thr,s}
\sup \abs{\sfrac{\del^{s_\one}}{\del \th^{s_\one}} (J\tilde J \ch_l)}
\sup\abs{\sfrac{\del^{s}}{\del \th^{s}} A} 
.\cr}\eqn $$
Thus, for $l \ge 4$, 
$$
\abs{\sfrac{\del^s}{\del\th^s}\YY_{\ulj,l}^{b,\ulnu}}\le 2 \ga\; 
M^{-j_\one\nu_\one-j_\two\nu_\two-\nu_\thr j_\thr}\; 
\True{|i\om_\one-\rh_\one| \le M^{j_\one}} \;
\True{|i\om_\two-\rh_\two| \le M^{j_\two}}
\eqn $$
and therefore
$$
\abs{\int d\om_\one d\rh_\one d\om_\two d\rh_\two
\sum\limits_{l=4}^N \sfrac{\del^s}{\del\th^s}\YY_{\ulj,l}^{b,\ulnu}} \le 
2 \pi^2 N \ga \; 
M^{(2-\nu_\one) j_\one+(2-\nu_\two) j_\two+(1-\nu_\thr) j_\thr}
.\eqn $$
So the contribution from the region away from the singularities
actually fulfills a better bound than \queq{\Xuljb} (recall that
$N=N(\de)$ is fixed independently of the scales). 

We now turn to the singular region: let $l \in \{ 1,2,3\}$.
Here the dependence of $\SP{3}{\nu_\thr}$ on $\th$ cannot be removed,
but we shall use the specific form of the singularity proven in 
Lemma \Unif\ to give bounds. We first do a Taylor expansion 
in $\rh_\one$ and $\rh_\two$ to reduce the problem to all vectors
being on the Fermi surface (if $\rh \ne 0$, we expand in $\rh_\one-\rh$
and $\rh_\two-\rh$ instead of $\rh_\one$ and $\rh_\two$.
This does not make any difference because $\rh$
is assumed to be less than $M^{j_\one}$ in the statement of the
Lemma, because all constants are uniform in $\rho$ for 
$|\rh| < \de$, and because we take a derivative with respect
to $\th$, and not with respect to $\rh$. 
We may therefore specialize to $\rh=0$ without
loss of generality). This gives 
$$
e\left(L_b(\p(\rh_\one,\th_\one),\p(\rh_\two,\th_\two),
\p(0,\th))\right) = \et_b (\th_\one, \th_\two, \th )
+\rh_\one \ v_\one + \rh_\two \ v_\two
\eqn $$
with 
$$
v_i(\th,\rh_\one,\rh_\two,\th_\one,\th_\two) =
\int\limits_0^1 d\al \; \nabla
e\left(L_b(\p(\al\rh_\one,\th_\one),\p(\al\rh_\two,\th_\two),
\p(0,\th))\right) \cdot \del_\rh\p(\al\rh_i,\th_i))
.\eqn $$
$v_\one$ and $v_\two$ are $C^{k-1}$ in $\th$. 
We change variables from $\ulth$ to $\ult=\ulth-c_{bl}(\th)$.
The Jacobian of this change of variables is one. 
By construction of $\ch_l$, and because $|\th-\th^{(0)}| < r_\zer$,
$\supp \ch_l \subset U_{3\de}(c_{bl}(\th))$.
Thus, we may use Lemma \Unif\ to change variables from $\ult$ to $x,y$.
By \queq{\noJac}, the Jacobian of this change of variables is 
again one. Calling $(\th^*_\one,\th^*_\two)=c_{bl}(\th)$
and $\Ga_{bl}^{-1}(x,y,\th) = (t_\one(\th,x,y),t_\two(\th,x,y),\th)$,
the integral for $\YY$ is
$$\eqalign{
\YY_{\ulj,l}^{b,\ulnu} (\om,\p(\rh,\th),\om_\one, \ldots, \rh_\two)
& = 
\int dx\; dy \; J(\rh_\one,\th^*_\one+t_\one(\th,x,y)) 
J(\rh_\two,\th^*_\two+t_\two(\th,x,y))
\ch_l(c_{bl}(\th)+\ult(\th,x,y))\cr
&  A(p_\one,p_\two,p) \; \SP{1}{\nu_\one} (p_\one)\; 
\SP{2}{\nu_\two} (p_\two) \;
\SP{3}{\nu_\thr} \left(\om_3^{(b)},\p(E_{bl},\th_3^{(b)} (x,y))\right)
\cr}\eqn $$
with $p_\one$ and $p_\two$ now being rewritten in terms of $(x,y)$
in the obvious way, i.e.\ 
$\p_\one = \p(\rh_\one, \th^*_\one+t_\one(\th,x,y))$ etc.,
and with
$$
E_{bl}= m_{bl}(x,y,\th) \ xy + \tilde v_\one \rh_\one + \tilde v_\two \rh_\two
\EQN\Eblxydef $$
and 
$$
\tilde v_i = v_i (\th,\rh_\one,\rh_\two,c_{bl}(\th)+\ult(\th,x,y))
.\eqn $$
By Lemma \Unif, the integrand is $C^{k-1}$ in $\th$. We apply
$s\le k-1$ derivatives. $\SP{1}{\nu_\one}$ and $\SP{2}{\nu_\two}$ depend on
$\th$ only through their angular variables $\th_\one$ and $\th_\two$,
so \queq{\thunif} applies there. Thus
$$\eqalign{
\abs{\sfrac{\del^s}{\del\th^s}\YY_{\ulj,l}^{b,\ulnu}} \le 
M^{-\nu_\one j_\one-\nu_\two j_\two} & \int dx\; dy\; 
\sum\limits_{s_\zer+\ldots+s_4=s} 
\frac{s!}{s_\zer! \ldots s_4!}
 \; \Ga_{1,0,s_\one} \Ga_{2,0,s_\two}\; 
\abs{\sfrac{\del^{s_\zer}}{\del\th^{s_\zer}} (J_\one J_\two \ch_l
A)} 
\cr & 
\sup\abs{\sfrac{\del^{s_4}}{\del \th^{s_4}} A} \; 
\abs{\sfrac{\del^{s_\thr}}{\del\th^{s_\thr}}
\SP{3}{\nu_\thr} \left(\om_r^{(b)}, \p(E_{bl},\th_3^{(b)} (x,y,\th))\right)}
.\cr}\eqn $$ 
As mentioned, $\th_3^{(b)}$ is $C^{k-1}$ in $\th$ by Lemma \contagain.
Whenever the $\th$ derivative acts on $\SP{3}{\nu_\thr}$ through
the dependence of $\th_3^{(b)}$ on $\th$, it acts
only tangentially, so \queq{\thunif} ensures that the scale behaviour
of such terms remains $M^{-j_\thr \nu_\thr}$. The dangerous terms are
those where $\SP{3}{\nu_\thr}$ gets differentiated because of
the dependence of $\rh_\thr$ on $\th$, i.e.\ when $\del_E \SP{3}{\nu_\thr} = 
\sfrac{\del }{\del E_{bl}}\SP{3}{\nu_\thr}$
or higher such derivatives occur, because the only bound we have
for that case is \queq{\onlyyou}. However, there are small factors
in the numerator because of the following. 
By Lemma \Unif\ $(ii)$, and since for $i=1,2$, 
$|\rh_i|<M^{j_i} \leq M^{j_\thr}$,
$$
\abs{\frac{\del^s E_{bl}}{\del \th^s}} \leq \Const M^{j_\thr}
\qquad \hbox{ for all } s \le k-2
\EQN\kaminzwo $$
and 
$$
\abs{\frac{\del^{k-1} E_{bl}}{\del \th^{k-1}}} 
\leq \Const M^{\sfrac{j_\thr}{2}}
.\EQN\kaminein $$
The $\del_E $ derivatives acting on the propagator $ \SP{3}{\nu_\thr} $
give terms of the form
$$
\left(\del_E^{n} \SP{3}{\nu_\thr} \right)\prod_{i=1}^n
\left(\del_{\th}^{\ell_i}E_{bl}\right)
$$
with $\sum_{i=1}^n\ell_i\le s \le k-1$.  
At most one $\ell_i$ 
can be $k-1$, and this case can occur only if $s=k-1$ and $n=1$.
If no $\ell_i$ is $k-1$, \queq{\kaminzwo} applies, and 
$\abs{\sfrac{\del^{s_\thr}}{\del\th^{s_\thr}}\SP{3}{\nu_\thr}}$
is bounded by 
$$
\Const M^{-j_3(n+\nu_\thr)}M^{j_3n}=\Const M^{-\nu_\thr j_3}
\EQN\fastscho$$
with the first factor coming from the $\left(\del_E^{n}\SP{3}{\nu_\thr} \right)$
and the second from $n$ $\del_{\th}^{\ell_i}E_{bl}$'s. 
If one of the $\ell_i$ is $k-1$, then $n=1$ and \queq{\kaminein} applies, and 
$\abs{\sfrac{\del^{s_\thr}}{\del\th^{s_\thr}}\SP{3}{\nu_\thr}}$
is bounded by 
$$
\Const M^{-j_3(1+\nu_\thr)}M^{\sfrac{j_\thr}{2}}= 
\Const M^{-({1\over 2}+\nu_\thr )j_3}
\EQN\fastschk$$
with the first factor coming from the $\left(\del_E\SP{3}{\nu_\thr} \right)$
and the second one from the one $\del_{\th}^{\ell_i}E_{bl}$
with $\ell_i=k-1$. 

The volume of the $(x,y)$--integration is bounded by
$$\eqalign{
\int\limits_{\Ga_{bl}(U_{3\de}(0,0))} 
dx\ dy\; \True{|m_{bl}(\th,x,y) xy| \le M^{j_\thr}} &\le
\int\limits_{\sqrt{x^2+y^2} < r_D} 
dx\ dy\; \True{|xy|\le \frac{2}{w_\zer} M^{j_\thr}} \cr
&\le
\Const |\log r_D|\; |j_\thr|\; M^{j_\thr}
\cr}\eqn $$
where $r_D$ is the radius of a disk containing $\Ga_{bl}(U_{3\de}(0,0))$. 
This multiplies \queq{\fastscho} and \queq{\fastschk} by a factor 
$|j_\thr|\ M^{j_\thr}$.\endproof

\sni
{\bf Proof of Lemma \jaythree\ for $d=2$:} 
We take one derivative right away and get
$$\eqalign{
\frac{\del}{\del \th}Y_{\ulj}^\pi (0,\p(0,\th)) & = 
\int dp_\one \int dp_\two
C_{j_\one}((p_\one)_\zer,e(\p_\one))\ 
C_{j_\two}((p_\two)_\zer,e(\p_\two)) \cr
&\left(
\del_\th P_\pi (p_\one,p_\two,p) C_{j_\thr} (\ze_b,e_b) +
P_\pi(p_\one,p_\two,p) \del_e 
C_{j_\thr} (\ze_b,e_b) \frac{\del e_b}{\del \th}
\right) 
\cr}\EQN\dochzitiert $$ 
where $b=\pi(3)$, and $\ze_b$ and $e_b$ are defined in \queq{\eazea}.
The integrand is still $C^{k-1}$ in $\th$.
Theorem \dreipro\ applies to both terms. In the first term, take
$$
A(p_\one,p_\two,p) = \del_\th P_\pi (p_\one,p_\two,p)
\eqn $$
and $\nu_\one=\nu_\two=\nu_\thr =1$, and 
$$
\SP{3}{\nu_\thr} (p) = C_{j_\thr} (p_\zer, e(\p))
.\eqn $$
Obviously, then, 
$$
\frac{\del}{\del \th}\SP{3}{\nu_\thr} (p_\zer, \p(\rh,\th)) = 0
\eqn $$
so \queq{\thunif} holds, and \queq{\onlyyou} holds with 
$\Ga_{3,1,s} = W_s$, $W_s$ given in \queq{\shellj}. 
So Theorem \dreipro\ applies and proves the bound for that term.
In the second term, choose 
$$
A(p_\one,p_\two,p) = P(p_\one,p_\two,p)\del_\th e_b
\EQN\myA $$
and $\nu_\one=\nu_\two =1$, $\nu_\thr=2$, and 
$$
\SP{3}{\nu_\thr} (p) = \del_e C_{j_\thr} (p_\zer, e)\vert_{e=e(\p)}
.\eqn $$
Again, the hypotheses of Theorem \dreipro\ are satisfied, and the
bound holds. 
\endproof

\Rem{\Constants} The above proof shows that $\de $ depends only on 
$w_\zer$, and $r_\zer $ is chosen depending only on $\de$ and $g_\zer$.
Thus these constants are fixed once $w_\zer$ is given, and hence 
uniform on the set of $e$'s that satisfy our hypotheses. 
Note also that $\abs{e}_k$
does occur in the bound: the $A$ of
\queq{\myA} already contains a derivative of $e$, and 
$|A|_{k-1}$ appears when Theorem \dreipro\ is applied with $s=k-1$.

\Rem{\thetavsrho} One may wonder where the consistency condition
($\sfrac{\del \et}{\del \th} =0$ whenever 
$\sfrac{\del \et}{\del \th_\one} = \sfrac{\del \et}{\del \th_\two} =0$)
enters in the above proof, and why the proof does not work
for derivatives with respect to $\rh$.  The problem
is that one cannot prove the analogue
of Lemma \Unif\ $(ii)$ for the $\rh$--derivative. 
Consider, e.g., the function
$$
H=e\left(\p(\rh_\one,\th_\one)+\p(\rh_\two,\th_\two)-\p(\rh,\th)\right)
.\eqn $$
Proceeding in exactly the same way as above would mean first writing 
$$
H= e\left(\p(0,\th_\one)+\p(0,\th_\two) -\p(\rh,\th)\right)
+\rh_\one \tilde v_\one + \rh_\two \tilde v_\two
.\eqn $$
The $\rh_\one$ and $\rh_\two$--terms from the Taylor expansion
still provide enough small factors for \queq{\kaminzwo}
and \queq{\kaminein}, but the analogue of the function $m_{bl} xy$ 
appearing in 
\queq{\Eblxydef} does not: the critical points 
$\th_\one^*(\rh,\th),\th_\two^*(\rh,\th)$ of the function
$e\left(\p(0,\th_\one)+\p(0,\th_\two) -\p(\rh,\th)\right)$
now depend on $\rh$ (their existence is proven
in Lemma B.3), and most importantly, the function need not vanish 
at those critical points, so that \queq{\noconsterm} gets 
replaced by
$$
 e\left(\p(0,\th_\one)+\p(0,\th_\two) -\p(\rh,\th)\right)= f_\zer
(\rh,\th) + (\th_\one-\th_\one^*) (\th_\two-\th_\two^*) 
\tilde m(\rh,\th,\th_\one,\th_\two)
\eqn $$
with 
$$
f_\zer (\rh,\th) = e\left(\p(0,\th_\one^*(\rh,\th)) +
\p(0,\th_\two^* (\rh,\th)) -\p(\rh,\th)\right)
\eqn $$
($f_\zer$ is analogous to the function $f_\zer (\q)$ appearing
in $(B.48)$). The crucial point is that the $\rh$--derivative
of $f_\zer$ is not small. In fact, it is near to maximal
(and hence $O(1)$) since
$\th$-- and $\rh$--lines are transversal to each other in $\p$--space,
so whenever $\del_\th\p$ is orthogonal to $\nabla e$, 
$\del_\rh \p$ will point almost in the same direction as $\nabla e$. 

The above problem cannot simply be circumvented by expanding $\rh_\one$
and $\rh_\two$ around $\rh$, because in 
$$
H= e\left(\p(\rh,\th_\one)+\p(\rh,\th_\two) -\p(\rh,\th)\right)
+(\rh_\one-\rh) \tilde v_\one + (\rh_\two-\rh) \tilde v_\two
,\eqn $$
the $\rh$--derivative can now also act on the prefactor of $\tilde
v_\one$, and $\abs{v_\one}$ is again not zero, but near to maximal
at the $\th$--critical points.

\sect{H\" older Continuity}
\noindent
In this section, we prove H\" older continuity of the second
order counterterm under the assumption that $\hat v$ and $e$
have the same properties. The main reason why we can show this
additional regularity is that, by the above 
theorems, there is still a decay of almost $M^{j/2}$ left in the scale
sums, so the usual counting of derivatives by factors $M^{-j}$
suggests that one can still afford almost half a derivative,
i.e., H\" older continuity with any exponent 
$\be < \sfrac{1}{2}$. The proof will be a not very difficult
add--on to the proof of  Theorem \dreipro. Basically, we use
that the highest derivative can only appear linearly, 
take the differences required in the H\" older inequality,
and use the differencing formula
$$
\prod\limits_{i=1}^n \ph(\xi_i) -
\prod\limits_{i=1}^n \ph(\xi_i') =
\sum\limits_{k=1}^n \left(\prod\limits_{i<k}\ph(\xi_i)\right)\;
\left(\ph(\xi_k)-\ph(\xi_k')\right)\;
\left(\prod\limits_{i>k}\ph(\xi_i')\right)
\EQN\diffi $$
to reexpress this as a sum over differences of each factor 
in the integrand. The differences are either estimated by 
Taylor expansion or by the according H\" older property of $e$
and $\hat v$. 

\The{\dreiHoel}{\Thsty Assume the hypotheses of Theorem $\dreipro$.
Let $0 < \be < \sfrac{1}{2}$ and assume that $e \in C^{k,\be}$,
$A\in C^{k-1,\be}$, and $S_{l,j_l}^{(\nu_l)} \in C^{k-1,\be}$.
That is, there are constants $H_e(\be)>0$, $H_A(\be)>0$ and 
$H_S(\be)>0$ such that for all multiindices $\al$: if $|\al|=k$,
for all $\p,\p'\in \cB$
$$
\abs{D^\al e (\p) - D^\al e(\p')} \le H_e \abs{\p-\p'}^\be
\EQN\eHoelAss $$
and if $|\al| = k-1$, 
$$
\abs{D^\al A (p_\one,p_\two,p) - D^\al A(p_\one',p_\two',p')}
\le H_A \abs{(p_\one,p_\two,p)-(p_\one',p_\two',p')}^\be
,\EQN\AHoelAss $$
and for $\abs{p-p'} \le M^j$,
$$
\abs{D^\al S_{l,j_l}^{(\nu_l)} (p) - D^\al S_{l,j_l}^{(\nu_l)} (p')}
\le H_S \abs{p-p'}^\be \; 
M^{-j_l (\nu_l+k)} \; 
\True{\abs{ip_\zer - e(\p)} \le G_\zer M^{j_l}}
\EQN\SHoelAss $$
Finally, we assume that
for $|\rh-\rh'| \le M^{j_l}$ and $|\th-\th'| \le M^{j_l}$,
$$\eqalign{
& \abs{\left(\sfrac{\del}{\del \th}\right)^{k-1} S_{l,j_l}^{(\nu_l)} (p_\zer,
\p(\rh,\th)) - \left(\sfrac{\del}{\del \th}\right)^{k-1} S_{l,j_l}^{(\nu_l)} (p_\zer,
\p(\rh',\th'))} \cr
& \qquad \le H_S \abs{\p(\rh,\th)-\p(\rh',\th')}^\be \; 
M^{-j_l (\nu_l + 1)} \; 
\True{\abs{ip_\zer - \rh} \le 4 M^{j_l}}
\cr}\EQN\SHoelAssp $$
Then the function $X_{\ulj}^{b,\ulnu} $, defined in \queq{\Xuljbulnudef},
satisfies: for all $\rh,\rh'$ with $\max\{|\rh|,|\rh'|\} \le M^{j_\one}$,
all $\th$ and $\th'$, and all $|p_\zer | \le M^{j_\one}$, 
$$\eqalign{
&\abs{\left(\sfrac{\del}{\del \th}\right)^{k-1}X_{\ulj}^{b,\ulnu} (p_\zer,\p(\rh,\th))
-\left(\sfrac{\del}{\del \th}\right)^{k-1}X_{\ulj}^{b,\ulnu} 
(p_\zer,\p(\rh',\th'))} \cr
& \qquad \le \tilde Q \abs{\p(\rh,\th) - \p(\rh',\th')}^\be 
M^{j_\one(2-\nu_\one-\be)+j_\two(2-\nu_\two) + j_\thr({1\over
2} - \nu_\thr) } |j_\thr|
.\cr}\EQN\jHoel $$
If $\nu_\one = 1$ and $\nu_\two+\nu_\thr \le 3$, the scale
sum 
$$
\sum\limits_{\ulj \in \cM} \left(\sfrac{\del}{\del \th}\right)^{k-1}
X_{\ulj}^{b,\ulnu} 
\EQN\MHoel $$
converges absolutely to a uniformly H\" older continuous function
$X_\cM$ for $\cM$ any of $\Om, T,D$ (defined in \queq{\allthosesets}).
}

\Proof Let 
$$
\Ga(p,p') = 
\left(\sfrac{\del}{\del \th}\right)^{k-1}X_{\ulj}^{b,\ulnu} 
(p_\zer,\p(\rh,\th))
-\left(\sfrac{\del}{\del \th}\right)^{k-1}X_{\ulj}^{b,\ulnu} 
(p_\zer,\p(\rh',\th'))
.\eqn $$
If $\abs{\p - \p'} \ge M^{j_\one}$, then by \queq{\Xuljb}
$$\eqalign{
\abs{\frac{\Ga(p,p')}{|\p-\p'|^\be}} & \le M^{-\be j_\one} 
Q_s \abs{A}_s \; M^{j_\one(2-\nu_\one)+j_\two(2-\nu_\two)}\;
|j_\thr|\ M^{({1 \over 2}-\nu_\thr) j_\thr } 
\cr}\eqn $$
which is the stated bound. Thus, it suffices
to consider the case $\abs{p-p'} \le M^{j_\one}$ in the proof.

The regular and singular region
are chosen as in the proof of Theorem 3.5. We prove the H\" older
bound for the contribution from the regular region. The
contribution from the singular region is bounded in exactly the
same way.  Let $x=(\rh_\one,\rh_\two,\om_\one,\om_\two)$.
We take $k-1$ derivatives with respect to $\th$
of \queq{\YYyy}, to get 
$$\eqalign{
\sfrac{\del^{k-1}}{\del \th^{k-1}} 
\YY_{\ulj,l}^{b,\ulnu} (p,x) &= 
\int d\th_\one J(\rh_\one,\th_\one)
\SP{1}{\nu_\one}(p_\one)
\int de \; 
\sum\limits_{s_1+s_2+s_3+s_4=k-1} \frac{(k-1)!}{s_1!s_2!s_3!s_4!}
\left(\sfrac{\del\;}{\del\th}\right)^{s_\one} (J\tilde J \ch_l)
\cr
&
\left(\left(\sfrac{\del\;}{\del\th}\right)^{s_4} A \right)\;
\left(\left(\sfrac{\del\;}{\del\th}\right)^{s_\two} \SP{2}{\nu_\two}(p_\two)\right)
\left(\sfrac{\del\;}{\del\th}\right)^{s_\thr} 
\SP{3}{\nu_\thr}\left(\om_\thr^{(b)},\p(e,\th_\thr^{(b)} (\th,e))\right)
\cr}\eqn $$
where in the integrand $p_\one = (\om_\one,\p(\rh_\one,\th_\one))$,
$p_\two = (\om_\two, \p(\rh_\two, \th_\two(\th,e))$, and
$p=(\om,\p(\rh,\th))$.
We take the difference
$$
\De = \sfrac{\del^{k-1}}{\del \th^{k-1}}\YY_{\ulj,l}^{b,\ulnu} (p,x)
-
\sfrac{\del^{k-1}}{\del \th^{k-1}}\YY_{\ulj,l}^{b,\ulnu} (p',x)
\eqn $$
and split $Z=Z_\one+Z_\two$ and accordingly, $\De=\De_\one+\De_\two$,
by regrouping the terms in the sum over $(s_\one, \ldots, s_4)$.
Those 4--tuples that have $s_r < k-1$ for all $r\in \{ 1,2,3,4\}$
contribute to $\De_\one$; the others, where one of the $s_r$
equals $k-1$ and the others are zero, contribute to $\De_\two$.

In $\De_\one$, no derivative of order $k-1$ acts. Therefore we
may use Taylor expansion to bound $\De$. It gives a factor 
$p-p'$, but, of course, the derivative can now act on all 
$p$--dependent factors of the integrand. It produces at worst
a factor bounded by $\Const M^{-j_\one}$ (since some of the factors
have bounded derivatives, and since in the others, 
we can use $j_\one \le j_\two \le j_\thr$).
This combines with $\abs{p-p'}\le M^{j_\one}$ to 
$$
M^{-j_\one} \abs{p-p'} = M^{-j_\one} \abs{p-p'}^{1-\be} \;
\abs{p-p'}^{\be} \le M^{-\be j_\one}\abs{p-p'}^{\be} 
\EQN\TayHoel $$
which proves \queq{\jHoel} for the integral of $\De_\one$ over
$\om_\one,\rh_\one,\om_\two$ and $\rh_\two$.

In $\De_\two$, we use \queq{\diffi} to rearrange the integrand
for $\De_\two$. The function $\ph_k$ appearing in the difference
on the right side of \queq{\diffi}
is $\sfrac{\del^{k-1}}{\del \th^{k-1}}$ of $J$, $\tilde J$, $A$,
or one of the $\SP{r}{\nu_r}$ (or a function on which 
$\sfrac{\del^{k-1}}{\del \th^{k-1}}$ did not act; this
case is treated as in the last paragraph).
Suppose it is $\tilde J$ (the other cases are easier). 
By \queq{\tilJac}, $\sfrac{\del^{k-1}}{\del \th^{k-1}}\tilde
J$ is a sum of terms of the form 
$$
\frac{1}{({\del e}/{\del \th_\two})^{1+l}}
\prod\limits_{\mu = 1}^m \del_\th^{r_\mu} \del_{\th_\two} e
\eqn $$
with $l \ge 1$, $m \ge 1$, and where $r_\one + \ldots + r_m =
k-1$. If all $r_\mu$ are strictly less than $k-1$, a Taylor expansion,
combined with \queq{\TayHoel}, does the job. Let one of the $r_\mu$
equal $k-1$ (then $m=1$ and $l=2$). Applying \queq{\diffi} and
the uniform H\" older property \queq{\eHoelAss} of $e$ gives the factor
$\abs{p-p'}^\be$ and proves the statement. The strategy is the same
for all other terms -- the only change is that \queq{\AHoelAss}
and \queq{\SHoelAss} (and in the singular region, 
\queq{\SHoelAssp}) are used, 
and that, when the derivative acts
on propagators, \queq{\TayHoel} is again used, in the form
$M^{-j/2} \abs{p-p'} \le M^{j({1\over 2} -\be)} \abs{p-p'}^\be$.
Convergence of the scale sum for $\be < \sfrac{1}{2}$ follows
by doing the scale sums over $j_\thr$, $j_\two$, and $j_\one$,
as in Section 3.2. \endproof

\noindent{\bf Proof of H\" older continuity of $\del^k K_\two$:}
It suffices to bound the function $Y_{\ulj}^\pi $ in a way such
that the scale sum still converges. Choose the functions $A$
and $\SP{k}{\nu_k}$ as done after \queq{\dochzitiert}. 
They satisfy \queq{\SHoelAss}, \queq{\AHoelAss}
and \queq{\SHoelAssp}. The scale sum converges because 
the condition $\nu_\one=1$, and $\nu_\two+\nu_\thr \le 3$ 
is satisfied. \endproof

\noindent{\bf Proof of H\" older continuity of $\sfrac{\del}{\del
p}\Si_\two$:} This proof is a trivial variation of the previous
one. Recall that by the volume bound, Theorem \bestvol, we got
\queq{\thrthty}. Applying \queq{\diffi} to the integral for
$\sfrac{\del}{\del p}\Si_\two$ and proceeding as above, 
H\" older continuity of any degree $\ga <1$ follows 
by the same argument as above because setting $s=1$ in 
\queq{\thrthty} leaves a decay factor $\abs{j}M^j$ in the 
sum which can control almost one derivative. \endproof

\sect{Higher Dimensions}
\noindent
As in two dimensions, the part of the integrand for $Y_j^\pi$
that depends on the external momentum is $\om_{j_\thr,b}$, as given
by \queq{\omjdef}, only that now the integrals over $\th_\one$
and $\th_\two$ run over $S^{d-1}$ instead of $[0,2\pi ]$.
As for $d=2$, we attempt a change of variables to make $C_{j_\thr}$
independent of $\p$. Again, this is possible in part of the integration
region. Near the singularities of the Jacobian, we employ a
strategy  different from that of the two--dimensional case, 
and we actually show
that not only the tangential, but also the radial derivatives
of second order exist. We do not prove a statement about higher
derivatives for $d \ge 3$.

We give an outline of this strategy before going into the details. 
The scale behaviour of $Y_j^{\pi}$ can be bounded by 
$$
\abs{Y_{\ulj}^\pi}_s \le \Const M^{j_\one+j_\two} 
\max\limits_{b}\abs{\om_{j_\thr,b}}_s
\eqn $$
The scale sum obtained by this bound will converge if we can
show that 
$$
\abs{\om_{j_\thr,b}}_\two \le \Const M^{-j_\thr (2-\veps)}
\EQN\wishom $$
for some $\veps > 0$. The main idea is that by strict convexity,
the singularities of the Jacobian on $S$ are isolated points
$c_{bl}(\th)$, and that for $d \ge 3$, they thus have codimension $d-1
\ge 2$ on $S$. We make the regular and singular regions scale--dependent.
Let $\th^{(0)}\in S^{d-1}$ be fixed and $0 < \al < 1$. 
Instead of the scale--independent neighbourhood given by the $\de$
of Lemma \Unif, we take the singular region as an $M^{\al j_\thr}$--%
neighbourhood $U(\al,j_\thr)$ of $c_{bl}(\th^{(0)})$. To take
derivatives, we then vary $\rh$ and $\th$ only in $M^{\al j_\thr}$--%
neighbourhoods of $0$ and $\th^{(0)}$ (which suffices to calculate
derivatives). The advantage of making $\de $ depend on $j_\thr$
is that the smallness of $U(\al,j_\thr)$ provides additional
small factors: 
$$\eqalign{
\abs{\int\limits_{U(\al,j_\thr)}d \th_\one\;
\int\limits_{U(\al,j_\thr)}d \th_\two\;
\left(\frac{\del}{\del p}\right)^2 C_{j_\thr}} & \le\Const M^{-3j_\thr}
\int\limits_{U(\al,j_\thr)}d \th_\one\;
\int\limits_{U(\al,j_\thr)}d \th_\two \cr
& \le \Const M^{(-3+2(d-1)\al)j_\thr}
\cr}\eqn $$
because each of $\th_\one$ and $\th_\two$ is confined to a 
$(d-1)$--dimensional ball $U(\al,j_\thr)$ around the critical value.
\queq{\wishom} will thus hold if we choose $\al > \sfrac{1}{2(d-1)}$.
However, we cannot choose $\al$ as large as we want because the
price we pay for making the neighbourhood $U$ depend on $\al$
is that the bound for the Jacobian and its derivatives also become
scale--dependent. Since $S$ is strictly convex, the angle between
the normal vectors increases at least linearly with the distance
on $S$, and we can show that in the regular region,
for $s \le 1$,
$$
\abs{\tilde J}_s \le \Const M^{-\al j_\thr (1+2s)}
\EQN\tJclaim $$
(as before, we cannot take $|\tilde J|_\two$ because 
$\tilde J$ is only
$C^1$). Taking one derivative before and one after the change
of variables, as done previously, we get the bound
$$\eqalign{
\abs{\int d\th_\one \int d\tilde e\; \frac{\del }{\del \tilde
e} C_{j_\thr} (\ze,\tilde e) \frac{\del}{\del p}
\tilde J } & \le 
\Const M^{-2j_\thr} M^{-3 \al j_\thr}
\int d\th_\one \int d\tilde e\; \True{|\tilde e| \le M^{j_\thr}}
\cr
& \le \Const M^{-2 j_\thr} M^{(1-3\al)j_\thr}
.\cr}\eqn $$
In the detailed argument, there are more contributions, but 
they obey the same bound. 
For this contribution, \queq{\wishom} holds if $\al < \sfrac{1}{3}$.
It is now obvious that this leaves no region for $\al$ in $d=2$.
But for any $d \ge 3$, one has a window 
$ \al \in (\sfrac{1}{4}, \sfrac{1}{3})$
for $\al $ to obtain \queq{\wishom}. 

Thus, at this point, one sees that the two--dimensional case
is more singular than the higher--dimensional one, although,
superficially, the Fermi surface has the same codimension in
all dimensions $d \ge 1$. The neighbourhood of singularities
of the Jacobian does depend on the dimension. 

We now fill in the details of this argument. We again have to
make sure that the critical points, which depend on the external
momentum $\q$, do not leave the fixed neighbourhood of $\q^{(0)}$
as long as $\abs{\q-\q^{(0)}} < M^{\al j_\thr}$, so that we can
split the integration region in a way that does not depend on
$\q$. 

\goodbreak

\Lem{\jomei}{\Lesty Let $v_\th, v_\ph \in \{ \pm 1\}$ and
$\q \in \cB $ be fixed, and 
define $\si_\ka = \set{(\rh_\th,\rh_\ph)}{
|\rh_\th| \le \ka  \hbox{ and } |\rh_\ph| \le \ka }$
and  
$$
E(\q,\rh_\th,\rh_\ph,\th,\ph) = 
e\left(\q+v_\th \p(\rh_\th,\th) + v_\ph \p(\rh_\ph,\ph)\right)
.\EQN\Edef $$

\leftit{$(i)$} There exists a $\ka > 0$ such that for each fixed $(\rh_\th, 
\rh_\ph) \in \si_\ka $
and all $(\th,\ph) \in S^{d-1} \times S^{d-1}$ with 
$\abs{E(\q,\rh_\th,\rh_\ph,\th,\ph)}\le 2\ka$, the equation 
$$
\del_{\th_i} E = \del_{\ph_i} E = 0 \qquad \forall i \in \nat{d-1}
\eqn $$
has at most four solutions $(\th,\ph) = \left(
\th_{cr}^b(\q), \ph_{cr}^b(\q)\right)$. $\ka $ can be chosen 
so small that if there is a solution at $\rh_\th=\rh_\ph=0$ 
and at a given $\q$, then there is a solution for all  
$(\rh_\th,\rh_\ph,\q')\in \si_\ka \times U_\ka (\q)$. 
The solutions are $C^1$ in $\q$, $\rh_\th$ and $\rh_\ph$. There is  
$L>0$ such that for all $(\rh_\th,\rh_\ph,\q')\in \si_\ka \times U_\ka (\q)$,
$$
\abs{\left(\ph_{cr}^b(\rh_\th,\rh_\ph,\q'), \th_{cr}^b(\rh_\th,\rh_\ph,\q')\right)-
\left(\ph_{cr}^b(\rh_\th,\rh_\ph,\q), 
\th_{cr}^b(\rh_\th,\rh_\ph,\q)\right)} \le L \abs{\q'-\q}
.\EQN\Lipco $$

\leftit{$(ii)$} 
There are $K_\one\ge 1$ and $K_\two\ge 1$ such that
for all $\veps_\thr \le \sfrac{\ka}{2}$,
all $(\rh_\th,\rh_\ph,\p)\in \si_{\veps_\thr} 
\times U_{\veps_\thr} (\q)$ 
and all $(\th,\ph)$ with 
$\abs{E(\p,\rh_\th,\rh_\ph,\th,\ph)} \le \veps_\thr$, either
$$
\sum\limits_{i=1}^{d-1} \left( \abs{\del_{\th_i} 
E(\p,\rh_\th,\rh_\ph,\th,\ph)} +
\abs{\del_{\ph_i} E(\p,\rh_\th,\rh_\ph,\th,\ph)} \right) \ge \frac{\veps_\thr}{K_\one}
\EQN\FAR $$
or there is $ b \in \nat{4} $ such that 
$$
\abs{\th-\th_{cr}^b(\q)} \le K_\two \veps_\thr \quad \hbox{ and } \quad
\abs{\ph-\ph_{cr}^b(\q)} \le K_\two \veps_\thr 
.\EQN\NEAR $$
} 

\Proof See Appendix B. \endproof

\The{\oisonacha}{\Thsty 
For all $d\ge 3$ and all $v_\th, v_\ph \in \{ \pm 1\}$,
there is a constant $Q_3$
such that for all $P \in C^2(\R\times \cB,\C)$ with $\abs{P}_\two
< \infty$,
$$\eqalign{
\om_{j_\thr} &= \om_{j_\thr} (\ze,\p,z_\ph,z_\th,\rh_\ph,\rh_\th,v_\ph,v_\th,P)\cr
&=
\int\limits_{S^{d-1}} d\th \; J(\rh_\th,\th)\;
\int\limits_{S^{d-1}} d\ph \; J(\rh_\ph,\ph)\;
P(p_\th,p_\ph,(\ze,\p)) \; 
C_{j_\thr} (\ze, E(\p,\rh_\th,\rh_\ph,\th,\ph))
\cr}\eqn $$ 
satisfies
$$
\abs{\om_{j_\thr}}_\two \le Q_3 M^{-j_\thr (2-\ga)}
\EQN\shobo $$
with some $\ga \ge \sfrac{1}{8}$. Here the derivatives 
with respect to $\p$ can be taken in any direction
(i.e., $\sfrac{\del}{\del \rh}$
is included), and the notation is 
$$
p_\ph = (z_\ph,\p(\rh_\ph,\ph)), \qquad
p_\th = (z_\th,\p(\rh_\th,\th)) 
\eqn $$

}
\Proof  Let $\al \in (0,1)$, and fix $\q \in \cB$.
Without loss of generality, we may assume $\q \in \cP_\ka$.
Let $\veps_\thr = M^{\al j_\thr}$ and let $\p$ be such that 
$\abs{\p-\q} \le \veps_\thr$. Since $\veps_\thr \ge M^{j_\thr}$,
the support properties of $C_{j_\thr}$ imply 
$\abs{E(\p,\rh_\th,\rh_\ph,\th,\ph)} \le \veps_\thr$.
As usual, we take one derivative with respect to $\p$ right away,
and get
$$
\om_{j_\thr,\mu}' = \sfrac{\del}{\del p_\mu}
\om_{j_\thr} = 
\int\limits_{S^{d-1}} d\th \; J(\rh_\th,\th)\;
\int\limits_{S^{d-1}} d\ph \; J(\rh_\ph,\ph)\;
\sfrac{\del}{\del p_\mu}\left(P(p_\th,p_\ph,(\ze,\p)) \; 
C_{j_\thr} (\ze, E(\p,\rh_\th,\rh_\ph,\th,\ph))\right)
\eqn $$
Since for a finite number of $j_\thr$'s
the bound holds trivially if the constant is chosen large enough, 
we can assume that $\veps_\thr < \sfrac{\ka}{2}$, where $\ka$ is as in 
Lemma \jomei. Since $\max\{ j_\one,j_\two\} \le j_\thr$, 
the hypothesis of Lemma \jomei\ $(ii)$ is fulfilled. 
Let $K_\one $ and $K_\two$ be as in Lemma \jomei\ $(ii)$. 
For $\de > 0$ let 
\def\Sing{\tilde S}
$$
\Sing (\de) = \bigcup\limits_{b=1}^4 \set{(\th,\ph)}{
\abs{\th-\th_{cr}^b(\q)} \le K_\two \de, \quad
\abs{\ph-\ph_{cr}^b(\q)} \le K_\two \de }
\EQN\Singfour $$
Fix $\de $ such that the four sets in this union are disjoint.
Let $j_\thr $ be so small that $\veps_\thr < \de/2$. Split the 
integration region 
\def\Reg{\tilde R}
$$
S^{d-1} \times S^{d-1} = \Sing (\de) \dotcup \Reg(\de)
.\eqn $$
In $\Reg(\de)$ a change of variables as for $d=2$ is possible
because we are at a fixed, scale--independent distance $\de /2$
from the critical  points. For this reason, the Jacobian is also 
bounded by a constant that depends only on $\de$, and the statement
of the theorem follows as in the two--dimensional case.

The singular region is the union \queq{\Singfour}
of four disjoint sets, and 
corrrespondingly, $\om_{j_\thr,\mu}'=\sum_b \om_{j_\thr,b,\mu}'$.
We may consider every $b$ separately. 
We subdivide the $\de$--neighbourhood of the critical point further, 
as follows. Let $\ch_\one$, 
$\ch_\two \in C^\infty (\R_\zer^+,[0,1])$ be a partition of unity
on $\R_\zer^+$, with
$\supp \ch_\one = [0,4]$ and $\supp \ch_\two = [1,\infty)$. 
Insert 
$$
1=\ch_\one \left( \frac{(\th-\th_{cr}^b(\q))^2 +
 (\ph-\ph_{cr}^b(\q))^2}{(2K_\two \veps_\thr)^{2} } \right) 
+ \ch_\two \left( \frac{(\th-\th_{cr}^b(\q))^2 +
 (\ph-\ph_{cr}^b(\q))^2}{(2K_\two \veps_\thr)^{2} } \right) 
\eqn $$
in the integral. This gives two contributions, which we denote
by $\om_{j_\thr,b,\mu,1}'$ and $\om_{j_\thr,b,\mu,2}'$. 

Because of the factor $\ch_\one$, the integrand for $\om_{j_\thr,b,\mu,1}'$
is zero unless $\abs{\th-\th_{cr}^b(\q)} \le 4 K_\two \veps_\thr$
and $\abs{\ph-\ph_{cr}^b(\q)} \le 4 K_\two \veps_\thr$. Thus
$$\eqalign{
\sfrac{\del}{\del p_\nu} \om_{j_\thr,b,\mu,1}' =
\int d\th \int d\ph J(\rh_\th,\th) J(\rh_\ph,\ph)
&\ch_\one \left( (2K_\two \veps_\thr)^{-2} \left(
(\th-\th_{cr}^b(\q))^2 + (\ph-\ph_{cr}^b(\q))^2\right) \right)
\cr
&\frac{\del^2}{\del p_\mu \del p_\nu} 
\left(P(p_\th,p_\ph,(\ze,\p)) \; 
C_{j_\thr} (\ze, E(\p,\rh_\th,\rh_\ph,\th,\ph))\right)
\cr}\eqn $$
By the properties of $P$ and \queq{\shellj}, we have
$$\eqalign{
\abs{\frac{\del}{\del p_\nu} \om_{j_\thr,b,\mu,1}' } 
& \le 
\Const  M^{-3j_\thr} \; 
\int\limits_{\abs{\th-\th_{cr}^b(\q)}\le 4 K_\two \veps_\thr }d\th
\int\limits_{\abs{\ph-\ph_{cr}^b(\q)}\le 4 K_\two \veps_\thr }d\ph
\cr
& \le \Const M^{-2j_\thr} \; M^{-j_\thr + 2(d-1)\al j_\thr}
\cr}\eqn $$
where the constant contains $\abs{e}_\two$, 
$\abs{P}_\two$, and ${\abs{J}_\zer}^2$.

Because of the factor $\ch_\one$, the integrand for 
$\om_{j_\thr,b,\mu,2}'$
vanishes unless $\abs{\th-\th_{cr}^b(\q)} \ge 2 K_\two \veps_\thr$
or $\abs{\ph-\ph_{cr}^b(\q)} \ge 2 K_\two \veps_\thr$, which, 
by Lemma \jomei\ $(ii)$, implies that \queq{\FAR} holds.
Without loss of generality, we may assume that 
$\abs{\del_{\th_\one} E } \ge \sfrac{\veps_\thr}{2dK_\one}$
throughout that region (otherwise subdivide into pieces where such a 
condition holds for $\abs{\del_{\th_i} E }$ or $\abs{\del_{\ph_i} E }$ 
for some $i$). Change variables from $\th_\one$ to $\tilde e
= E$, with $E$ given by \queq{\Edef}, so that
$$
\th_\one=\th_\one(\p,\tilde e,\th_\two,\ldots,\th_{d-1},\ph,\rh_\th,\rh_\ph)
.\eqn $$
We shall suppress the other arguments in $\th_\one=\th_\one(\p,\tilde e)$
to keep the notation manageable, and write
$\th(\p,\tilde e) = (\th_\one(\p,\tilde e),\th_\two,\ldots,\th_{d-1})$.
The Jacobian
$$
\tilde J = \frac{1}{\del_{\th_\one} E(\p,\rh_\th,\rh_\ph,\th(\p,\tilde e), \ph)}
\eqn $$
is $C^1$ and it satisfies 
$$
\abs{\tilde J}_\zer \le \frac{2 d K_\one}{\veps_\thr} 
\le \Const M^{-\al j_\thr}
\EQN\tJbone $$
Moreover, differentiating 
$$
\tilde e = E(\p,\rh_\th,\rh_\ph,\th(\p,\tilde e), \ph)
,\EQN\tileE $$
with respect to $p_\nu$, we see that
$$
\frac{\del \th_\one}{\del p_\nu} = 
-\; \frac{\sfrac{\del E}{\del p_\nu}}{\sfrac{\del E}{\del \th_\one}}
\eqn $$
and therefore 
$$
\abs{\frac{\del \th_\one}{\del p_\nu}} \le 
\abs{e}_\one \frac{2 d K_\one}{\veps_\thr}
\le \Const M^{-\al j_\thr}
\EQN\tJbtwo $$
and
$$
\abs{\frac{\del \tilde J}{\del p_\nu}} \le \Const
\left(\frac{2 d K_\one}{\veps_\thr}\right)^3
\le \Const M^{-3\al j_\thr}
\EQN\tJbthr $$
(this proves \queq{\tJclaim}). After the change of variables, we have
$$\eqalign{
\om_{j_\thr,b,\mu,2}' = \int\limits_\R d\tilde e \; \del_{\tilde e} C_{j_\thr}(\ze,\tilde e)
\int d\ph\; J(\rh_\ph,\ph)\;
\int d\th_\two \ldots d\th_{d-1}
J(\rh_\th,\th(\p,\tilde e))\; \sfrac{\del E}{\del p_\mu}\;
\tilde J(\ze,\tilde e) 
\cr
P(p(\rh_\th,\th(\p,\tilde e)),p(\rh_\ph,\ph), (\ze,\p))\quad
\ch_\two \left( \frac{(\th(\p,\tilde e)-\th_{cr}^b(\q))^2 +
 (\ph-\ph_{cr}^b(\q))^2}{(2K_\two \veps_\thr)^{2} } \right)
\cr}\eqn $$
plus a term where the derivative $\sfrac{\del}{\del p_\mu}$ acted on 
$P$, not on $C_{j_\thr}$,
and which therefore obeys a bound that is by a factor $M^{j_\thr}$ better
than the one we are about to prove.
The second derivative $\sfrac{\del}{\del p_\nu}$ can now act on $\tilde J$, 
$\ch_\two$, $J(\rh_\th,\th(\p,\tilde e)$, $\sfrac{\del E}{\del p_\mu}$,
and $P$, but not on $C_{j_\thr}$. 
By \queq{\tJbone}, \queq{\tJbtwo}, and \queq{\tJbthr}, 
its effect is in all cases bounded by 
$\Const {\veps_\thr}^{-3}$. The support properties of 
$C_{j_\thr}$ restrict $\tilde e$ to an interval of length $M^{j_\thr}$.
Thus
$$\eqalign{
\abs{\om_{j_\thr,b,\mu,2}'} & \le \Const \int\limits_\R d\tilde e \;
\True{\abs{\tilde e} \le M^{j_\thr}} M^{-2j_\thr}  \; M^{-3\al j_\thr} \cr
& \le \Const M^{-2j_\thr}\; M^{(1-3\al) j_\thr}
.\cr}\eqn $$
To fulfill \queq{\shobo}, we have to have $2\al (d-1) -1 > 0$ and
$1-3\al > 0$, i.e., 
$$
\frac{1}{2(d-1)} < \al < \frac{1}{3}
\eqn $$
Taking $\al = \sfrac{7}{24}$, we get $1-3\al=\sfrac{1}{8}$ and
$2(d-1)\al -1 \ge \sfrac{1}{6}$.
\endproof

\Rem{\ismazbled} 
Proving H\" older continuity under the hypotheses that 
the second derivatives of
$e$ and $\hat v$ are also H\" older continuous is 
a straightforward adaption of the proof given for $d=2$ in 
Section 3.4.
Recall that in second order, $P$ is given by
\queq{\voilaP}, so it fulfills the hypothesis of Theorem \oisonacha,
and hence Theorem 1.2 $(ii)$ follows. 
Using the methods of Chapter 3 of \FST, one 
can show the same for $p_\zer$--derivatives, i.e., that 
$\Si$ is $C^2$ in $p_\zer $ for all $d \ge 3$. This requires
additional integration by parts and a resummation of the 
partition of unity over the scales, so we will not include this proof here.

\def\WL{wicked ladder} 
\chap{The Wicked Ladders}
\noindent
In this section, we generalize the results proven above to 
the class of graphs shown in Figure \nextfig{\figufou}. 
These graphs are constructed by 
joining two legs of a four--legged ladder diagram by a Wick line
to get a two--legged diagram
(although, of course, the scales associated to these lines will
in general not make them Wick lines). 
For this reason, and for other reasons that should 
become clearer in the following discussion, we call them the `\WL s'.
Alternatively, one may call these graphs the RPA graphs since their sum 
contains the RPA resummation for the self--energy. 

\Def{\genRPA} The generalized RPA self--energy $\Si_{\rm RPA}$
and the counterterm $K_{\rm RPA}$ are the formal power series
in $\la$ given by the sum of all first and second order graphs and all 
graphs of the form shown in Figure $\figufou$, with an arbitrary
number of bubbles, and where the vertices have the vertex function
$\hat v$ of \queq{\twobodyint}.
\ssni
Thus this class of graphs includes both the contributions from
those that are usually called RPA graphs and contributions from the ladders
to the self--energy. 
The motivation for looking at these graphs is that they give the most
singular contributions to the self--energy (this is proven in \THREE).
We shall prove in this section that if $e$ and $\hat v$ are $C^k$, 
then the same holds for tangential derivatives of the value of the \WL s.
More precisely, for the particle--hole \WL, shown in Figure $\figufou$ $(a)$,
we show this for any $k\ge 2$. For the particle--particle \WL, 
shown in Figure $\figufou$ $(b)$, we show this for all $k\ge 2$ 
{\it if the Fermi surface obeys \SYmm}. If the Fermi surface 
is nonsymmetric, we prove the statement for the particle--particle \WL\
only for $k=2$.
 
The method used is similar to that of the 
second order case: we split the integration region 
into a regular and a singular region, apply a simple change of variables in 
the regular region and analyze the critical points in the singular 
region in detail. This generalization turns out to be less
straightforward than one might expect.  For this class of graphs
there is already an essential complication in the critical point analysis.
Thus, the analysis of critical points is 
fragile in that it changes a lot if the graphs get more complicated.
The method used in \THREE\ to deal with all other graphs is much
more robust.
 
\herefig{nondol}

Without loss of generality, we may put $\hat v=1$, since it is
$C^k$ by \AOne, so that terms arising when 
derivatives act on $\hat v$ are bounded uniformly in the scales 
by \AOne. Also, we put the same scale $j$ on all the 
propagators. The value of the graph shown in Figure $\figufou$ is then
$$
Val(G)(\p)
=\int\left(\prod\limits_{r=1}^{n+1} \db p_r C_j (\om_r, e(\p_r))\right)
\; \prod\limits_{s=2}^{n+1} C_j(\tilde\om_s, e(\ve{t}_s))
\EQN\WLvar $$
with $$\eqalign{
\tilde\om_s & = \cases{\om_\one-\om+\om_s & Fig. $\figufou$ $(a)$\cr
                 \om_\one+\om-\om_s & Fig. $\figufou$ $(b)$\cr} \cr
\ve{t}_s & = \cases{\p_\one-\p+\p_s & Fig. $\figufou$ $(a)$\cr
                    \p_\one+\p-\p_s & Fig. $\figufou$ $(b)$.\cr} 
\cr}\EQN\WLval$$
\def\ulth{\Th}
Again, the derivative with respect to the external momentum may act on 
propagators associated to
lines of the graph (with the choice of spanning tree of Figure $\figufou$, this may 
affect many lines; this spanning tree is, however, convenient in 
the following because of its symmetry). There are $n+1$ angular variables
that can be used to make changes of variables, as in the second order case.
But, again, there are critical points where such a change of variables 
is not possible. 

To see where these critical points are, we again look first 
at the case where all $\p_i$ are on $S$. Let $g(\p) = \nabla e(\p)$,
$\ta(\p)=\del_\th \p $, and for $k \in \{2,\ldots,n+1\}$ let
$$
g_{1k} = \cases{g(\p_\one-\p+\p_k) & Fig. $\figufou$ $(a)$\cr
                g(\p_\one+\p-\p_k) & Fig. $\figufou$ $(b)$.\cr}
\eqn $$
Moreover, let $\ta_k = \ta(\p_k)$ and $\ta = \ta(\p)$. 
Suppose we make a change of variables $\ulth = 
(\th_\one, \ldots, \th_{n+1})^t$  $\to \tilde{\ulth}(\ulth,\th) = 
(\tilde\th_\one, \ldots, \tilde\th_{n+1})^t$ (the superscript $t$
is to mean the transpose so that these are column vectors).
The condition that the propagator for the $k^{th}$ line in the tree
is independent of the 
external momentum $\p = \p(0,\th)$ is
$$
\frac{\del}{\del\th} e\left( \p(0,\tilde\th_\one({\ulth},\th)) \mp 
\p(0,\th) \pm \p(0,\tilde\th_k ({\ulth},\th))\right) =
g_{1k}\cdot\ta_\one \frac{\del\tilde\th_\one}{\del\th}
\pm g_{1k}\cdot\ta_k \frac{\del\tilde\th_k}{\del\th}
\mp g_{1k}\cdot \ta =0
.\eqn $$
The system of equations  ensuring that all propagators in the tree are
independent of $\th$ is
$$
\Ga^{(n)}(\tilde{\ulth}(\ulth,\th),\th) \cdot \frac{\del \tilde{\ulth}}{\del \th} = \ga(\tilde{\ulth}(\ulth,\th),\th)
\EQN\singDE $$
with 
$$
\Ga^{(n)}(\ulth,\th) = \left(
\matrix{ 
g_{\one\two}\cdot \ta_\one & \pm g_{\one\two}\cdot \ta_\two 
& 0 & 0 &\ldots & 0\cr
g_{\one\thr}\cdot \ta_\one & 0 & \pm g_{\one\thr}\cdot \ta_\thr & 0 
& \ldots & 0 \cr
g_{1 4}\cdot \ta_\one & 0 & 0 & \pm g_{14}\cdot \ta_4 
& \ldots & 0 \cr
\vdots & \vdots & \vdots & & \ddots & \vdots \cr
g_{\one \, n+1}\cdot \ta_\one & 0 & 0 & \ldots & 0 & 
\pm g_{\one \, n+1}\cdot \ta_{n+1}\cr}
\right)
\EQN\matcrit $$
and
$$
\ga(\ulth,\th) = \pm (g_{12}\cdot\ta,\ldots,g_{1,n+1}\cdot\ta)^t
.\eqn $$
By definition, the critical points are those points where the matrix
$\Ga^{(n)}$ (which has $n$ rows and $n+1$ columns) has 
rank$(\Ga^{(n)}) \leq n-1$. Away from these points, 
\queq{\singDE} is easily solved by first setting one of 
$\sfrac{\del\tilde\th_\one}{\del\th},\ldots,
 \sfrac{\del\tilde\th_{n+1}}{\del\th}$ equal to zero.
This effectively deletes one column from $\Ga^{(n)}$. 
The resulting matrix is then invertible, and the corresponding 
system of differential equations is soluble with initial condition 
$\tilde\th_i=\th_i$ (this case is discussed in detail below).

Since one could also have chosen a different spanning tree, 
we briefly motivate our choice. At a first glance, it may seem 
a much better strategy to put the line that carries momentum $\p_\one$ 
into the spanning tree, because then only this line depends on the external
momentum. This would then give, e.g.\ for $(a)$, 
$\p_\one = \p+\p_\two -\q_\one$, 
where $\q_\one $ is now another loop momentum. Assuming that we are
in the regular region, i.e.\ away from the critical points of 
$e( \p+\p_\two -\q_\one)$, one can then do the same change of variables 
as in the second order case to make the propagator on line 1 independent
of $\p$. However, this time the dependence on the external $\th$ goes not
only into a Jacobian, but, through the changed variables, also into one of
the propagators associated to the other lines of the spanning tree,
so that the derivative w.r.t.\ $\th$ may still degrade the scaling 
behaviour in a dangerous way. Then, one can do another change of 
variables to get rid of the dependence on $\th $ of the next line 
in the spanning tree, etc. Although such a procedure is possible,
it gets rather complicated already in the three--loop case ($n=2$).
It is much easier to look at the entire system of equations at once.
The simple form of the matrix $\Ga^{(n)}$ is due to our symmetric 
choice of the spanning tree.

\sni
We now classify the critical points for the system of equations \queq{\singDE}.

\Rem{\Garank}
For $\Ga^{(n)}$ to have rank strictly less than $n$, 
it is necessary that either 
\ssni
(1) one of columns $2$ through $n+1$ vanishes, say $g_{1k}\cdot\ta_k=0$,
and the corresponding entry in column one, $g_{1k}\cdot\ta_1$, also vanishes,
or
\ssni
(2) at least two of columns $2$ through $n+1$ vanish.
\ssni
We deal with case of the particle--hole \WL\ (Figure $\figufou$ $(a)$) first.
\Lem{\acrit}{\Lesty For the particle--hole \WL, if
rank $\Ga^{(n)} \le n-1$ and all momenta are on $S$ 
then one of the following holds.

\leftit{$(i)$} $\p_\one=\p$; then rank $\Ga^{(n)}\le 1$ and all 
columns of $\Ga^{(n)}$ except for the first vanish.

\leftit{$(ii)$} $\p_\one=\ve{a}(\p)$, and there is 
$\emptyset \ne C \subset\{2,\ldots,n+1\}$ such that for all $k \in C$,
$\p_k=\p$, and for all $\l \not\in C$, $\p_l\ne\p$; then rank $\Ga^{(n)}
=n-|C|$, and 

\leftit{} \null\hroom all columns with index $k \in C$ vanish: $\Ga^{(n)}_{mk}=0$ 
for all $m\in \{2,\ldots,n+1\}$, 

\leftit{} \null\hroom all rows with index $k-1$, $k\in C$, vanish: 
$\Ga^{(n)}_{k-1,m}=0$ for all $m\in \nat{n}$,

\leftit{} \null\hroom all columns with index $k\in \{2,\ldots,n+1\}\setminus C$
are nonzero.

\noindent Whenever row $k$ of $\Ga^{(n)}$ vanishes, the corresponding 
component $\ga_k$ of the vector $\ga $ on the right hand side of 
\queq{\matcrit} also vanishes. In case $(i)$, the right hand side $\ga$
coincides with the first column of $\Ga^{(n)}$.
}

\Proof We go through th cases mentioned in Remark \Garank.
Case (1): The two equations $g_{1k}\cdot \ta_k=0$ and
$g_{1k}\cdot\ta_\one=0$
are precisely the system of equations discussed in the second order case,
only now $\p_k$ appears instead of $\p_\two$. Therefore
$\p_\one=\p_k=\p$ or $\p_\one=\ve{a}(\p)$, $\p_k=\p$ or $\p_\one=\p$,
$\p_k=\ve{a}(\p)$ holds, and in particular, $\p_\one-\p+\p_k\in\{\p,
\ve{a}(\p)\}$ so that $g_{1k}\cdot \ta$ vanishes. 
If $\p_\one=\p$, then $g_{1\ell}\cdot\ta_\ell=
g(\p_\ell)\cdot\ta(\p_\ell) =0$ holds for all $\ell\ge 2$, and 
this case is stated as item $(i)$ in the Lemma. 
If $\p_\one=\ve{a}(\p)$, then $\p_k=\p$ and 
$\p_\one-\p+\p_k = \ve{a}(\p)$. Therefore, for any additional $\ell\in
\{ 2, \ldots, n+1\}$, the vanishing of column number $\ell$, $g_{1\ell}\cdot\ta_\ell=0$, means 
$g(\ve{a}(\p)-\p+\p_\ell) \cdot \ta(\p_\ell) =0$
which implies $\ve{a}(\p)-\p+\p_\ell=\p_\ell$ or
$\ve{a}(\p)-\p+\p_\ell=\ve{a}(\p_\ell)$. The first equation is impossible
since $\ve{a}(\p) \ne \p$. The second yields, by Lemma \byAfive, $\p_\ell=\p$.
Let $C=\{ \ell\in \{2,\ldots,n+1\}: \p_\ell=\p\}$. For any $k \in C$,
$g_{1k}\cdot\ta_\one = g(\ve{a}(\p))\cdot \ta(\ve{a}(\p))=0$, therefore
row $k-1$ vanishes for all $k \in C$. If $k\not\in C$, 
$g_{1k}\cdot\ta_k \ne 0$, so statement $(ii)$ of the Lemma holds.

Case (2): there is $C\subset \{ 2, \ldots, n+1\}$, $|C|\ge 2$, such that 
for all $k\in C$, $g_{1k}\cdot\ta_k =0$. This forces
$\p_\one-\p+\p_k \in \{\p_k,\ve{a}(\p_k)\}$. Hence either 
$\p_\one=\p$ or, by Lemma \byAfive, $\p_\one=\ve{a}(\p)$ and $\p_k=\p$
for all $k \in C$. The remaining statements about the columns and rows 
of $\Ga^{(n)}$ follow as in case (1). 

In case $(i)$, only the first column of $\Ga^{(n)}$ is nonzero. The right 
hand side is identical to the first column of $\Ga^{(n)}$.
 In case $(ii)$, vanishing of row number $k$
means $\p_k=\p$, so $g_{1k}\cdot\ta=0$ at those points is obvious.
\endproof

\noindent
The last statement of the Lemma means that the system of equations 
remains consistent at the critical points because the right hand side 
remains in the span of the columns of $\Ga^{(n)}$.

The condition that all momenta are on the Fermi surface can again be 
relaxed to a condition that all momenta be in a scale--independent 
neighbourhood of $S$, i.e.\ $|\rh|=|e(\p)|\le r_\zer$, by the following 
Lemma.

\Lem{\weitergehts}{\Lesty For any $\veps > 0$, there are $r_\zer> 0$ and
$s>0$, depending on $|e|_\two$ and $u_\zer$, such that if $|e(\p)|<r_\zer$ 
in all propagators and if $\abs{\th-\th^{(0)}}<s$, then all critical points
of $\Ga^{(n)}$ are in an $\veps$--neighbourhood of the critical points 
of $\rh=0$, $\th=\th^{(0)}$ given in Lemma \acrit.}

\Proof Again, we order the proof according to the cases (1) and (2) 
of Remark \Garank. As mentioned, 
the critical point condition in case (1) is just a relabelling 
of that of the second--order case. Therefore, the critical points for 
$\th_\one$ and $\th_k$ are $C^1$ in $\th$ and the statement of the 
Lemma follows immediately from Lemma \contagain. If $\p_\one=\p$,
that's it since the other momenta are not fixed in $(i)$ of Lemma \acrit.
Let $\p_\one $ be near to $\ve{a}(\p)$, and $\th $ near to $\th^{(0)}$. 
For 
$\ell\in\{2,\ldots,n+1\}\setminus \{ k\}$, we perform a Taylor expansion 
of $g_{1\ell}\cdot\ta_\ell$
in $\rh_\one$, $\rh_\ell$, $\th - \th^{(0)}$, and $\th_\one-a(\th^{(0)})$, 
to obtain
$$\eqalign{
\Big\vert\nabla 
e\big(\p(\rh_\one,\th_\one) & - \p(0,\th) + \p(\rh_\ell,\th_\ell)\big)
\cdot \del_\th\p(\rh_\ell,\th_\ell)-
\nabla e\big(\p(0,a(\th^{(0)}))-\p(0,\th^{(0)})+\p(0,\th_\ell)\big)
\cdot \del_\th\p(0,\th_\ell)\Big\vert\cr
& \le \Const \left( |\rh_\ell|+|\rh_\one|+
|\th-\th^{(0)}|+|\th_\one-a(\th^{(0)})|\right)
\cr}\eqn $$
where the constant is a bound for the Taylor remainder. It depends on 
$|e|_\two$ and $u_\zer$ because the Taylor 
remainder contains the second derivative of $e$ and objects like 
$\del_\rh\p$ and $\del_\th \p$. It follows that for $r_\zer $ and 
$\th - \th^{(0)}$ small enough, $|g_{1\ell}\cdot\ta_\ell| \ge g_\one > 0$ 
unless $\th_\ell$ is in an $\veps$--neighbourhood of $\th^{(0)}$.

For case (2), we do a similar Taylor expansion argument, but now 
applied to 
$$
\nabla e \left( \p(\rh_\one,\th_\one)-\p(0,\th)+\p(\rh_k,\th_k))
\cdot \del_\th\p(\rh_k,\th_k)\right) =0
\eqn $$
to get
$$
\nabla e(\p(0,\th_\one)-\p(0,\th)+\p(0,\th_k))\cdot \del_\th\p(0,\th_k)
+\rh_\one\ph_\one +\rh_k\ph_k = 0
\eqn $$
with $\ph_i$ continuous functions coming from Taylor expansion.
By the Lemma \acrit, the first term in the sum only vanishes when
$\th=\th_\one$, or if $\th=a(\th_\one)$ and $\th_k=\th$. Away from
 these points, it is nonzero and therefore 
for small enough $|\rh_\one|$ and $|\rh_k|$, all critical points are 
in a neighbourhood of $\th^{(0)}$. \endproof

\Lem{\areg}{\Lesty Let $k \ge 2$, $d=2$, and assume \AOne{k,0},
\ATwo{k,0}, and \AThr--\AFiv.
Then the contribution of 
the particle--hole \WL\ to the counterterm function $K$ is $C^k$.}

\Proof The strategy to control the derivative is now similar to that of 
the second--order case: in the region away from the critical points,
we do a change of variables that makes all propagators independent 
of the external momentum. Near the critical points, we show that every
increase in the power of the denominator is accompanied by a small factor
in the numerator. Again as before, we take one derivative with respect
to the external momentum right away, so that $k-1$ derivatives remain 
to be taken after the manipulations.  

Fix $\th^{(0)}$, and let $\th$ be in a sufficiently small neighbourhood
of $\th^{(0)}$. By Lemma \acrit, any critical point 
of $\Ga^{(n)}$ at $\th$ is near to one of $\Ga^{(n)}$ at $\th^{(0)}$.
Thus we may define the regular region dependent only on $\th^{(0)}$.
For $\de > 0$ and $a>0$, the regular region is defined as
$$
\cR_\de (\th^{(0)},a) = 
\set{\ulth=(\th_\one, \ldots, \th_{n+1})}{\forall
\ulth' \hbox{ with } |\ulth'-\ulth| \leq \de, \hbox{ and }
\forall |\th-\th^{(0)}| \le a,  
\hbox{ rank }\Ga^{(n)}(\rh,\ulth,\th)=n}
\eqn $$
Here we wrote $\Ga^{(n)}(\rh,\ulth,\th)$ to indicate that now  
the momenta on the lines of the graph do not have to be exactly on
the Fermi surface, but they only have to be near it. 
By Lemma \weitergehts, the complement of the regular region is 
a neighbourhood of the critical points of $\Ga^{(n)}=
\Ga^{(n)}(0,\ulth,\th)$ given in Lemma \acrit.
For $k \in \nat{n+1}$, let $\Ga^{(n)}_k$ be the $n\times n$ 
submatrix of $\Ga^{(n)}$ obtained by deleting column number $k$. 
We cover $\cR_\de $ with patches chosen such that in every patch, 
there is a $k\in\nat{n+1}$ such that rank $\Ga^{(n)}_k = n$. 
Since $\cR_\de $ is compact, the covering can be chosen to contain 
only finitely many such patches. Let $P$ be one of these patches and
rank $\Ga^{(n)}_k (\ulth,\th) = n$ for all $\ulth \in P$. 
Then $\Ga^{(n)}_k$ is invertible, since it is a square
matrix of maximal rank and the inverse is $C^{k-1}$ in all variables.
Delete $\tilde\th_k$ from the list of variables, and write 
$\tilde{\ulth}^{(k)} = (\tilde \th_\one, \ldots, \tilde\th_{k-1}, 
 \tilde\th_{k+1}, \ldots, \tilde\th_{n+1})^t$. The change of variables
is then given by by the solution to the initial value problem
$$\eqalign{
\frac{\del \tilde{\ulth}^{(k)}}{\del \th} & =
\Ga^{(n)}_k (\tilde{\ulth}^{(k)},\th)^{-1} 
\cdot \ga(\tilde{\ulth}^{(k)},\th)\cr
 \tilde{\ulth}^{(k)}(\ulth^{(k)},\th^{(0)}) & = \ulth^{(k)}
.\cr}\eqn $$
The solution is a $C^{k}$ function of all variables because the 
right side of the differential equation is $C^{k-1}$. Thus the Jacobian is 
$C^{k-1}$ in all variables. After this change of variables, all 
the dependence on the external angle $\th$ is in the Jacobian and
in the support function of the patch. The patch itself is independent 
of $\th$ since by construction
it depends only on $\th^{(0)}$. Thus, taking all the remaining
derivatives is harmless. Moreover, the support restriction on the 
propagators reduces the integration regions for the $\tilde \th_k$ 
in a way that cancels the large factor from the first derivative we took.

It remains to bound the contribution from the neighbourhood of the 
set where the rank of $\Ga^{(n)}$ is nonmaximal. We split the proof 
in the two cases, according to the characterization of critical points
given in Lemma \acrit.

\ssni 
$(i)$ When $\p_\one=\p$, only the first column in $\Ga^{(n)}$ is nonzero.
So, when $\p_\one \approx \p$, instead of a change of variables, 
we shall use that
the derivative on a propagator on scale $j < 0$ produces a small factor $M^j$
that cancels the extra big factor $M^{-j}$ from the denominator. We can be brief 
about this because the argument is very similar to that of the second order 
case. By Taylor expansion in the $\rh$ variables, we can reduce the problem 
to $\rh_\one=\rh_k=0$. 
Let $\p_\one $ be near to $\p$, and $\p_\two, \ldots \p_{n+1}$
be arbitrary. Since $\p_\one \approx \p$, we change variables from $\th_\one $
to $t=\th_\one-\th$. Since $e(\p(\th_k))=0$, the usual Taylor expansion yields
$$\eqalign{
e(\p(0,\th_\one)-\p(0,\th)+\p(\th_k)) &= 
t \ph(\th,t,\th_k) \cr
\ph(\th,t,\th_k)  & = \int\limits_0^1 d \al\; 
g(\p(\th+\al t) - \p(\th) + \p(\th_k)) \cdot \ta(\p(\th+\al t)) 
\cr}\eqn $$  
If $\p_k$ is not in a small neighbourhood of $\p(\th^{(0)})$ or of
$\p(a(\th^{(0)}))$, then $\ph(\th,0,\th_k) \ne 0$, and therefore 
for all $|t|<\de'$ ($\de'$ independent of the scales),
$|\ph(\th,t,\th_k)|\ge \underline{\ph} > 0$, so
$|e(\p_\one-\p+\p_k)| \le M^j$ implies 
$$
|t| \le \Const M^j
\EQN\volforever $$
Thus, as in the second order case, 
every derivative acting on $C_j$ produces not only a large factor 
$M^{-j}$, but also a small factor $|t|$ that cancels it, e.g.\
$$
\abs{\frac{\del}{\del\th}C_j(p_\zer,e(\p(\th+t)-\p(\th)+\p(\th_k)))}
\le |t|\; \abs{\frac{\del \ph}{\del\th}} \; |\del_\two C_j|
\le \Const M^j M^{-2j}
\eqn $$
does not change the scaling behaviour of the propagator. 
The factor $M^j$ that got lost when the very first derivative 
was taken is regained, up to a factor $|j|$, by the volume improvement
in the integration over $t$ coming from the restriction 
\queq{\volforever}.

For any $k\ge 2$ for which $\th_k $ is near to $\th^{(0)}$ (and hence 
near to $\th$), change variables from $\th_k$ to $y=\th_k-\th$. 
Then the $\th$--dependence in the propagator is in the function
$$
\et(\th,x,y)=e\big(\p(\th+x)+\p(\th+y)-\p(\th)\big)
.\eqn $$
The analysis of the singularity is now identical to that to the 
second--order case, and in particular, Lemma \Unif\ applies. 
Thus for every increase in the power of the denominator, there is
a corresponding small factor in the numerator. The details of the 
bound are as in the second--order case.

For any $k\ge 2$ for which $\th_k $ is near to $a(\th^{(0)})$, hence 
near to $a(\th)$, change variables from $\th_k$ to $y=\th_k-a(\th)$. 
Then the $\th$--dependence in the propagator is in the function
$$
\et(\th,x,y)=e\big(\p(\th+x)+\p(a(\th)+y)-\p(\th)\big)
.\eqn $$
The analysis of the singularity is now identical to the 
second--order case, in particular, Lemma \Unif\ applies. 
All the details are as in the second--order case.

There remains the case $(ii)$: $\p_\one = a(\p)$, 
and $\p_k = \p$ for $k \in C\subset \{ 2, \ldots , n+1\}$.
We change variables from $\th_\one $ to $\tilde\th_\one=\th_\one -a(\th)$.
Then $\sfrac{\del\tilde\th_\one}{\del \th_\one}=1$, and the 
first column of $\Ga^{(n)}$ is moved to the right hand side of the 
equation. For $k \in C$, we change variables from $\th_k$ to $\tilde\th_k=\th_k-\th$ and move the corresponding columns of $\Ga^{(n)}$
to the  right hand side of the equation. 
We delete row number $k$ for every $k \in C$.
The square matrix left over of $\Ga^{(n)}$ after this procedure 
has maximal rank by construction and Lemma \acrit, 
so the change of variables 
is there as in the regular region.  We solve the resulting 
system of differential equations for $\tilde\th_k$, $k\not\in C$.
For $k\in C$, 
$$
e(\p_\one-\p+\p_k) = e(\p(a(\th)+x)-\p(\th)+\p(\th+y))
\eqn $$
so Lemma \Unif\ applies. 
\endproof

We now turn to the particle-particle \WL\ (see Figure $\figufou$
$(b)$).
The analogue of Lemma \acrit\ is in the symmetric case
\Lem{\bcrit}{\Lesty Assume \SYmm.
For  the particle-particle \WL, if rank $\Ga^{(n)} \le n-1$ and  
all momenta are on $S$ then one of the following holds.

\leftit{$(i)$} $\p_\one=-\p$; then rank $\Ga^{(n)}\le 1$ and all 
columns of $\Ga^{(n)}$ except for the first vanish.

\leftit{$(ii)$} $\p_\one=\p$, and there is 
$\emptyset \ne C \subset\{2,\ldots,n+1\}$ such that for all $k \in C$,
$\p_k=\p$, and for all $\l \not\in C$, $\p_l\ne\p$; then rank $\Ga^{(n)}
=n-|C|$, and 

\leftit{} \null\hroom all columns with index $k \in C$ vanish:
 $\Ga^{(n)}_{mk}=0$ 
for all $m\in \{2,\ldots,n+1\}$, 

\leftit{} \null\hroom all rows with index $k-1$, $k\in C$, vanish: 
$\Ga^{(n)}_{k-1,m}=0$ for all $m\in \nat{n}$,

\leftit{} \null\hroom all columns with index $k\in \{2,\ldots,n+1\}\setminus C$
are nonzero.

\noindent Whenever row $k$ of $\Ga^{(n)}$ vanishes, the corresponding 
component $\ga_k$ of the vector $\ga $ on the right hand side of 
\queq{\matcrit} also vanishes.  In case $(i)$, the right hand side $\ga$
coincides with the first column of $\Ga^{(n)}$.
}

\Proof To start, we do not assume \SYmm.
Case (1) is as in second order, so $\p_\one=\p_k=\p$ or
$\p_\one=\ve{a}(\p)$, $\p_k=\p$, or $\p_\one=\p_k=\ve{a}(\p)$. 
In case (2), the condition that $g_{1k}\cdot \ta_k=0$ for $k\in C$,
$|C|\ge 2$, forces $\p_\one+\p-\p_k \in \{ \p_k, \ve{a}(\p_k)\}$
for all $k \in C$. This is possible only if either 
$\p_\one=\p$, $\p_k=\p$ for all $k\in C$, or if 
$$
\p_\one+\p=\p_k+\ve{a}(\p_k) \quad \hbox{ for all } \; k \in C
.\EQN\derhaken $$
Note that in the asymmetric case it does not follow from \queq{\derhaken}
that $\p_k$ or $\p_\one$ have to be $\p$ or $\ve{a}(\p)$: although there 
is the solution $\p_\one=\p_k$, $\p=\ve{a}(\p_k)$, there can be 
other solutions, due to the asymmetry of $S$. At these additional
solutions, the system of equations \queq{\matcrit} is inconsistent.

We now assume \SYmm. Then $\ve{a}(\p_k)=-\p_k$, and \queq{\derhaken} implies
$\p_\one = -\p$. Looking back at all cases, we see that they are covered
by statements $(i)$ and $(ii)$ of the Lemma. The statements about the rank
and the vanishing of columns and rows of $\Ga^{(n)}$, and about the consistency
at the critical points follow by inspection of the matrix. \endproof

\Lem{\breg}{\Lesty Let $k \ge 2$, $d=2$, and assume \AOne{k,0},
\ATwo{k,0}, \AThr--\AFiv, and \SYmm.
Then the contribution of the particle--particle \WL\ to the counterterm 
function $K$ is $C^k$.}

\ssni
The proof of this Lemma, as well as that of the analogue of Lemma \weitergehts,
is an obvious variation of that of Lemma \areg, and we leave it to the 
reader. We instead turn to the nonsymmetric case in which the potential
inconsistencies of \queq{\matcrit} at solutions of \queq{\derhaken}
caused problems. The main point there is the same that leads to the 
Fermi liquid behaviour of such models: The particle--particle bubble, 
which in the symmetric case prevents the convergence of perturbation 
theory because of its singularity at zero transfer momentum, has no 
singularities. Under our assumptions, this can be deduced from the 
following Lemma.

\Lem{\bubvol}{\Lesty Let $d=2$, $\hxp > 0$, and
\ATwo{2,\hxp}--\AFoup\ hold and assume that \SYmm\ 
does not hold. Then there is a constant $Q_B>0$ such that
for all $\veps_\two\ge \veps_\one>0$ and for all $\q \in \cB$
$$
\sup\limits_{|\rh|\le \veps_\one}
\int d\th\; \True{|e(-\p(\rh,\th))+\q)|\le \veps_\two} \le Q_B \; 
{\veps_\two}^{1\over 3}
\EQN\bubimp $$
and therefore
$$
\int\limits_{\cB} d^d\ve{p}\; 
\True{|e(-\p+\q)| \le \veps_\two} \True{|e(\p)|\le \veps_\one}
\le 2\abs{J}_\zer Q_B \; 
\veps_\one {\veps_\two}^{1\over 3}
\EQN\bubimptwo $$
where $J$ is the Jacobian defined in \queq{\JacobianJ}.
}

\Proof See Appendix C. \endproof

\noindent
It is important in the proof that the curvature of $S$ at $\p$ and
$\ve{a}(\p)$ differs except at a finite number of points; see Appendix C. 
Given Lemma \bubvol,
the boundedness of the particle--particle bubble is obvious because
instead of the ordinary power counting behaviour $M^{0j}$ of a \NOL\
four--legged diagram, Lemma \bubvol\ implies a bound by $M^{j/3}$ which 
gives a convergent scale sum. Similarly, the proof that the 
particle--particle \WL\ is $C^2$ is now an easy consequence of  
volume estimates; it does not require any analysis of critical points.
In particular, assumption \AFiv\ is not needed because it was needed
neither in the proof of Theorem \bestvol\ nor in that of Lemma \bubvol.

\The{\goodWL}{\Thsty Let $d\ge 2$, $\hxp > 0$, and 
\AOne{2,\hxp}, \ATwo{2,\hxp}, and \AThr--\AFoup\ hold, 
and assume that \SYmm\ 
does not hold. Let $G$ be the particle--particle \WL\ 
shown in Figure $4\; (b)$, and denote $Val(G)(\p(\rh,\th))$, 
defined in \queq{\WLval}, by $R_j(\rh,\th)$. 
Then the scale sum 
$\sum\limits_{j<0} R_j(\rh,\th)$ is $C^2$ both in $\rh$ and $\th$.}

\Proof The idea of the proof is simple: by Lemma \bubvol\ there is
a volume gain in any of the bubbles uniformly in the transfer momentum. 
There are enough loops to extract both that gain in one bubble
and the volume gain from two overlapping loops. 
This gives an improvement factor $M^{4j/3}$, hence enough decay to take two 
derivatives, no matter whether they are taken in $\rh$ or in $\th$ 
direction.  

To do the details, it is most convenient to choose the 
spanning tree for $G$ as shown in Figure \nextfig{\WLfig} because then the 
derivative acts only on the propagator of line $\ell_\one$, and we avoid some 
uninteresting combinatorics (in Figure $\WLfig$, the additional bubbles are all 
put into a subgraph drawn as the shaded disk). Taking two derivatives, using 
\queq{\shellj}, and doing the $p_\zer$--integrals in the usual way, 
we get (denoting $\bbbone_j(e(\p))=\True{|e(\p)|\le M^j}$)
$$
|D^2 R_j(\rh,\th)| = W_\two {W_\zer}^{2n} \; M^{-2nj}\; M^{-3j} (2M^j)^{n+1}
\int\prod\limits_{l \not\in L(T)} d^d\p_l
\; \prod\limits_{l \in L(G)} \bbbone_j(e(\p_l))
\eqn $$
where the $M^{-2nj}$ comes from the sup norm of the propagators 
in the bubbles, the $M^{-3j}$ comes from the second derivative 
of the propagator on line $\ell_\one$, and the 
factor $(2M^j)^{n+1}$ comes from the $p_\zer$--integrations.
We call the $n+1$ loop variables $\p_\one, \ldots, \p_{n+1}$, and 
introduce $(\rh_l,\th_l)$ as integration variables such that
$\p_k=\p(\rh_k,\th_k)$, and do the $\rh$--integrations. Every $\rh_k$
produces a factor $\Const M^j$ when integrated. Thus
$$
\abs{R_j}_\two \le \nu M^{-j}
\sup\limits_{\rh_\one, \ldots, \rh_{n+1}\atop |\rh_k|\le M^j} 
\int\prod\limits_{i=1}^{n+1} d\th_i\; 
\prod\limits_{\ell \in L(T)} \bbbone_j(e(\q_\ell))
\eqn $$
where the momenta $\q_\ell$ for $\ell \in L(T)$ are given as linear combinations
of the loop momenta $\p(\rh_\one,\th_\one)$, $\ldots$,  
$\p(\rh_{n+1},\th_{n+1})$, and the external momentum $\p(\rh,\th)$, 
and where $\nu$ contains all the constants.
We may choose the labelling of the loop momenta as indicated in Figure $\WLfig$.
Since $n\ge 2$, i.e.\ there are at least
two bubbles, so the product over $\ell \in L(T)$ contains 
at least the two factors $\bbbone_j(e(\q_{\ell_\one})) $ and 
$\bbbone_j(e(\q_{\ell_\two}))$, where $\ell_\one $ and $\ell_\two $ are 
as in Figure \WLfig. $\q_{\ell_\one}=\p_\one+\p_\two-\p$ and 
$\q_{\ell_\two}=\p_\one+\p_\two-\p_\thr$, so
$$\eqalign{
\abs{R_j}_\two & \le \Const M^{-j}
\sup\limits_{|\rh_\one|\le M^j, |\rh_{\two}|\le M^j}
\int d\th_\one d\th_\two\; 
\bbbone_j(e(\p_\one+\p_\two-\p)) \; 
\sup\limits_{|\rh_\thr|\le M^j}\int d\th_\thr \; 
\bbbone_j(e( -\p_\thr+\p_\one+\p_\two))
.\cr}\eqn $$
By Lemma \bubvol, the integral over $\th_\thr$ is bounded uniformly in 
$\p_\one $ and $\p_\two$, and by a Taylor expansion in $\rh_\one$ and 
$\rh_\two$, as used to derive \queq{\famous}, we get
$$
\abs{R_j}_\two  \le \Const M^{-j}\; 
Q_B M^{j/3}\; \cW\left((1+2\sfrac{|e|_1}{u_\zer})M^{j} \right)
\le \Const \; |j| \; M^{j/3}
\eqn $$
so that
$$
\sum\limits_{j<0} \abs{R_j(\rh,\th)}_\two \le
\Const \sum\limits_{j<0} |j|\; M^{j/3}
\le \Const \frac{1}{1-M^{-1/6}}
.\eqndprf\eqn $$

\herefig{wicked}
\def\ulth{{\underline{\th}}}

\ssni
\Rem{\klaerend} The reason why the problematic critical points 
 of \queq{\derhaken} did not appear in second order is simply 
that in second order, the case (2) mentioned before Lemma \acrit\
cannot occur because there are only two loop momenta. 
The conditions of case (1) exclude \queq{\derhaken}.

\Rem{\mognetmea} For $d \ge 3$, the proof is similar to that
of Theorem \oisonacha\ because, again, one has restrictions to
a neighbourhood of the critical points. Lemma \jomei\ applies
directly to do the argument. 
           
\goodbreak
\endpage

\appendix{A}{The $C^2$ Morse Lemmas}
\noindent
In this Appendix, we prove the $C^2$ Morse Lemmas that we need
to prove Theorem \bestvol. 
The proof of the Morse Lemma for smooth functions
is in many textbooks, but our functions are only $C^2$,
which makes the proof less straightforward.
One proof of the $C^2$ Morse lemma can be found in \quref{MW}.
For convenience of the reader, we include another proof here.

\ssni
\sect{Hyperbolic case}
Let $\nu(\ph_1,\ph_2)$ be a $C^2$ function in a neighbourhood of $(0,0)$
obeying
$$\meqalign{
\nu(0,0)&=0\cr
\del_1\nu(0,0)&=0  &&  \del_1\nu(0,0)&=0 \cr
\del_1^2\nu(0,0)&=-1  &&  \del_2^2\nu(0,0)&=1  
&&\del_1\del_2\nu(0,0)=0 
\cr}\eqn $$
\Lem{\Zeroo}{\Lesty There exists a neighbourhood $\cN$ of $0$ and two $C^1$
functions $\psi_\pm(\ph_1)$  such that
$$
\nu\big(\ph_1,\psi_\pm(\ph_1)\big)=0\qquad{\rm for\ all\ }\ph_1\in\cN
\eqn $$
and 
$$\eqalign{
\psi_\pm(0)&=0\cr
\psi'_\pm(0)&=\pm 1\cr
}\eqn $$
}
\The{\MH}{\Thsty 
There exist $C^1$ functions $x(\ph_1,\ph_2)$ and $y(\ph_1,\ph_2)$
in a neighbourhood of $(0,0)$ such that
$$
\nu(\ph_1,\ph_2)=x(\ph_1,\ph_2)\,y(\ph_1,\ph_2)
\eqn $$
and 
$$\meqalign{
x(0,0)&=0   && y(0,0)&=0\cr
\del_1x(0,0)&=1  &&  \del_2x(0,0)&=-1 \cr
\del_1y(0,0)&=-1/2  &&  \del_2y(0,0)&=1/2 \cr
}\eqn $$
}
\sni
The proofs of both Lemma \Zeroo\ and Theorem \MH\ use the representations
$$\eqalign{
\nu(\ph_1,\ph_2)&=-\half\al(\ph_1,\ph_2)\ph_1^2
+\be(\ph_1,\ph_2)\ph_1\ph_2
+\half\ga(\ph_1,\ph_2)\ph_2^2
\cr}\EQN\Rone $$
and
$$\eqalign{
\del_1\nu(\ph_1,\ph_2)&=-\tilde\al(\ph_1,\ph_2)\ph_1
+\tilde\be(\ph_1,\ph_2)\ph_2\cr
\del_2\nu(\ph_1,\ph_2)&=\tilde\be(\ph_1,\ph_2)\ph_1
+\tilde\ga(\ph_1,\ph_2)\ph_2
\cr}\EQN\Rtwo $$
where all the coefficients on the right hand side are continuous
and where
$$\deqalign{
\al(0,0)&=\tilde\al(0,0)&=1\cr
\ga(0,0)&=\tilde\ga(0,0)&=1\cr
\be(0,0)&=\tilde\be(0,0)&=0\cr
}\eqn $$
We shall use the notation $o(1)$ to denote a generic and unimportant
continuous function that vanishes at $(0,0)$. Hence, for example,
we shall write $\al(\ph_1,\ph_2)=1+o(1)$. To prove \queq{\Rone} one uses
$$\eqalign{
\nu(\ph_1,\ph_2)
&=\int_0^1 dt\ (1-t)\sfrac{d^2\hfill}{dt^2}\nu(t\ph_1,t\ph_2)\cr
&=\int_0^1 dt\ (1-t)\left[\del_1^2\nu(t\ph_1,t\ph_2)\ph_1^2
+2\del_1\del_2\nu(t\ph_1,t\ph_2)\ph_1\ph_2
+\del_2^2\nu(t\ph_1,t\ph_2)\ph_2^2\right]\cr
}\eqn $$
Hence
$$\eqalign{
\al(\ph_1,\ph_2)&=-2\int_0^1 dt\ (1-t)\del_1^2\nu(t\ph_1,t\ph_2)\cr
\be(\ph_1,\ph_2)&= 2\int_0^1 dt\ (1-t)\del_1\del_2\nu(t\ph_1,t\ph_2)\cr
\ga(\ph_1,\ph_2)&=2\int_0^1 dt\ (1-t)\del_2^2\nu(t\ph_1,t\ph_2)\cr
}\eqn $$
Similarly, to prove \queq{\Rtwo} one uses
$$\eqalign{
\del_1\nu(\ph_1,\ph_2)
&=\int_0^1 dt\ \sfrac{d\hfill}{dt}\del_1\nu(t\ph_1,t\ph_2)\cr
&=\int_0^1 dt\ \left[\del_1^2\nu(t\ph_1,t\ph_2)\ph_1
+\del_1\del_2\nu(t\ph_1,t\ph_2)\ph_2\right]\cr
\del_2\nu(\ph_1,\ph_2)
&=\int_0^1 dt\ \sfrac{d\hfill}{dt}\del_2\nu(t\ph_1,t\ph_2)\cr
&=\int_0^1 dt\ \left[\del_1\del_2\nu(t\ph_1,t\ph_2)\ph_1
+\del_2^2\nu(t\ph_1,t\ph_2)\ph_2\right]\cr
}\eqn $$
to yield
$$\eqalign{
\tilde\al(\ph_1,\ph_2)&=-\int_0^1 dt\ \del_1^2\nu(t\ph_1,t\ph_2)\cr
\tilde\be(\ph_1,\ph_2)&=\int_0^1 dt\ \del_1\del_2\nu(t\ph_1,t\ph_2)\cr
\tilde\ga(\ph_1,\ph_2)&=\int_0^1 dt\ \del_2^2\nu(t\ph_1,t\ph_2)\cr
}\eqn $$
\sni
{\it Proof of Lemma \Zeroo:}
First fix any sufficiently small $\ph_1$. If $\al,\be,\ga$ were constants
$\nu(\ph_1,\ph_2)=0$ would have two solutions, namely
$$
\ph_2=\frac{-\be\ph_1\pm\sqrt{\be^2\ph_1^2+\al\ga\ph_1^2}}{\ga}
\eqn $$
So define
$$
\Psi_\pm(\ph_1,\ph_2)
=\frac{\sqrt{\al(\ph_1,\ph_2)\ga(\ph_1,\ph_2)+\be^2(\ph_1,\ph_2)}
\ \mp\be(\ph_1,\ph_2)}{\ga(\ph_1,\ph_2)}=1+o(1)
\eqn $$
We have
$$
\nu(\ph_1,\ph_2)=\half\ga(\ph_1,\ph_2)
\left[\ph_2-\Psi_+(\ph_1,\ph_2)\ph_1\right]
\left[\ph_2+\Psi_-(\ph_1,\ph_2)\ph_1\right]
\eqn $$
Note that the factor $\half\ga(\ph_1,\ph_2)$ never vanishes, the next factor
obeys
$$\eqalign{
\ph_2-\Psi_+(\ph_1,\ph_2)\ph_1<0\qquad{\rm if\ \ }
\ph_2<\ph_1\cases{\min\Psi_+& if $\ph_1>0$\cr
                     \max\Psi_+& if $\ph_1<0$\cr}\cr
\ph_2-\Psi_+(\ph_1,\ph_2)\ph_1>0\qquad{\rm if\ \ }
\ph_2>\ph_1\cases{\max\Psi_+& if $\ph_1>0$\cr
                     \min\Psi_+& if $\ph_1<0$\cr}\cr
}\eqn $$
and the last factor obeys
$$\eqalign{
\ph_2+\Psi_-(\ph_1,\ph_2)\ph_1>0\qquad{\rm if\ \ }
\ph_2>-\ph_1\cases{\min\Psi_-& if $\ph_1>0$\cr
                     \max\Psi_-& if $\ph_1<0$\cr}\cr
\ph_2+\Psi_-(\ph_1,\ph_2)\ph_1<0\qquad{\rm if\ \ }
\ph_2<-\ph_1\cases{\max\Psi_-& if $\ph_1>0$\cr
                     \min\Psi_-& if $\ph_1<0$\cr}\cr
}\eqn $$
If necessary, restrict the neighbourhood so that $\min\Psi_\pm$ and 
$\max\Psi_\pm$ all lie in $[1/2,3/2]$.
Then, still for fixed $\ph_1$, $\nu(\ph_1,\ph_2)$ necessarily changes 
sign at least once between $\ \ph_1\min\Psi_+\ $ and $\ \ph_1\max\Psi_+\ $ and
at least once between $\ -\ph_1\min\Psi_-\ $ and $\ -\ph_1\max\Psi_-\ $. But
in a neighbourhood of the origin $\del_2^2\nu>0$, so that $\nu$,
viewed as a function of $\ph_2$, is strictly convex and can have at most two
zeroes. So it has exactly two zeroes, one in each of the aforementioned
intervals. Call the zeroes $\psi_\pm(\ph_1)$. 

The zeroes obey
$$\eqalign{
\psi_+(\ph_1)&=\Psi_+\big(\ph_1,\psi_+(\ph_1)\big)\ph_1\cr
\psi_-(\ph_1)&=-\Psi_-\big(\ph_1,\psi_-(\ph_1)\big)\ph_1
,\cr}\eqn $$
so
$$
\nu(\ph_1,\ph_2)= \half \ga(\ph_1,\ph_2)\  
(\ph_2-\ps_+(\ph_1))\ (\ph_2+\ps_-(\ph_1))
.\eqn $$
Since, to be in the aforementioned intervals, 
$|\psi_\pm(\ph_1)|\le |\ph_1|\max\Psi_\pm$ 
and since $\Psi_\pm=1+o(1)$, we have that $\psi_\pm$ is differentiable at $\ph_1=0$
and obeys
$$\eqalign{
\psi_\pm(0)&=0\cr
\psi_\pm'(0)&=\pm 1\cr
}\eqn $$
As $\nu\big(\ph_1,\psi_\pm(\ph_1)\big)=0$ we have for all $\ph_1\ne 0$
$$
\psi_\pm'(\ph_1)=-\frac{\del_1\nu\big(\ph_1,\psi_\pm(\ph_1)\big)}
{\del_2\nu\big(\ph_1,\psi_\pm(\ph_1)\big)}
\eqn $$
Note that for the denominator $\del_2\nu\big(\ph_1,\psi_\pm(\ph_1)\big)$
to vanish $\psi_\pm(\phi)$ must be a double zero of $\nu(\ph_1,\ \cdot\ )$
and we know that this cannot happen for $\ph_1\ne 0$. Hence $\psi_\pm$
is $C^1$ away from $\ph_1=0$. It only remains to verify
the continuity of $\psi'_\pm$ at $\ph_1=0$. From \queq{\Rtwo}
$$\eqalign{
\psi_\pm'(\ph_1)
&=-\frac{\del_1\nu\big(\ph_1,\psi_\pm(\ph_1)\big)}
{\del_2\nu\big(\ph_1,\psi_\pm(\ph_1)\big)}\cr
&=-\frac{-\tilde\al\big(\ph_1,\psi_\pm(\ph_1)\big)\ph_1
+\tilde\be\big(\ph_1,\psi_\pm(\ph_1)\big)\psi_\pm(\ph_1)}
{\tilde\be\big(\ph_1,\psi_\pm(\ph_1)\big)\ph_1
+\tilde\ga\big(\ph_1,\psi_\pm(\ph_1)\big)\psi_\pm(\ph_1)}\cr
&=-\frac{-[1+o(1)]\ph_1\pm o(1)\Psi_\pm\big(\ph_1,\psi_\pm(\ph_1)\big)\ph_1}
{o(1)\ph_1\pm[1+o(1)]\Psi_\pm\big(\ph_1,\psi_\pm(\ph_1)\big)\ph_1}\cr
&=\pm 1+o(1)\cr
}\eqn $$
\endproof
\pni
{\it Proof of Theorem \MH:}
Define
$$\eqalign{
x(\ph_1,\ph_2)&=\ph_2-\psi_+(\ph_1)\cr
\noalign{\vskip.1in}
y(\ph_1,\ph_2)&=\int\limits_0^1 d\al\; \del_\two \nu(\ph_\one,
(1-\al) \ps_+(\ph_\one) + \al \ph_\two)
.\cr}\eqn $$
Then 
$\ 
\nu(\ph_1,\ph_2)=x(\ph_1,\ph_2)\,y(\ph_1,\ph_2)
\ $
holds by Taylor expansion. That $x$ is $C^1$, that  
$\ x(0,0)=0\ $, that $\ \del_1x(0,0)=-1\ $ and that 
$\ \del_2x(0,0)=1\ $ are all obvious, and so are the statements for $y$.
\endproof

\ssni
\sect{Elliptic case}
Let $\nu(\ph_1,\ph_2)$ be a $C^2$ function in a neighbourhood of $(0,0)$
obeying
$$\meqalign{
\nu(0,0)&=0\cr
\del_1\nu(0,0)&=0  &&  \del_1\nu(0,0)&=0 \cr
\del_1^2\nu(0,0)&=1  &&  \del_2^2\nu(0,0)&=1  
&&\del_1\del_2\nu(0,0)=0 \cr
}\eqn $$
\The{\ME}{\Thsty Define $\th=\tan^{-1}\sfrac{\ph_2}{\ph_2}$ to be the usual
polar angle and $R=\nu(\ph_1,\ph_2)$. Then $R(0,0)=0$, $R$ is increasing on 
each fixed ray $\th=\Const$ and the Jacobian
$$
\frac{\del(R,\th)}{\del(\ph_1,\ph_2)}=1+o(1)
\eqn $$
Define
$$\eqalign{
x(\ph_1,\ph_2)=\sqrt{2R(\ph_1,\ph_2)}\cos\th(\ph_1,\ph_2)\cr
y(\ph_1,\ph_2)=\sqrt{2R(\ph_1,\ph_2)}\sin\th(\ph_1,\ph_2)\cr
}\eqn $$
Then 
$$
\nu(\ph_1,\ph_2)=\half\big(x(\ph_1,\ph_2)^2+y(\ph_1,\ph_2)^2\big)
\eqn $$
and the Jacobian
$$
\frac{\del(x,y)}{\del(\ph_1,\ph_2)}=1+o(1)
\eqn $$
}
\Rem{\dummy}$R$ plays the role of $\half r^2$ in the usual polar coordinates.
\Proof
We now have the representations (with new notation)
$$\eqalign{
\nu(\ph_1,\ph_2)&=\half\al(\ph_1,\ph_2)\ph_1^2
+\be(\ph_1,\ph_2)\ph_1\ph_2
+\half\ga(\ph_1,\ph_2)\ph_2^2\cr
\del_1\nu(\ph_1,\ph_2)&=\hat\al(\ph_1,\ph_2)\ph_1
+\hat\be(\ph_1,\ph_2)\ph_2\cr
\del_2\nu(\ph_1,\ph_2)&=\hat\be(\ph_1,\ph_2)\ph_1
+\hat\ga(\ph_1,\ph_2)\ph_2\cr
}\eqn $$
where
$$\deqalign{
\al(\ph_1,\ph_2)&=2\int_0^1 dt\ (1-t)\del_1^2\nu(t\ph_1,t\ph_2)
&=1+o(1)\cr
\be(\ph_1,\ph_2)&= 2\int_0^1 dt\ (1-t)\del_1\del_2\nu(t\ph_1,t\ph_2)
&=o(1)\cr
\ga(\ph_1,\ph_2)&=2\int_0^1 dt\ (1-t)\del_2^2\nu(t\ph_1,t\ph_2)
&=1+o(1)\cr
\hat\al(\ph_1,\ph_2)&=\int_0^1 dt\ \del_1^2\nu(t\ph_1,t\ph_2)
&=1+o(1)\cr
\hat\be(\ph_1,\ph_2)&=\int_0^1 dt\ \del_1\del_2\nu(t\ph_1,t\ph_2)
&=o(1)\cr
\hat\ga(\ph_1,\ph_2)&=\int_0^1 dt\ \del_2^2\nu(t\ph_1,t\ph_2)
&=1+o(1)\cr
}\eqn $$
That $R(0,0)=0$ and  $R$ is increasing on each fixed ray follows easily
from 
$$\eqalign{
\nu(\ph_1,\ph_2)&=\half\left(\ph_1^2+\ph_2^2\right)
 +o(1)\ph_1^2+o(1)\ph_1\ph_2+o(1)\ph_2^2\cr
\ph_1\del_1\nu(\ph_1,\ph_2)+\ph_2\del_2\nu(\ph_1,\ph_2)
&=\ph_1^2+\ph_2^2 +o(1)\ph_1^2+o(1)\ph_1\ph_2+o(1)\ph_2^2\cr
}\eqn $$
The Jacobian
$$\eqalign{
\pmatrix{\sfrac{\del R}{\del\ph_1}&
\sfrac{\del R}{\del\ph_2}\cr
\sfrac{\del \th}{\del\ph_1}&
\sfrac{\del \th}{\del\ph_2}}
&=\pmatrix{\hat\al(\ph_1,\ph_2)\ph_1+\hat\be(\ph_1,\ph_2)\ph_2&
\hat\be(\ph_1,\ph_2)\ph_1+\hat\ga(\ph_1,\ph_2)\ph_2\cr
\sfrac{-\ph_2}{\ph_1^2+\ph_2^2}&
\sfrac{\ph_1}{\ph_1^2+\ph_2^2}\cr}
\cr
&=\frac{\hat\al(\ph_1,\ph_2)\ph_1^2+2\hat\be(\ph_1,\ph_2)\ph_1\ph_2
+\hat\ga(\ph_1,\ph_2)\ph_2^2}{\ph_1^2+\ph_2^2}\cr
&=1+\frac{[\hat\al(\ph_1,\ph_2)-1]\ph_1^2+2\hat\be(\ph_1,\ph_2)\ph_1\ph_2
+[\hat\ga(\ph_1,\ph_2)-1]\ph_2^2}{\ph_1^2+\ph_2^2}\cr
&=1+o(1)\cr
}\eqn $$

For the transition from $R,\th$ to $x,y$
$$\eqalign{
\pmatrix{\sfrac{\del x}{\del R}&
\sfrac{\del y}{\del R}\cr
\sfrac{\del x}{\del\th}&
\sfrac{\del y}{\del\th}}
&=\pmatrix{\sfrac{1}{\sqrt{2R}}\cos\th&\sfrac{1}{\sqrt{2R}}\sin\th\cr
-\sqrt{2R}\sin\th&\sqrt{2R}\cos\th\cr
}=1
}\eqn $$
\endproof

\appendix{B}{Sharp Volume Bounds}
\noindent
In this Appendix we prove Theorem \bestvol\ 
under the assumptions \ATwo{2,0} and \AThr\ 
(obviously, \ATwo{k,\hxp} implies \ATwo{2,0} for all $k \ge 2$
and $\hxp \ge 0$).
The assumption \AFiv\ will not be used in this proof.
It will become clear in the proof that this estimate is best possible.
Before going into the details we outline the strategy. Away from the 
critical points (in $\th_\one$, $\th_\two$) of the map 
$\eta = e(v_\one \p(0,\th_\one) + v_\two \p (0,\th_\two ) + \q )$, 
the estimate is easily shown by a change of variables from one of the 
$\th_i$ to $e$. Near the critical points, a detailed analysis
of the singularity is required for getting the optimal improvement factor. 
We first determine for which $\q$ critical points are possible at all. 
Then we show that at any critical point the second derivative is a 
nonsingular matrix, thus  
the function $\eta$ is either of type $x^2 + y^2$ or of type $xy$
(the factor $\log \veps $ comes from the second case). 
We then use the Morse Lemma proven in Appendix A to calculate the 
volume improvement effect.

Note that $\q \in \cB$ is not restricted to be near to $S$, so that
we can make use of strict convexity only in the sense that
on a fixed level set of $e$ near to $S$, 
there are only two solutions of the equation $\del_\th \p = \ve{v}$
for any $0 \neq \ve{v} \in \R^d$, 
and that the curvature radius is finite at every point of $S$.
A convenient property of boundaries $S$ of strictly convex sets
in $\R^d$, namely that no three different points of $S$ 
can lie on the same straight line, does not hold on the torus $\cB$.
The reason we could use it in Chapter 3 was that all momenta
involved were on or near to $S$, and that we assumed \AFiv. 
The supremum in the definition of 
$\cW$ is over all $\q \in \cB$, not only those near to $S$, 
therefore \AFiv\ would be of no use here. 
Most of the following Lemmas 
deal with the complications due to $\cB$ being a torus.  

\sect{Two Dimensions}
\noindent
Let $d=2$, use the coordinates $\rh $ and $\th$ defined in Section
2.2, and let 
$$
n(\rh,\th ) = \frac{\nabla e (\p (\rh,\th ) )} 
{\abs{\nabla e (\p (\rh,\th ) )}}
.\eqn $$
For brevity, we write $n(\th ) = n(0,\th)$.
The set 
$$
E=\set{(\th_\one, \th_\two ) \in S\times S}
{n(\th_\one) \times n(\th_\two )=0}
\eqn $$
is the zero set of the map 
$$
H: S \times S \to \R, \quad 
(\th_\one, \th_\two ) \mapsto (n(\th_\one ) \times n(\th_\two ))\cdot e_\thr
\eqn $$
(where the vector product is inherited from $\R^3$ and $e_\thr $ denotes the 
unit vector in $3$--direction).
$H \in C^1 (S\times S, \R)$, and 
$$
\nabla H = \left( \frac{\del H }{\del \th_\one},
\frac{\del H }{\del \th_\two} \right) = 
\left( e_\thr\cdot (\del_\th n (\th_\one ) \times n(\th_\two )) , 
e_\thr\cdot (n(\th_\one ) \times \del_\th n(\th_\two ))\right)
\eqn $$
is nonzero for all $\ulth = (\th_\one, \th_\two ) \in E$
(because $n$ is a unit vector, $\del_\th n \perp n$, 
and because the Fermi surface has nonzero curvature, 
$\del_\th n \neq 0$).
Thus $E$ is a $C^{1}$--submanifold of $S\times S$, of codimension $1$.
Moreover, $E$ is compact, thus covered  by finitely many balls in 
$S \times S$. Thus there is $\de >0$ such that $U_{\de} (E)$
is contained in this finite covering. Since $\cR_\de = S \times S \setminus 
U_{\de} (E)$ is also compact, $|H(\ulth)| \geq \vphi (\de ) > 0$
holds for all $\ulth \not\in U_{\de} (E) $, with $\vphi$ some positive
function ($\vphi (x) > 0 $ for $x > 0$). We shall choose $\de > 0 $
later (independent of $\veps$; this is the $\veps$ of Theorem \bestvol). Split
$$
\cW(\veps ) = E_\de (\veps ) + R_\de (\veps )
\eqn $$
according to the decomposition $S\times S = U_\de (E) \dotcup \cR_\de$.
It was shown in Appendix A of \FST\ that the contribution $R_\de $
from the regular region is bounded by 
$$
R_\de (\veps ) \leq \Const {\veps \over \vphi (\de ) }
\eqn $$
The $1/\vphi(\de )$ comes from a Jacobian. To bound $E_\de$, 
the contribution from the region where this Jacobian can become singular, 
we collect some more consequences of the geometry of the Fermi surface.

\Lem{\CriP} {\Lesty Let $\p = \p (\rh, \th ) \in \cU (S)$. Then all   
solutions $\Q $ of
$$
\nabla e(\Q ) \cdot \del_\th \p =0
\eqn $$
that satisfy $\Q \in \cU (S)$ are given as $\Q = \p (\rh_\one, 
\vth^{(k)} (\rh, \th, \rh_\one ))$ where for $k \in \{ 1,2\}$, 
$\vth^{(k)} $ are $C^1$ in $(\rh, \th, \rh_\one)$ and at $\rh_\one = \rh$, 
$$\eqalign{
\vth^{(1)} (\rh, \th, \rh) & = \th \cr
\vth^{(2)} (\rh, \th, \rh) & = a_\rh (\th)
.\cr}\EQN\atrho $$
At fixed $\rh $ and $\th $, the curve $\rh_\one\mapsto\vth^{(k)}
(\rh,\th,\rh_\one)$ is transversal to $S_{\rh_\one}$. 
}
\Proof Since $\Q \in \cU (S)$, we can write $\Q = \p (\rh_\one, \th_\one )$.
Fix $\rh$ and $\th$.
Since $S_{\rh_\one} = \set{\p (\rh_\one, \th_\one)}{\th_\one \in [0,2\pi ]}$ is 
strictly convex and a $C^1$--manifold, there are, for each fixed $\rh_\one$, 
exactly two values of $\th_\one$ for which 
$\nabla e(\p (\rh_\one, \th_\one )) \cdot \del_\th \p (\rh, \th ) =0$.
For $\rh_\one = \rh $ they are given by $\p (\rh, \th)$ and 
$\p (\rh, a_\rh(\th))$. Fix $(\rh, \th)$. The map
$\ph : (\rh_\one, \th_\one ) \mapsto 
\nabla e(\p (\rh_\one, \th_\one ))\cdot \del_\th \p (\rh, \th )$ satisfies 
$\ph (\rh, \th) =0 $ and $\ph (\rh, a_\rh (\th ))=0$. 
The derivative
$$
\frac{\del \ph }{\del \th_\one} (\rh, \th, \rh_\one , \th_\one ) 
= 
\left(\del_\th \p (\rh_\one ,\th_\one ), e''(\p (\rh_\one, \th_\one ))
\del_\th \p (\rh ,\th )\right)
\eqn $$
is continuous. At $\th_\one=\th$ and $\rh_\one =\rh$, it is equal to 
$w(\p(\rh,\th))$ (defined in \SYmm), hence nonzero. Thus in a 
neighbourhood of $\rh_\one = \rh$ there is a function 
$\vth^{(1)}$ depending on $(\rh, \th, \rh_\one) $ such that  
$\ph (\rh_\one, \vth^{(1)}(\rh, \th,\rh_\one )) =0$. Similarly, one constructs
the solution $\vth^{(2)}$.
$\ph$ is $C^1$ in $\rh_\one,\rh , \th $, so the $\vth^{(k)}$ are also 
$C^1$.

Transversality holds because $\th_\one $ moves $\p (\rh_\one, \th_\one )$
tangentially to $S_{\rh_\one}$ and  $\sfrac{\del\ph}{\del \th_\one}
\neq 0$. \endproof

Note that by compactness of $\cB$ and $S$, the size of the neighbourhoods 
can be chosen uniform in $(\rh,\th )$. We may therefore assume that $\cU (S)$ 
is such that the solution curves of the Lemma \CriP\ 
exist for all $\p \in \cU (S)$.

Lemma \CriP\ allows us to determine the set $\cP_\ka$ of those $\q \in \cB $ 
for which critical points of the map
$$
\et:\quad \ulth = (\th_\one,\th_\two) \mapsto \et (\ulth , \q ) = 
e(v_\one \p (0, \th_\one ) + v_\two \p (0, \th_\two ) + \q )
\EQN\Etadef $$
are possible. For $0<\ka < r_\zer $ let 
$$
\cP_\ka = \bigcup\limits_{k=1}^2
\set{\p (\rh,\vth^{(k)} (0,\th_\one,\rh )) 
- v_\one \p (0, \th_\one ) - v_\two \p (0, \th_\two )}
{(\th_\one, \th_\two ) \in E, \; \abs{\rh } \leq \ka }
.\EQN\cPkadef $$
The set $\set{\p (\rh,\vth^{(1)} (0,\th,\rh )) +\p (0, \th )+ \p (0, \th )}
{\abs{\rh } \leq \ka }$, which is one of the four sets in the union
making up $\cP_\ka $, is sketched in Figure \nextfig{\figufft}. The square indicates the 
boundary of the fundamental region for the torus $\cB$, so the shaded region
should be thought of as folded back into this fundamental region by 
periodicity. 

\herefig{critp}

\Lem{\noCP} {\Lesty If $\q \not\in \cP_\ka $, then 
$\abs{\et(\ulth , \q)} \leq \ka $ implies 
$\sfrac{\del \et }{\del \th_\one} \neq 0$ or
$\sfrac{\del \et }{\del \th_\two} \neq 0$.}

\Proof 
Let 
$$
\Q = v_\one \p (0, \th_\one ) + v_\two \p (0, \th_\two ) +\q
\EQN\Qdef $$
The condition 
$$
\frac{\del \et}{\del \th_i} =
\nabla e (\Q ) \cdot \del_\th \p (0,\th_i) = 0 
\quad \hbox{ for } i=1 \hbox{ and } i=2
\EQN\CriCond $$
implies $\ulth \in E$: if $\abs{e(\Q)} \leq \ka $, $\nabla e(\Q ) \neq 0$,
so $\del_\th \p (0, \th_\one )$ and $\del_\th \p (0, \th_\two )$
are both orthogonal to the same nonzero vector. Since $d=2$, they must be 
collinear, thus $\th_\two \in \{ \th_\one , a(\th_\one ) \}$ 
by strict convexity.

If there is a critical point, i.e.\ a solution to \queq{\CriCond} 
with $\abs{e(\Q )} \leq \ka $, $\Q$ is determined as in Lemma \CriP, 
so $\q \in \cP_\ka$ by \queq{\Qdef}. \endproof  

Fix $\ka > 0$. Without loss, we may assume that $\veps \leq \sfrac{\ka}{2}$
since the estimate in Theorem \bestvol\ is trivially true for $\veps $ 
bounded away from zero by choice of the constant.

Since $\cP_\veps \subset \cP_{\ka/2} {\subset \atop \neq} \cP_\ka$ and
by continuity of $\sfrac{\del \et}{\del \th_i}$ and compactness,
there is $\de_\one >0$ such that for all $\ulth $ in the support 
of the integral \queq{\cWdef} for $\cW$, 
$$
\inf\limits_{\q \in \overline{\cB \setminus \cP_\ka}}\;
\inf\limits_{\ulth \in E}\;
\max\left\{ 
\abs{\frac{\del \et }{\del \th_\one}} ,
\abs{\frac{\del \et }{\del \th_\two}}\right\} \geq \de_\one 
.\eqn $$
Thus the $\de > 0$  introduced previously can be chosen so small
(independently of $\veps $) that for all $\q \not\in\cP_\ka$ and 
all $\ulth \in U_\de (E)$
$$
\abs{\frac{\del \et}{\del \th_i}} \geq \de_\one 
\quad \hbox{ for } i=1 \hbox{ or } i=2.
\eqn $$
Thus there are $R_\one, R_\two$ such that $U_\de (E) = R_\one \dotcup R_\two$
and $|\sfrac{\del \et}{\del \th_i}|\geq \de_\one$ on $R_i$. Obviously, 
$R_\one $ and $R_\two $ are similar, so we deal with $R_\one$.
By a change of variables from $\th_\one $ to $\et$, 
$$\eqalign{
\sup\limits_{\q \in \overline{\cB \setminus \cP_\ka}} \quad
\int\limits_{R_\one} d\th_\one \, d\th_\two \; 
\True{\abs{\et(\th_\one, \th_\two, \q ) } \leq \veps } & \leq
2 \pi \frac{1}{\de_\one} \int d\et\; \True{\abs{\et} \leq \veps }\leq \cr
& \leq 4 \pi\frac{\veps}{\de_\one}
,\cr}\eqn $$
the factor of $2\pi$ coming from the $\th_\two $ integration.
Thus, to prove Theorem \bestvol\ it suffices to bound 
$$
W_\ka (\veps ) = 
\max\limits_{v_i \in \{ \pm 1\}} \quad
\sup\limits_{\q \in \cP_\ka(v_\one, v_\two)}
\quad
\int\limits_{U_\de (E)} d\th_\one \, d\th_\two \,
\True{\abs{\et(\ulth, \q)} \leq \veps}
.\EQN\itsuff $$

\Lem{\qCP} {\Lesty  For fixed $\q \in \cP_\ka $, 
the solutions $\ulth=\ulth^{(cr)}$ of \queq{\CriCond} are isolated.
The critical points $\ulth^{cr}$ are $C^1$ functions of $\q$. } 

\Proof {\it Case $\p_\two=\p_\one $:} the right side of the equation 
$$
\q = \p (\rh, \vth^{(k)} (0,\th, \rh ) )
- v_\one \p (0,\th) - v_\two \p (0, \th )
\eqn $$
is a $C^1 $ function of $(\rh, \th)$. We now show
that this function is invertible.The derivatives are
$$\eqalign{
\frac{\del \q}{\del \rh } & = 
{\del_\rh \p} (\rh, \vth^{(k)} (0,\th_\one, \rh )) 
+ \del_\th \p (\rh, \vth^{(k)} (0,\th_\one, \rh )) 
\frac{\del \vth^{(k)}}{\del \rh} \cr
\frac{\del \q}{\del \th } & = \del_\th 
\p (\rh, \vth^{(k)} (0,\th_\one, \rh ))
\frac{\del \vth^{(k)}}{\del\th} 
- v_\one \del_\th \p (0,\th ) 
- v_\two \del_\th \p (0,\th )
.\cr}\EQN\qDerivs $$
We know that $\rh $ is in a small neighbourhood of zero by Lemma \CriP, 
so by continuity in $\rh$, it suffices to prove that the derivative is
nonsingular at $\rh =0$.\sni 
{\it Case $\vth^{(1)}$:}
At $\rh =0$,  $\vth^{(1)}=\th$, so 
$$\eqalign{
\frac{\del \q}{\del \rh } & = 
\del_\rh \p (\th ) + \del_\th \p (\th )
\frac{\del \vth^{(1)}}{\del \rh} \cr
\frac{\del \q}{\del \th } & = \del_\th \p (\th ) 
\left( \frac{\del \vth^{(1)}}{\del\th} - v_\one  - v_\two \right)
.\cr}\eqn $$
From \queq{\atrho}, we see that 
$$
\frac{\del \vth^{(1)}}{\del\th}(\rh,\th,\rh) =1
\EQN\Btwon $$
so $ \sfrac{\del \vth^{(1)}}{\del\th}(\rh,\th,\rh)-v_\one -v_\two $
is $-1$, $1$ or $3$.  $\sfrac{\del \vth^{(1)}}{\del\th}$ is continuous, 
so there is $r_\one $ such that for $|\rh_\one-\rh|<r_\one$, 
$$
\abs{\frac{\del \vth^{(1)}}{\del\th}(\rh,\th,\rh_\one)-v_\one -v_\two}
\geq \sfrac{1}{2}
\eqn $$
It is now obvious that rank
$(\sfrac{\del\q}{\del\rh}, \sfrac{\del\q}{\del\th}) =2$: 
the first column is nonzero because $\del_\rh \p (0,\th)$ is nonzero, and 
$\del_\rh \p (0,\th )$ and $\del_\th \p (0,\th)$ are linearly independent. 
The second column is a nonzero multiple of $\del_\th \p (0,\th)$. Thus 
the column vectors are linearly independent.

{\it Case $\vth^{(2)}$:} Recall that $\vth^{(2)}$ is the solution with
$\vth^{(2)}(\rh,\th,\rh )= a_\rh (\th)$.
Hence
$$
\frac{\del \vth^{(2)}}{\del \th } (\rh, \th , \rh ) =
\frac{\del a_\rh (\th)}{\del \th}
.\EQN\Btwth $$
Thus, by \queq{\qDerivs}, at $\rh=0$
$$\eqalign{
\frac{\del \q}{\del \rh } & = 
\del_\rh \p (a(\th)) + \del_\th \p (\th )
\frac{\del \vth^{(2)}}{\del \rh}(0,\th,\rh=0) \cr
\frac{\del \q}{\del \th } & = \del_\th \p (a(\th )) 
 \frac{\del \vth^{(2)}}{\del\th}(0,\th,0) - v_\one \del_\th \p (\th )
  - v_\two  \del_\th \p (\th )\cr
& =  \del_\th \p (\th ) \left( -\frac{\del a}{\del\th}(\th) 
-v_\one -v_\two \right)
.\cr}\eqn $$
By \queq{\antidel}, $\sfrac{\del \q}{\del \th}\neq 0$. So, similarly to the 
first case, the two rows are linearly independent. 
\sni
{\it Case $\th_\two = a(\th_\one)$:} The argument is similar to the above, 
only that now in the case $\vth^{(1)}$,
$$
\frac{\del\q}{\del\th}(0,\th) = \del_\th\p(0,\th) 
\left(\frac{\del\vth^{(1)}}{\del\th}(0,\th,0)-v_\one +v_\two 
\frac{\del a}{\del\th}(\th) \right) 
\eqn $$
and in the case $\vth^{(2)}$,
$$
\frac{\del\q}{\del\th}(0,\th) = \del_\th\p(0,\th) 
\left(-\frac{\del\vth^{(2)}}{\del\th}(0,\th,0)
-v_\one +v_\two  \frac{\del a}{\del\th}(\th) \right) 
.\eqn $$
\endproof

\noindent
\Rem{\compcont} By compactness of $\overline{\cP_\ka }$, the critical 
points are uniformly continuous functions of $\q$. Thus, given $s>0$,
there is $r>0$ such that for all $\q \in \cP_\ka$ and all $\q' \in U_r (\q)$,
$\abs{\ulth^{cr} (\q' ) -\ulth^{cr} (\q )} < s/2$. Moreover, there is 
a finite set of points $\q^{(1)}, \ldots , \q^{(n)}$ such that 
$\cP_\ka \subset \bigcup\limits_{i=1}^n U_r (\q^{(i)})$. 

\sni
{\bf Proof of Theorem \bestvol\ for $d=2$:} 
It suffices to prove \queq{\itsuff} for any given $i$. 
$s$ will be chosen at the end. $n$ depends on $s$. 
Let $i \in \nat{n}$, $\q \in U_r (\q^{(i)} )$, and let
$$
K_i = E \setminus \bigcup\limits_{l=1}^4 U_s (\ulth^{cr} (\q^{(i)}))
.\eqn $$
If $\ulth \not\in K_i$, $\sfrac{\del \et }{\del \th_\one } (\ulth, \q )\neq 0$
or  $\sfrac{\del \et }{\del \th_\two} (\ulth, \q )\neq 0$
because by choice of $r$, all critical points to $\q \in U_r(\q^{(i)})$
are in $U_{s/2}(\ulth^{cr}(\q^{(i)}))$. 
Thus, by a similar argument as in the case $\q \not\in \cP_\ka $, 
$$
\sup\limits_{\q \in  U_r (\q^{(i)}) } \quad
\int\limits_{K_i} d\th_\one \, d\th_\two \; 
\True{\abs{\et(\th_\one, \th_\two, \q ) } \leq \veps } \leq \Const\frac{\veps}{\de_\two}
\eqn $$
where $\de_\two $ does not depend on $\veps $.

Finally, let $\ult=\ult(\q)$ be one of the critical points to $\q$ and 
$\ulth \in U_s (\ult )$. We now have to bound
$$
W_\ka^{(i)} (\veps ) = \;
\max\limits_{v_\one,v_\two =\pm1}\;
\sup\limits_{\q \in U_r(\q^{(i)})\cap\cP_\ka}
\;
\int\limits_{U_s(\ult(\q))} 
d\th_\one d\th_\two\;
\True{\abs{\et(\ulth,\q)}\le \veps}
.\eqn $$
By Taylor expansion, 
$$
\et(\ulth , \q ) = \et (\ult , \q ) + 
\left(\ulth -\ult  , D(\ult , \ulth -\ult , \q ) 
(\ulth - \ult )\right) 
\eqn $$
with 
$$
D(\ult , \ulph, \q ) = \int\limits_0^1 d\al \; (1-\al) \quad 
\left(
\matrix{ {\del_\one}^2 \et (\ult + \al \ulph, \q ) & 
\del_\one\del_\two  \et (\ult + \al \ulph, \q ) \cr
\del_\two\del_\one  \et (\ult + \al \ulph, \q ) & 
{\del_\two}^2 \et (\ult + \al \ulph, \q ) \cr} \right)
.\eqn $$
Let 
$$
D_\zer=D_\zer(\q)=D(\ult(\q),0,0,\q)
.\eqn $$

\Lem{\Dgood}{\Lesty For all $\q \in U_r(\q^{(i)})$, $\abs{\det D_\zer (\q)}
\ge \sfrac{1}{4}w_\zer^2$.}

\Proof  By \queq{\Qdef} and \queq{\Etadef},
$$\eqalign{
\del_\one^2 \et (\ulth , \q) &= 
(\del_\th \p_\one,e''(\Q )\del_\th \p_\one ) + 
v_\one \nabla e(\Q)\cdot \del_\th^2\p_\one \cr
\del_\two^2 \et (\ulth , \q) &= 
(\del_\th \p_\two,e''(\Q )\del_\th \p_\two ) + 
v_\two \nabla e(\Q)\cdot \del_\th^2\p_\two \cr
\del_\one\del_\two \et (\ulth, \q) &= v_\one v_\two 
(\del_\th\p_\two, e''(\Q)\del_\th\p_\one)=
\del_\two\del_\one \et (\ulth, \q)
.\cr}\eqn $$
$\eta$ is $C^2$, so $D$ is continuous in all its arguments. 
We set $\ulth$ to its critical value $\ult$. 
Then $\Q = \p (\rh_\thr, \vth^{(k)}(0,\th_\one,\rh_\thr))$.
By continuity of $D$, it suffices to show that 
$|\det D_\zer|\ge \sfrac{3}{8}w_\zer^2$ for $\rh_\thr = 0$.
Thus we can put $\rh_\thr = 0$. Then
$\vth^{(1)}=\th_\one$ and $\vth^{(2)}=a(\th_\one)$.
Let $\q \in U_r(\q^{(i)})\cap \cP_\ka $, and let $\ult $ be the 
critical point for $\q$. 
Recall that $w(\p) = (\del_\th \p , e''(\p ) \del_\th \p)$, 
and that $\q = \Q - v_\one \p_\one - v_\two \p_\two $.
\ssni
{\it Case $\vth^{(1)}$:} $\Q = \p (0,\th_\one)=\p_\one$.
If $\th_\two=\th_\one$, i.e.\ $\p_\two=\p_\one =\p$, 
$$
D_\zer=\left(\matrix{ w(\p) (1-v_\one) & v_\one v_\two w(\p) \cr
v_\one v_\two w(\p) &  w(\p) (1-v_\two ) \cr }\right)
\eqn $$
If $v_\one=v_\two=-1$, $\det D_\zer = 3 w(\p )^2 > 0$, and $\q=3\p$.
Otherwise, $\det D_\zer = -w(\p)^2 < 0$.

If $\th_\two=a(\th_\one)$, we use \queq{\deltwop} and \queq{\Trick}, to get
$$
\nabla e(\p(0,\th)) \cdot  (\del_\th^2 \p)(0,a(\th)) = 
- \frac{1}{\frac{\del a}{\del \th}(\th) } 
\nabla e(\p(0,\th)) \cdot  (\del_\th^2 \p)(0,\th) =
\frac{1}{\frac{\del a}{\del \th}(\th) }\; w(\p(0,\th)) 
.\eqn $$
Thus we can rewrite $D_\zer $ as
$$
D_\zer =w(\p_\one)\;\left(\matrix{(1-v_\one) &
 -v_\one v_\two   \cr & \cr
 -v_\one v_\two  &
\left(1 + v_\two\; \frac{1}{
\sfrac{\del a}{\del\th}(\th_\one)}\right)
\cr} \right)
\eqn $$
If $v_\one =1$, $\det D_\zer = -w(\p_\one)^2$.
If $v_\one=-1$, 
$$
\det D_\zer =w(\p_\one)^2
\left( 2 \left(1+\frac{v_\two}{\sfrac{\del a}{\del \th}}\right) -1\right)
\eqn $$
so if $v_\two =1$, $\det D_\zer\geq\sfrac{25}{9}w(\p_\one)^2$. If $v_\two=-1$, 
$$
\left(1+\frac{v_\two}{\sfrac{\del a}{\del \th}}\right) =
\left(1-\frac{1}{\sfrac{\del a}{\del \th}}\right) \le \frac{1}{9}
\eqn $$
by \queq{\antidel}, so 
$\det D_\zer \le -\sfrac{3}{4}w(\p_\one)^2$. 
\ssni
{\it Case $\vth^{(2)}$:} $\Q = \p (0,a(\th_\one))=\ve{a}(\p_\one)$.
If $\p_\two = \p_\one$, 
$$
D_\zer = \left(\matrix{ 
(\del_\th \p_\one, e''(a(\p_\one))\del_\th \p_\one) + 
v_\one \nabla e(a(\p_\one)) \cdot \del_\th^2 \p_\one &
v_\one v_\two (\del_\th \p_\two, e''(a(\p_\one))\del_\th \p_\one) \cr
(\del_\th \p_\one, e''(a(\p_\one))\del_\th \p_\two) & 
(\del_\th \p_\one, e''(a(\p_\one))\del_\th \p_\one) + 
v_\one \nabla e(a(\p_\one)) \cdot \del_\th^2 \p_\one \cr}\right)
\eqn $$
We use \queq{\antidef} and 
$$\eqalign{
\nabla e(\p(a(\th))) \cdot (\del_\th^2 \p) (\th) &=
-\; \sfrac{\del a}{\del \th}\;  
\nabla e(\p(a(\th))) \cdot (\del_\th^2 \p) (a(\th)) = \cr
&= \sfrac{\del a}{\del \th} w(\p (a(\th ))) 
\cr}\eqn $$
to rewrite this as 
$$
D_\zer = w(a(\p_\one)) 
\left(\matrix{
1+v_\one \sfrac{\del a}{\del \th}(\th_\one) & 
v_\one v_\two \cr
v_\one v_\two &
1+v_\two \sfrac{\del a}{\del \th}(\th_\one) 
\cr}\right)
.\eqn $$
For all $v_\one=\pm 1$ and $v_\two = \pm 1$, 
$|\det D_\zer| \ge \sfrac{3}{8}w_\zer^2$ by \queq{\antidel}.

If $\p_\two = \ve{a}(\p_\one)$, 
$$
D_\zer = w(a(\p_\one)) \left(\matrix{
 (1+ v_\one  \sfrac{\del a}{\del\th} (\th_\one)) &
-v_\one v_\two \cr
-v_\one v_\two &
1-v_\two \cr}\right)
\eqn $$
For all $v_\one=\pm 1$ and $v_\two = \pm 1$, 
$|\det D_\zer| \ge \sfrac{3}{4}w_\zer^2$ by \queq{\antidel}.
\endproof

\pni
Fix $\q \in U_r (\q^{(i)})$. 
Let $\cO(\q)$ be the rotation such that $\cO(\q) D_\zer(\q) \cO(\q)^{-1}=
\tilde D_\zer = \hbox{ diag } \{ d_\one,d_\two\}$. Rotate
$\ulth-\ult \to \tilde{\ulph} = \cO(\q) (\ulth-\ult)$. This change 
of variables is $C^\infty$ and the Jacobian is one. Denoting 
$\tilde D(\ult,\tilde{\ulph},\q) = \cO(\q) D(\ult,\cO(\q)^{-1}\tilde{\ulph},\q)
\cO(\q)^{-1}$, we are going to bound 
$$
\int\limits_{U_s(0,0)} d\tilde \ph_\one d\tilde \ph_\two\;
\True{\abs{\et(\t(\q),\q)+(\tilde{\ulph}, \tilde D(\ult,\tilde{\ulph},\q) 
\tilde{\ulph})}\le \veps}
\eqn $$
uniformly in $\q$. We rescale 
$$
(\tilde \ph_\one,\tilde \ph_\two) \to (\ph_\one,\ph_\two) =
(|d_\one|^{\sfrac{1}{2}} \tilde \ph_\one,
(|d_\two|^{\sfrac{1}{2}} \tilde \ph_\two)
.\eqn $$
This transformation is again $C^\infty$, and its Jacobian is 
$$
\abs{\det D_\zer}^{-\sfrac{1}{2}} \leq \frac{2}{|w_\zer|}
.\eqn $$
Thus, in the new coordinates, 
$$
\et(\ulth,\q) - \et(\ult(\q),\q) = \nu (\ph_\one,\ph_\two)
\eqn $$
with a function $\nu \in C^2$, since $\et \in C^2$ 
and the coordinate change from 
$\ulth$ to $\ulph$ is $C^\infty$. Moreover, 
$$\eqalign{
\nu(0,0)&=0,\quad \del_i\nu (0,0)=0, \quad \del_\one\del_\two\nu (0,0)=0,\cr
\del_\one^2\nu(0,0) &= \pm 1, \quad 
\del_\one^2\nu(0,0) = \pm 1
,\cr}\eqn $$
and $\ulth \in U_s(\t)$ implies $\ulph \in U_\si(0,0)$ for some $\si(s)>0$. 
Thus we have proven the statement if we can bound
$$
\int\limits_{U_\si(0,0)} d\ph_\one d\ph_\two\; 
\True{\abs{f_\zer + \nu(\ph_\one,\ph_\two)} \leq \veps }
\EQN\almostdone $$
uniformly in $f_\zer$ (here $f_\zer = \et(\ult(\q),\q)$).
Because of the absolute value in \queq{\almostdone}, we may assume 
$\del_\one^2\nu = 1$, $\del_\two^2 \nu = \pm 1$. If $\del_\two^2 \nu =1$,
by Theorem \ME\ there is a change of variables 
$(\ph_\one,\ph_\two) \to (R,\al)$
with Jacobian bounded by $2$ if $s$ is chosen small enough, such that 
$\nu(\ph_\one,\ph_\two) = R$. Thus \queq{\almostdone} is bounded by
$$
2 \int dR\ d\al \True{|f_\zer + R | \leq \veps } \le 8\pi \ \veps 
.\eqn $$
If $\del_\two^2 \nu = -1$, by Theorem \MH, there is a change of variables
$(\ph_\one,\ph_\two) \to (x,y)$, such that $\nu(\ph_\one,\ph_\two)=xy$, and
therefore \queq{\almostdone} is bounded by 
$$
\int\limits_{U_{\tilde \si} (0,0)} dx\ dy\; 
\True{|f_\zer + xy| \leq \veps} \leq 
\Const \;|\log \tilde \si|\; \veps\; |\log \veps|
\eqn $$
where $\tilde \si$ is such that the image of $U_\si(0,0)$
under the last change of variables is contained
in $U_{\tilde \si}(0,0)$. This completes the proof of Theorem \bestvol\
in two dimensions.

\sect{Higher Dimensions}
\noindent
In this Section, we prove Theorem 1.1 for $d \ge 3$. The method
of proof is a generalization of that for $d=2$. The main change
is that instead of numbers, matrices of dimension $d-1$ appear.
\Lem{\nondthr}{\bf (Nondegeneracy of Critical Points) \Lesty
Fix any $\q\in\R^d$ and  $v_\th,v_\phi\in\{\pm 1\}$. 
Suppose that $(\th_\zer,\phi_\zer)\in S^{d-1}\times S^{d-1}$ 
is a critical point of $F(\th,\phi)=e\big(\q+v_\th\p(0,\th)+v_\phi\p(0,\phi)\big)$ 
and that $e(\q+v_\th\p(0,\th_\zer)+v_\phi\p(0,\phi_\zer))=0$.  
Then $\partial^2F(\th,\phi)$
has a nonzero determinant for $\th=\th_\zer,\ \phi=\phi_\zer$.
}
\Proof Abuse notation by replacing $\p(0,\th)$ with $\p(\th)$.
Define $\al_\zer$ by $\p(\al_\zer)=\q+v_\th\p(\th_\zer)+v_\phi\p(\phi_\zer)$.
That $(\th_\zer,\phi_\zer)$ is a critical point of $F$ means
$$
\nabla e\big(\p(\al_\zer)\big)\cdot\partial_i\p(\th_\zer)
=\nabla e\big(\p(\al_\zer)\big)\cdot\partial_j\p(\phi_\zer)
=0 
\eqn $$
for all $i,j \in \nat{d-1}$.
Hence the normal vectors to the Fermi surface at $\p(\th_\zer)$ and $\p(\phi_\zer)$
must be parallel to  the normal vector to the Fermi surface at $\p(\al_\zer)$.
In other words, there are nonzero numbers $c_\th,c_\phi$ such that
$$\eqalign{
\nabla e\big(\p(\al_\zer)\big)&=c_\th \nabla e\big(\p(\th_\zer)\big)\cr
\nabla e\big(\p(\al_\zer)\big)&= c_\phi \nabla e\big(\p(\phi_\zer)\big)\cr
}\EQN\cthdef $$
For all $\al,i,j$
$$\eqalign{
&e\big(\p(\al)\big)=0\cr
&\nabla e\big(\p(\al)\big)\cdot\partial_i\p(\al)=0\cr
&\left(   \partial_j\p(\al),\partial^2 e\big(\p(\al)\big)\partial_i\p(\al)\right)   
+\nabla e\big(\p(\al)\big)\cdot\partial_j\partial_i\p(\al)=0
}\eqn $$
Hence
$$\eqalign{
\partial_{\th_i}\partial_{\th_j}F(\th_\zer,\phi_\zer)
&=\left(   \partial_j\p(\th_\zer),\partial^2 e\big(\p(\al_\zer)\big)\partial_i\p(\th_\zer)\right)   
+v_\th\nabla e\big(\p(\al_\zer)\big)\cdot\partial_j\partial_i\p(\th_\zer)\cr
&=\left(   \partial_j\p(\th_\zer),\partial^2 e\big(\p(\al_\zer)\big)\partial_i\p(\th_\zer)\right)   
+v_\th c_\th\nabla e\big(\p(\th_\zer)\big)\cdot\partial_j\partial_i\p(\th_\zer)\cr
&=\left(   \partial_j\p(\th_\zer),\partial^2 e\big(\p(\al_\zer)\big)\partial_i\p(\th_\zer)\right)   
-v_\th c_\th
\left(   \partial_j\p(\th_\zer),\partial^2 e\big(\p(\th_\zer)\big)\partial_i\p(\th_\zer)\right)   \cr
\partial_{\phi_i}\partial_{\phi_j}F(\phi_\zer,\phi_\zer)
&=\left(   \partial_j\p(\phi_\zer),\partial^2 e\big(\p(\al_\zer)\big)\partial_i\p(\phi_\zer)\right)   
+v_\phi\nabla e\big(\p(\al_\zer)\big)\cdot\partial_j\partial_i\p(\phi_\zer)\cr
&=\left(   \partial_j\p(\phi_\zer),\partial^2 e\big(\p(\al_\zer)\big)\partial_i\p(\phi_\zer)\right)   
+v_\phi c_\phi\nabla e\big(\p(\phi_\zer)\big)\cdot\partial_j\partial_i\p(\phi_\zer)\cr
&=\left(   \partial_j\p(\phi_\zer),\partial^2 e\big(\p(\al_\zer)\big)\partial_i\p(\phi_\zer)\right)   
-v_\phi c_\phi\left(   \partial_j\p(\phi_\zer),\partial^2 e\big(\p(\phi_\zer)\big)\partial_i\p(\phi_\zer)\right)   \cr
\partial_{\th_i}\partial_{\phi_j}F(\th_\zer,\phi_\zer)
&=v_\th v_\phi\left(   \partial_j\p(\phi_\zer),\partial^2 e\big(\p(\al_\zer)\big)\partial_i\p(\th_\zer)\right)   \cr
}\eqn $$
Define the $(d-1)\times (d-1)$ matrices
$$\eqalign{
M_\al&=\left[\left(   \partial_j\p(\th_\zer),\partial^2 e\big(\p(\al_\zer)\big)\partial_i\p(\th_\zer)\right)   \right]_{1\le i,j\le d-1}\cr
M_\th&=\left[\left(   \partial_j\p(\th_\zer),\partial^2 e\big(\p(\th_\zer)\big)\partial_i\p(\th_\zer)\right)   \right]_{1\le i,j\le d-1}\cr
M_\phi&=\left[\left(   \partial_j\p(\th_\zer),\partial^2 e\big(\p(\phi_\zer)\big)\partial_i\p(\th_\zer)\right)   \right]_{1\le i,j\le d-1}\cr
}\eqn $$
Because $\partial_1\p(\th_\zer),\partial_2\p(\th_\zer)$ and
$\partial_1\p(\phi_\zer),\partial_2\p(\phi_\zer)$ span the same space,
there is a nonsingular $(d-1)\times (d-1)$ matrix $U$ such that
$$
\left[\partial_1\p(\phi_\zer)\ \partial_2\p(\phi_\zer)\ \cdots\ \partial_{d-1}\p(\phi_\zer)\right]
=\left[\partial_1\p(\th_\zer)\ \partial_2\p(\th_\zer)\ \cdots\ \partial_{d-1}\p(\th_\zer)\right]U
\eqn $$
Hence the $2(d-1)\times 2(d-1)$ matrix $\partial^2 F(\th_\zer,\phi_\zer)$ blocks
$$
\partial^2 F(\th_\zer,\phi_\zer)=\pmatrix{M_\al -v_\th c_\th M_\th
&v_\th v_\phi M_\al U\cr v_\th v_\phi U^t M_\al& U^tM_\al U-v_\phi c_\phi U^tM_\phi U \cr}
\eqn $$ 
Multiplying this on the right by $\pmatrix{\bbbone&0\cr 0&U^{-1}}$ 
and on the left by $\pmatrix{\bbbone&0\cr 0&{(U^t)}^{-1}}$, we see that this has nonzero determinant if and only if 
$$
\pmatrix{M_\al -v_\th c_\th M_\th
&v_\th v_\phi M_\al\cr v_\th v_\phi M_\al& M_\al -v_\phi c_\phi M_\phi \cr}
\eqn $$
has a nonzero determinant or equivalently, if and only if
$$\eqalign{
(M_\al -v_\th c_\th M_\th)x+v_\th v_\phi M_\al y&=0\cr
 v_\th v_\phi M_\al x+ (M_\al -v_\phi c_\phi M_\phi)y&=0\cr
}\eqn $$
has a unique solution. Solving for $y$ in the first equation, substituting
in the second gives
$$
v_\th v_\phi M_\al x- (M_\al -v_\phi c_\phi M_\phi)v_\th v_\phi M_\al^{-1}
(M_\al -v_\th c_\th M_\th)x
=0
\eqn $$
The matrix $M_\al$ is invertible because $S$ is strictly convex.
Multiplying through by $v_\th v_\phi M_\al^{-1}$ gives
$$
\left[\bbbone- (\bbbone -v_\phi c_\phi M_\al^{-1} M_\phi)
(\bbbone -v_\th c_\th M_\al^{-1} M_\th)\right]x
=0
\eqn $$
Thus $\partial^2 F(\th_\zer,\phi_\zer)$ has nonzero determinant if and only if
$$\eqalign{
&\left[\bbbone- (\bbbone -v_\phi c_\phi M_\al^{-1} M_\phi)
(\bbbone -v_\th c_\th M_\al^{-1} M_\th)\right]\cr
&\hskip 1in =v_\phi c_\phi M_\al^{-1} M_\phi
+v_\th c_\th M_\al^{-1} M_\th
-v_\th v_\phi   c_\th c_\phi M_\al^{-1} M_\th M_\al^{-1} M_\phi\cr
}\EQN\nonzd $$
is invertible. By strict convexity and \queq{\cthdef}, 
$\al_\zer \in \{ \th_\zer, a(\th_\zer)\}$ and
$\ph_\zer \in \{ \th_\zer, a(\th_\zer)\}$. 
If $e$ is symmetric, this implies that $M_\al=M_\th=M_\ph$ and
$c_\th=c_\ph=1$, so
the right hand side of \queq{\nonzd} is $(v_\ph+v_\th-v_\ph v_\th)\bbbone$
which is invertible since the sum of three signs can never vanish. 
If $e$ is asymmetric, we use \AFou\ 
to say that, possibly after a change of basis (which we can do by
multiplying on the right by a $V$ and on the left by its $V^{-1}$), 
each of the three matrices on the right hand side are of the
form $\pm\bbbone$ plus a matrix whose norm is at most $\sfrac{1}{5},
\sfrac{1}{5},(1+\sfrac{1}{5})^2-1$, respectively, for the three matrices.
As
$$
\sfrac{1}{5}+\sfrac{1}{5}+(1+\sfrac{1}{5})^2-1=\sfrac{4}{5}+\sfrac{1}{25}<1
\eqn $$
the right hand side is invertible.
\endproof

\Pro{\dummy}{There is a constant  $\ \Const \ $ such that for all
$d\ge 3$ and all $j_1,j_2,j_3<0$ 
$$\eqalign{
&{\rm Vol}\set{(\k_1,\k_2)\in\R  ^{2d}}{|e(\k_1)|\le M^{j_1},
|e(\k_2)|\le M^{j_2},|e(\pm\k_1\pm\k_2+\q)|\le M^{j_3}}\cr
&\hskip2.5in\le \Const  M^{j_1}M^{j_2}M^{j_3}
.\cr}\eqn $$
}
\Proof We may assume without loss of generality that $j_3=\max\{j_1,j_2,j_3\}$.
Otherwise make a change of variables with $\p_1= \pm\k_1\pm\k_2+\q,\ \p_2=\k_2$.
By compactness of a closed neighbourhood of $S$, it suffices 
to show that for any $\k^{(0)}_1,\ \k^{(0)}_2,\ \q^{(0)}$ obeying
$$
e(\k^{(0)}_1)=e(\k^{(0)}_2)=e(\pm\k^{(0)}_1\pm \k^{(0)}_2+\q^{(0)})=0
\eqn $$
there is a constant $\ \Const \ $ and there are neighbourhoods 
$V_1,\ V_2,\ U$ of $\k^{(0)}_1,\ \k^{(0)}_2,\ \q^{(0)}$ respectively, 
such that for all $\q\in U$ and all $j<0$
$$\eqalign{
&{\rm Vol}\set{(\k_1,\k_2)\in V_1\times V_2}
{|e(\k_1)|\le M^{j_1},
|e(\k_2)|\le M^{j_2},|e(\pm\k_1\pm\k_2+\q)|\le M^{j_3}}\cr
&\hskip2.5in\le \Const  M^{j_1}M^{j_2}M^{j_3}
}\eqn $$
Since $M^{j_k} < r_\zer$ for $k \in \{ 1,2,3\}$,  
we can switch to the $\rho,\th$ coordinates. 
In these coordinates, the neighbourhoods
$V_1,\ V_2$ can be replaced by some $X_1\times Y_1$ and $X_2\times Y_2$.
Define
$$
E(\q,\th,\phi)=e\big(\pm\p(0,\th)\pm \p(0,\phi)+\q\big)
\eqn $$
Since, for all $\q$ in a neighbourhood of $\q^{(0)}$ and all 
$(\th,\phi)\in S^{d-1}\times S^{d-1}$,
$$\eqalign{
|E\big(\q,\th,\phi)|
&=|e\big(\pm\p(0,\th)\pm \p(0,\phi)+\q\big)
-e\big(\pm\p(\rho_1,\th)\pm \p(\rho_2,\phi)+\q\big)\cr
&\hskip.5in +e\big(\pm\p(\rho_1,\th)\pm \p(\rho_2,\phi)+\q)\big)|\cr
&\le\Const  |\rho_1|+\Const  |\rho_2|+|e\big(\pm\p(\rho_1,\th)\pm \p(\rho_2,\phi)+\q\big)|
,}\eqn $$
we have, for all $\q$ in a neighbourhood of $\q^{(0)}$,
$$\eqalign{
&{\rm Vol}\set{(\k_1,\k_2)\in V_1\times V_2}
{|e(\k_1)|\le M^{j_1},|e(\k_2)|\le M^{j_2},|e(\k_1\pm\k_2+\q)|\le M^{j_3}}\cr
&\le\Const  {\rm Vol}\set{(\th,\phi,\rho_1,\rho_2)\in X_1\times X_2\times Y_1\times Y_2}
{|\rho_1|\le M^{j_1},\ |\rho_2|\le M^{j_2},\cr
&\hskip1.5in \ |e\big(\p(\rho_1,\th)\pm \p(\rho_2,\phi)+\q\big)|\le M^{j_3}}\cr
&\le\Const  {\rm Vol}\set{(\th,\phi,\rho_1,\rho_2)\in X_1\times X_2\times Y_1\times Y_2}
{|\rho_1|\le M^{j_1},\ |\rho_2|\le M^{j_2},\cr
&\hskip1.5in \ |E(\q,\th,\phi)|\le \Const  M^{j_3}}\cr
&\le\Const  M^{j_1+j_2}{\rm Vol}\set{(\th,\phi)\in X_1\times X_2}
{|E(\q,\th,\phi)|\le \Const  M^{j_3}}\cr
}\eqn $$
Hence it suffices to prove that, for all $\q$ in a neighbourhood of $\q^{(0)}$,
$$
{\rm Vol}\set{(\th,\phi)\in X_1\times X_2}
{|E(\q,\th,\phi)|\le \Const  M^{j}}
\le\Const  M^{j} 
\eqn $$
In the event that $(\th_\zer,\phi_\zer)$ is not a critical point of 
$E(\q^{(0)},\th,\phi)$, this is trivial.

 So from now on suppose that $(\th_\zer,\phi_\zer)$ is a critical point of 
$E(\q^{(0)},\th,\phi)$. By the previous Lemma $\partial^2E$ is invertible at
$(\th_\zer,\phi_\zer)$.  So, by the implicit function theorem, for each $\q$ in
a neighbourhood of $\q^{(0)}$
$$\eqalign{
\nabla E(\q,\th,\phi)=0
}\eqn $$
(here $\nabla$ means the derivative with respect to $\th$ and
$\ph$)
has a unique solution $\big(\th_\zer(\q),\phi_\zer(\q)\big)$ in a neighbourhood
of $(\th_\zer,\phi_\zer)$ and this solution depends on $\q$ in a $C^{k-1}$ way.
Consequently, $\partial^2 E$ is also invertible at 
$\big(\th_\zer(\q),\phi_\zer(\q)\big)$. Let $d_1$ and $d_2$ be the number of positive
and negative eigenvalues of $\partial^2E(\th_\zer,\phi_\zer)$ respectively. Then
$d_1\ge0,\ d_2\ge 0,\ d_1+d_2=2(d-1)\ge 4$ since $d \ge 3$.
By a translation followed by a 
linear change of variables we may replace $E(\q,\th,\phi)$ by 
$F(u,v)=E(\q,\th(u,v),\phi(u,v))$ with $u$ running over a neighbourhood of
$0\in\R  ^{d_1}$, $v$ running over a neighbourhood of
$0\in\R  ^{d_2}$ and
$$\eqalign{
\partial_{u_i}F(0,0)&=0\cr
\partial_{v_i}F(0,0)&=0\cr
\partial_{u_i}\partial_{u_j}F(0,0)&=2\de_{i,j}\cr
\partial_{v_i}\partial_{v_j}F(0,0)&=-2\de_{i,j}\cr
\partial_{u_i}\partial_{v_j}F(0,0)&=0\cr
}\eqn $$
Note that $F(0,0)$ need not be zero.

The easy case is that with one of $d_1,d_2$ zero. Suppose that $d_2=0$.
Then, as in Section A.2 we can make a ``polar coordinate'' like
change of variables $u=u(r,\al)$, with $r$ being the square root of $F-F_\zer$
and $\al$ being the usual polar angles. Then $f(r,\al)=F(u(r,\al))=F_\zer+r^2$
and we are led to bound
$$\eqalign{
\int d\al\int dr\ r^{d_1+d_2-1}\chi\big(|F_\zer+r^2|\le\ep\big)
}\eqn $$
If $F_\zer\ge -2\ep$, then $r\le\Const \sqrt{\ep}$ and, as $d_1+d_2-1\ge 1$, the integral
is bounded by $\Const \ep$. If $F_\zer<-2\ep$, then $r$ must obey
$r^2=|F_\zer|+\veps$ for some $\veps\in[-\ep,\ep]$. As
$$\eqalign{
r&=\pm\sqrt{|F_\zer|+\veps}\cr
&=\pm\sqrt{|F_\zer|}\sqrt{1+\veps/|F_\zer|}\cr
&=\pm\sqrt{|F_\zer|}+O\big(\veps/\sqrt{|F_\zer|}\big)\cr
}\eqn $$
$r$ runs over an interval of length $O\big(\ep/\sqrt{|F_\zer|}\big)\le\Const \sqrt{\ep}$ centred on $\sqrt{|F_\zer|}\ge
\Const \sqrt{\ep}$. Hence the integral is bounded by
$$
|F_\zer|^{(d_1+d_2-1)/2}\sfrac{\ep}{\sqrt{|F_\zer|}}
\eqn $$
which, in turn is bounded by $\Const \ep$, uniformly in $F_\zer$ again
because $d_1+d_2-1\ge 1$.

The hard case is that with both $d_1$ and $d_2$ nonzero. Then we first
go to polar coordinates in $u$ and $v$ separately.
$$\eqalign{
r&=\sqrt{\sum  _{i=1}^{d_1} u_i^2}\cr
\al&=d_1-1{\rm\ angles\ for\ }u\cr 
\rho&=\sqrt{\sum  _{i=1}^{d_2} v_i^2}\cr
\be&=d_2-1{\rm\ angles\ for\ }v\cr 
}\eqn $$ 
The measure
$$
d^{d-1}\th\, d^{d-1}\phi =\Const  d^{d_1}u\, d^{d_2}v
=J(\al,\be)r^{d_1-1}\rho^{d_2-1}dr\,d\rho
\eqn $$
with $J(\al,\be)$ bounded above.
For each fixed $\al,\ \be$ the function
 $f(r,\rho)=F\big(u(r,\al),v(\rho,\be)\big)$ is $C^k$ in $(r,\rho)$ in a
neigbourhood of $(0,0)\in[0,\infty)^2$. Furthermore the first and
second derivatives of $f$ at the origin are
$$\eqalign{
f_r(0,0)&=0\cr
f_\rho(0,0)&=0\cr
f_{rr}(0,0)&=2\cr
f_{\rho \rho}(0,0)&=-2\cr
f_{r\rho}(0,0)&=0\cr
}\eqn $$
Hence there is, for $k\ge 2$,
a $C^1$ change of variables $x=x(r,\rho),\ y=y(r,\rho)$ such that
$$
f\big(r(x,y),\rho(x,y)\big)=F(0,0)+x^2-y^2
\eqn $$
and
$$
\frac{\partial(r,\rho)}{\partial(x,y)}(0,0)=\pmatrix{1&0\cr0&1\cr}
\eqn $$
The latter condition ensures that
$$\eqalign{
|r|&\le a_1|x|+b_1|y|\cr
|\rho|&\le b_2|x|+a_2|y|\cr
}\eqn $$
By choosing the neighbourhood sufficiently small, the constants $a_1,a_2$
and $b_1,b_2$ can be made arbitrarily close to one  and zero respectively.
Hence $r^{d_1-1}\rho^{d_2-1}$ is bounded by a finite sum of terms of the form
$|x|^{e_1}|y|^{e_2}$ with $e_1,e_2\ge 0,\ e_1+e_2=d_1+d_2-2=2d-4$ and
we are led to bound
$$\eqalign{
\int_{-1}^1 dx\,|x|^{e_1} \int_{-1}^1 dy\,|y|^{e_2}\ 
\chi\left(|F_\zer+x^2-y^2|\le\ep\right)
}\eqn $$
with $e_1,e_2\ge 0,\ e_1+e_2=d_1+d_2-2=2d-4$.

We may assume, without loss of generality that $F_\zer\ge 0$. Otherwise
just exchange the roles of $x$ and $y$. In the domain where $F_\zer+x^2\le 2\ep$
both $|x|$ and $|y|$ are bounded above by $O(\sqrt{\ep})$. This domain
domain contributes at most $O(\ep)$ to the total integral. In the domain
where $F_\zer+x^2\ge 2\ep$, $y$ must obey $y^2=F_\zer+x^2+\veps$ for some
$\veps\in[-\ep,\ep]$. So
$$\eqalign{
y&=\pm\sqrt{F_\zer+x^2+\veps}\cr
&=\pm\sqrt{F_\zer+x^2}\left(1+\sfrac{\veps}{F_\zer+x^2}\right)^{1/2}\cr
&=\pm\sqrt{F_\zer+x^2}\left(1+O\left(\sfrac{\ep}{F_\zer+x^2}\right)\right)\cr
&=\pm\sqrt{F_\zer+x^2}+O\left(\sfrac{\ep}{\sqrt{F_\zer+x^2}}\right)\cr
}\eqn $$
That is, $y$ runs over at most two intervals of length at most
$O\left(\sfrac{\ep}{\sqrt{F_\zer+x^2}}\right)\le O(\sqrt{\ep})$ whose centers have 
modulus $\sqrt{F_\zer+x^2}\ge\sqrt{2\ep}$. The contribution of this domain is
at most
$$\eqalign{
\Const \int dx\ |x|^{e_1}(\sqrt{F_\zer+x^2})^{e_2}\sfrac{\ep}{\sqrt{F_\zer+x^2}}
}\eqn $$
Since $e_1+e_2\ge 2$, this is bounded by $O(\ep)$,
uniformly in $F_\zer$. 
\endproof

\noindent
{\bf Proof of Lemma \jomei:} Let $\ka > 0$. We first note that $\cP_\ka$
can be defined as in \queq{\cPkadef}, and that Lemmas \CriP\ and \noCP\ 
carry over to $d\ge 3$ with trivial changes. The analogue of Lemma \qCP\
also holds because the matrix $\del^2 E$ is nonsingular by Lemma \nondthr,
and this implies by the implicit function theorem the existence and 
$C^1$ properties of the solutions stated in $(i)$. That there
are four solutions follows as in Lemma \findCP. All four solutions
given in the table after \queq{\therego} are possible because 
\AFiv\ has been relaxed.
Since the function is defined on the compact set $\cP_\ka \times
\left( [-\ka,\ka]\times S^{d-1}\right)^2$, it is uniformly $C^1$, 
so there is a global Lipschitz constant $L$ such that \queq{\Lipco}
holds. To prove $(ii)$, we may assume that $\q \in \cP_\ka$. 
Let $K_\two = L+1$ and $(\th,\ph)$ not obey \queq{\NEAR}.
Then for all $\abs{\q'-\q} \le \veps_\thr$, 
$$\eqalign{
\abs{\th-\th^b_{cr}(\q')} &\ge 
\abs{\abs{\th-\th^b_{cr}(\q)} -\abs{\th^b_{cr}(\q')-\th^b_{cr}(\q)}}\cr
& \ge K_\two \veps_\thr - L \abs{\q'-\q} \ge \veps_\thr
\cr}\eqn $$
and similarly $\abs{\ph-\ph^b_{cr}(\q')} \ge \veps_\thr$ (we suppress
$\rh_\th$ and $\rh_\ph$ in the notation since everything is uniform in 
these variables). Thus, by Taylor expansion, 
$$\eqalign{
\left(\matrix{\nabla_\th E\cr \nabla_\ph E}\right) 
&(\q',\th,\ph)  =
\del^2 E (\q',\th^b_{cr},\ph^b_{cr}) \; 
\left(\matrix{\th-\th^b_{cr}\cr \ph-\ph^b_{cr}}\right)\cr
&+\int\limits_0^1 dt \; 
\left(\del^2 E \left(\q',\th^b_{cr}+t(\th-\th^b_{cr}),
\ph^b_{cr}+t(\ph-\ph^b_{cr})\right) -
\del^2 E (\q',\th^b_{cr},\ph^b_{cr})\right)
\; 
\left(\matrix{\th-\th^b_{cr}\cr \ph-\ph^b_{cr}}\right)
\cr}\EQN\TAYLHUND $$
where $\th^b_{cr}=\th^b_{cr}(\q')$ and $\ph^b_{cr}=\ph^b_{cr}(\q')$.
Since the eigenvalues of $\del^2 E$ are nonzero and the parameters
run over a compact region, there is $K_\one >0$ such that
$$
\abs{\del^2 E(\q',\th^b_{cr},\ph^b_{cr}) \; v} \ge \frac{2}{K_\one} \abs{v}
\eqn $$
for all $v \in \R^{2(d-1)}$. By continuity, the operator norm
of the matrix in the remainder term in \queq{\TAYLHUND}
can be made smaller than $\sfrac{1}{K_\one}$. 
This implies that if \queq{\NEAR} does not hold,
\queq{\FAR} must hold. \endproof

\appendix{C}{One--loop volume bounds for asymmetric Fermi surfaces}

\noindent
In this appendix, we prove Lemma \bubvol. We mention again 
that Lemma \bubvol\ has two important consequences -- 
it implies differentiability of the particle--particle 
\WL\ contributions to the counterterm and the self--energy 
(defined in \queq{\WLvar} and \queq{\WLval}, and drawn in 
Figure $\figufou$ $(b)$)
in the asymmetric case (Theorem \goodWL), and it also implies that 
the particle--particle bubble 
$$
B(k,Q,q) = \lim\limits_{I \to -\infty}
\sum\limits_{I \le j,h<0} \; B_{jh}(k,Q,q)
\eqn $$
where
$$
B_{jh}(k,Q,q) =
\int\limits_{\R\times \cB} d^{d+1} p \;
C_j(p) \;\hat v(p-k)\; C_h(-p+Q) \;\hat v(q+p-Q)
\eqn $$
is a bounded function of the three momenta $(k,Q,q)$ if 
$e$ is asymmetric. If $e$ is symmetric, $B$ diverges 
for $\abs{Q} \to 0$ like 
$\log |Q|$ for all $k,q$; this leads to the Cooper instability
if the interaction is attractive.
Thus Theorem \goodWL\ and Lemma 2.41 of \FST\ imply that
no ladder subdiagrams can generate any factorials in the values
of individual graphs in the perturbation expansion, 
if $e$ is asymmetric and if \AFoup\ holds.  

We first do some easy reduction steps. 
By changing variables from $\p$ to $\rh,\th$, and bounding
$$\eqalign{
\int d^d\p \; \True{\abs{e(-\p+\q)} \le \veps_\two } \; 
\True{\abs{e(\p)}\le \veps_\one} & \le
\int\limits_{-\veps_\one}^{\veps_\one} d\rh
\int d\th \; J(\rh,\th) \; 
\True{\abs{e(-\p(\rh,\th)+\q)} \le \veps_\two} \cr
& \le 2 \veps_\one \abs{J}_\zer 
\sup\limits_{\abs{\rh} \le \veps_\one} 
\int d\th \True{\abs{e(-\p(\rh,\th)+\q)}\le \veps_\two}
\cr}\eqn $$
we see that \queq{\bubimptwo} follows from \queq{\bubimp},
with the constant stated in Lemma \bubvol. 

Since $\abs{\rh} \le \veps_\one\le\veps_\two$, we can do the
usual Taylor expansion 
$$\eqalign{
\abs{e(-\p(\rh,\th)+\q) -e(-\p(0,\th)+\q)} & =
\abs{\rh \int\limits_0^1 dt\; \nabla e (-\p(t\rh,\th)+\q)
\cdot \del_\rh\p(t\rh,\th)} \cr
& \le \veps_\one \abs{e}_\one \abs{\del_\rh \p}_\zer
\le \veps_\one \;\frac{\abs{e}_\one}{u_\zer}
.\cr}\eqn $$
Here we used that, since $\rh=e(\p(\rh,\th))$ and 
$\del_\rh\p(\rh,\th) = \abs{\del_\rh\p(\rh,\th)} u(\p(\rh,\th))$,
$$
1 = \nabla e(\p(\rh,\th))\cdot \del_\rh\p(\rh,\th) = 
\nabla e(\p(\rh,\th)) \cdot u(\p(\rh,\th)) 
\abs{\del_\rh \p(\rh,\th)}\ge u_\zer \abs{\del_\rh\p(\rh,\th)}
,\eqn $$
so $\abs{\del_\rh\p}_\zer \le \sfrac{1}{u_\zer}$. Therefore, 
for all $\abs{\rh} \le \veps_\one$, $\abs{e(-\p(\rh,\th)+\q)}
\le \veps_\two$ implies 
$$
\abs{e(-\p(0,\th)+\q)} \le \veps_\two + \sfrac{\abs{e}_\one}{u_\zer}
\veps_\one \le \veps_\two \left(1+ \sfrac{\abs{e}_\one}{u_\zer}\right)
.\eqn $$
Thus, if we show that there are $\al\ge \sfrac{1}{3}$ and 
 $\tilde Q_B > 0$ such that
$$
\int d\th \; \True{\abs{e(-\p(0,\th)+\q} \le \veps_\two } \le
\tilde Q_B \; {\veps_\two}^\al
\EQN\TOSHOW $$
then \queq{\bubimp} follows with $Q_B = (1+\sfrac{\abs{e}_\one}{u_\zer})
^\al \tilde Q_B$. 

\sect{Geometry of the Problem}
\pni
We now describe the geometrical picture behind the bounds. The
integral in \queq{\bubimptwo} is the $d$--dimensional volume
of the set $R_{\veps_\one} \cap (-R_{\veps_\two} + \q) 
\subset \cB$, where $R_\veps = \set{\p}{\abs{e(\p)} \le \veps}$
is a neighbourhood of the Fermi surface $S$ having thickness of order 
$\veps> 0$. 
From now on, we drop the subscript $2$ and denote $\veps_\two=\veps$.
Since \queq{\TOSHOW} holds trivially for all $\veps > \veps_\zer$,
if $\veps_\zer > 0 $ is fixed and  $\tilde Q_B$ is chosen 
dependent on $\veps_\zer$, 
we may assume that $\veps $ is small, and therefore we first
draw a picture of the intersection of $S$ with its translates,
in Figure \nextfig{\figCone} (one of them is drawn dotted to
make the distinction easier).
This corresponds to the reduction of \queq{\bubimptwo}
to \queq{\bubimp}.  The intersection may be transversal, as in $(a)$,
or tangential, as in $(b)$, $(c)$, and $(d)$. The translating
momentum $\q$ is $\q=2\p$ in case $(b)$, and $\q=\p+\ve{a}(\p)$
in case $(c)$ and $(d)$, where $\p$ is the point at which the tangential
intersection takes place. 

\herefig{Cfig1}

In Figure \nextfig{\figCtwo}, we redraw a neighbourhood of the intersection
point in coordinates where one of the surfaces appears as a straight
line. In the Figure, $\q$ is chosen such that exact tangency
happens, this is, of course, not the case in general.
In fact, the coordinates used in Figure $\figCtwo$ 
are our standard coordinates $\rh$ and $\th$, and 
Figure $\figCtwo$ simply contains the graph of the function
$$
g(\th) = e(\q-\p(0,\th))
\eqn $$
for four values of $\q$ (we suppress the dependence of $g$ on $\q$ in
the notation because $\q$ is fixed). The shaded region in 
Figure $\figCtwo$ is the region where $|g|\le\veps$, 
in other words, it is the support of $\True{|g(\th)|\le \veps}$.
It is obvious from the Figure that this support condition poses
a restriction on $\th$. We now discuss briefly why this figure
really captures the essential behaviour, and then turn to the
details of the proof.

\herefig{Cfig2}

We have 
$$\eqalign{
g'(\th) &= - \nabla e (\q-\p(0,\th)) \cdot \del_\th \p(0,\th)
\cr
g''(\th) &= 
\Big(\del_\th \p (0,\th), e''(\q-\p (0,\th)) \del_\th \p (0,\th)\Big)
- \nabla e (\q-\p(0,\th)) \cdot \del_\th^2 \p(0,\th)
.\cr}\EQN\Cnine $$
Using $\th^*$ to denote the value of $\th$ where $g(\th)=0$, the 
cases drawn in the Figures are
\leftit{$(a)$} $\q-\p(0,\th^*) \not\in \{ \p(0,\th^*),\p(0,a(\th^*))\}$, so
$g'(\th^*) \ne 0$. It is obvious that the condition
$\abs{g(\th)} \le \veps$, which is the support condition of
the integrand (indicated as the shaded region in Figure $\figCtwo$) 
restricts $\th$ to
an interval of length $\Const \veps $ around $\th^*$, 
so \queq{\TOSHOW} holds, with $\al =1$.

\leftit{$(b)$} $\q=2\p(0,\th^*)$. Then $\q-\p(0,\th^*)=\p(0,\th^*)$,
$g'(\th^*)=0$, and
$$
g''(\th^*) = 2 w(\p(0,\th^*))
\EQN\stern $$
by \queq{\wpdef} and \queq{\Trick}, so $\abs{g''(\th^*)} \ge 2
w_\zer>0$ by \queq{\wz}. Thus the function
is essentially quadratic and $\abs{g(\th)} \le \veps$ restricts
$\th$ to $\abs{\th-\th^*} \le \Const \sqrt{\veps}$, 
so \queq{\TOSHOW} holds with $\al=\sfrac{1}{2}$.
The factor $2$ in \queq{\stern} is intuitively clear from 
Figure $\figCone$ $(b)$: the curvatures of the two intersecting 
sets have the same magnitudes, but opposite signs. 
\ssni
The other case where $g'$ vanishes is 
$\q=\p(0,\th^*)+\p(0,a(\th^*))$. Then  $g'(\th^*)=0$, and
by \queq{\antidef}, \queq{\deltwop}, \queq{\wpdef}, and \queq{\Trick}, 
$$
g''(\th^*) = w(\p(0,a(\th^*))) \left(1-\frac{\del a}{\del \th}(\th^*)\right)
.\eqn $$
By \queq{\curvratio}, $\sfrac{\del a}{\del \th}$ is the 
curvature ratio at $\th^*$ and $a(\th^*)$. This is the point
where \AFoup\ comes in -- this ratio can equal one, and hence
$g''(\th)$ can vanish, only if $\th^*$ is one of 
$\th^{(1)}, \ldots, \th^{(N)}$.
So there are the cases
\ssni
\leftit{$(c)$} $\th^* \not\in U_\de(\th^{(1)}) \cup \ldots \cup 
U_\de(\th^{(N)})$ for some $\de >0$. 
Then $\abs{\sfrac{\del a}{\del \th}(\th^*)-1} \ge \ga(\de) >0$, 
and the bound
is as in case $(b)$, only with a $\de$--dependent constant $1/\ga(\de)$
that grows as $\de \to 0$.

\leftit{$(d)$} $\th^* \in U_\de(\th^{(1)}) \cup \ldots \cup 
U_\de(\th^{(N)})$. If $\th^*=\th^{(n)}$ for some $n \in \nat{N}$,
$g''(\th^*)=0$ because there is a 
tangential intersection at points with the same curvature. 
In a neighbourhood of these points,
$\abs{g''(\th)}$ gets arbitrarily small. This is the hard case of the proof. 
Here we shall use the growth condition in \AFoup\ to show that 
$g''$ grows at least linearly when one moves away from 
$\th^{(n)}$. This will imply \queq{\TOSHOW} with $\al=\sfrac{1}{3}$.  

\sect{Basic properties of the critical points}
\pni
We first collect some information about those $\th^*$ where
$g'(\th^*)=0$ and bound the contributions to the LHS of 
\queq{\TOSHOW} where $\th$ is away from these $\th^*$.
By Lemma \CriP, if $\ka > 0$ is small enough, the equation
$$
\nabla e(\p(r,\vth)) \cdot \del_\th\p(0,\th) = 0
\EQN\Cthtn $$
has, for fixed $r$ with $\abs{r} < \ka $ and fixed $\th \in \R/2\pi\Z$, 
exactly two solutions for $\vth$, given by $\vth=\vth^{(i)}(0,\th,r)$,
with 
$$\eqalign{
\vth^{(1)}(0,\th,0) & = \th \cr
\vth^{(2)}(0,\th,0) & = a(\th)
,\cr}\EQN\Cfotn $$
and both are $C^1$ functions of $(\th,r)$. Let 
$$
\cQ_\ka = \bigcup\limits_{i=1}^2 
\set{\p(r,\vth^{(i)}(0,\th,r)) + \p(0,\th) }{\abs{r} < \ka,
\th \in \R/2\pi\Z}
.\eqn $$

\Rem{\dummy} There is $\ka > 0$ and $m_\zer = m_\zer(\ka )>0$
such that for all $\q \not\in \cQ_\ka$ and all $\th \in \R/2\pi\Z$
satisfying $\abs{e(\q-\p(0,\th))} \le \ka/2$, 
$$
\abs{\frac{\del}{\del\th}
e(\q-\p(0,\th))} \ge m_\zer(\ka)
.\eqn $$
\Proof The function $F: (\cB \setminus \cQ_\ka )\times \R/2\pi\Z
\to \R$, $(\q,\th) \mapsto  \abs{\sfrac{\del}{\del\th}
e(\q-\p(0,\th))}$, is continuous. The set 
$$
X_\ka = \set{(\q,\th)\in (\cB \setminus \cQ_\ka )\times \R/2\pi\Z}{
\abs{e(\q-\p(0,\th))} \le \ka/2}
\eqn $$
is compact. By construction of $X_\ka$, $F$ has no zeros on this set, so
$m_\zer(\ka)> 0$ exists.
\endproof
\pni
If $\q \not \in \cQ_\ka$, then for all $\veps \le \sfrac{\ka}{2}$,
by a change of variables
$$
\int d\th \True{\abs{e(\q-\p(0,\th))} \le \veps} 
\le \frac{1}{m_\zer(\ka)} \int de\; \True{\abs{e}\le \veps}
= \frac{2}{m_\zer(\ka)}\veps
\EQN\veasy $$
Thus we may assume that $\q \in \cQ_\ka$, for some $\ka > 0$,
which will be fixed, independent of $\veps$, in the following. 

\sni
Let $b>0$, and split 
$$
\R/2\pi\Z = \bigcup\limits_{n=1}^N U_b(\th^{(n)}) \; 
\dotcup \cR_b
\EQN\domsplit $$
(by \AFoup, $b>0$ can be chosen such that $U_b(\th^{(n)})\cap
U_b(\th^{(n')})= \emptyset$ if $n \ne n'$). We fix $b>0$; then
\AFoup\ implies that there is $z_\zer>0$ such that for all $\th\in
\cR_b$,
$$
\abs{w(\p(0,a(\th))) \left(1-\frac{\del a}{\del \th}(\th)\right)} \ge
z_\zer
\EQN\zzeri $$

\Lem{\loci}{\Lesty Define the maps $F_\one$ and $F_\two$ by 
$$\eqalign{
F_i : \quad (-\ka,\ka) \times \R/2\pi\Z & \to \cB \cr
(r,\th) & \mapsto \p(0,\th) + \p\Big(r,\vth^{(i)}(0,\th,r)\Big)
.\cr}\eqn $$
There is $\ka_\zer=\ka_\zer(b) >0$ such that for all $0 < \ka
< \ka_\zer$, 
$$\eqalign{
\det F'_\one(r,\th) &\ne 0 \qquad \forall (r,\th) \in 
(-\ka,\ka) \times \R/2\pi\Z \cr
\det F'_\two(r,\th) &\ne 0 \qquad \forall (r,\th) \in 
(-\ka,\ka) \times \cR_b
\cr}\EQN\Csith $$
and such that for all $\th_\one \in  \R/2\pi\Z $,
$F_\one$ is invertible on $(-\ka,\ka) \times U_\ka (\th_\one)$,
and for all $\th_\one \in \cR_b$, $F_\two$ 
is invertible on $(-\ka,\ka) \times (U_\ka (\th_\one)\cap\cR_b)$.
}

\Proof $F_\one$ and $F_\two$ are $C^1$ functions. Their derivatives,
evaluated at $r=0$, are
$$\eqalign{
\sfrac{\del F_i}{\del r} &= \del_\rh\p(0,\vth^{(i)} (0,\th,0))
+
\del_\th\p(0, \vth^{(i)} (0,\th,0)) \del_\thr \vth^{(i)} (0,\th,0)
\cr
\sfrac{\del F_i}{\del \th} &= \del_\th\p(0,\th) + 
\del_\th\p(0,\vth^{(i)} (0,\th,0)) \del_\two \vth^{(i)} (0,\th,0)
.\cr}\eqn $$
By \queq{\atrho} and \queq{\antidef}, 
$$\eqalign{
\frac{\del F_\one}{\del \th}
 & = \del_\th\p(0,\th) \left(1+\del_\two\vth^{(1)} (0,\th,0)\right)
\cr
\frac{\del F_\two}{\del \th}
 & = \del_\th\p(0,\th) + \del_\th\p(0,a(\th)) \del_\two\vth^{(2)} (0,\th,0)
\cr
& = \del_\th\p(0,\th)\left( 1 - \del_\two\vth^{(2)} (0,\th,0)\right)
\cr}\eqn $$
so by \queq{\Btwon} and \queq{\Btwth},
$$\eqalign{
\frac{\del F_\one}{\del \th}
 & = 2 \del_\th\p(0,\th) 
\cr
\frac{\del F_\two}{\del \th}
 & = \del_\th\p(0,\th) \left( 1 - \sfrac{\del a}{\del \th}(\th)\right)
.\cr}\eqn $$
Since $\del_\rh\p$ and $\del_\th\p$ are linearly independent,
$\det F'_i (0,\th) \ne 0$ if $\sfrac{\del F_i}{\del \th} \ne
0$. Since $\abs{\del_\th\p(0,\th)}=1$ for all $\th \in \R/2\pi\Z$,
$\sfrac{\del F_1}{\del \th} \ne 0$ holds. Since $\cR_b$ is compact,
$\sfrac{\del a}{\del \th}$ is continuous on $\cR_b$, and 
$\sfrac{\del a}{\del \th}(\th) \ne 1$ for all $\th \in \cR_b$,
there is $\ga > 0$ such that
$$
\abs{\frac{\del a}{\del \th} (\th) -1} \ge \ga(b)
\qquad \forall \th \in \cR_b
.\eqn $$
By continuity of $F_i'$ in $r$, there is $\tilde \ka_\zer(b)>0$
such that $\abs{\det F_i' (r,\th)} \ne 0$ for all $r < \tilde
\ka_\zer(b)$. By the implicit function theorem, and by compactness
of $\R/2\pi\Z$ and $\cR_b$, there is $0 < \ka_\zer(b) \le 
\tilde \ka_\zer(b)$ such that the stated local invertibility holds for all
$\abs{r} < \ka_\zer(b)$. \endproof

\Rem{\dummy} In the corresponding two--loop statement, Lemma
\qCP, there are three summands, all close to $\pm 1$, which can
therefore never add up to zero. This is the reason why there, no restriction
to a set similar to $\cR_b$ was required, and why Lemma \qCP\
also holds in the symmetric case. Note that if \SYmm\ holds,
then $\cR_b = \emptyset$, and $\sfrac{\del F_\two}{\del \th}
\equiv 0$ (this corresponds to $\q=0$).

\Rem{\dummy} The local injectivity stated in Lemma \loci\ does not rule out
the existence of $(r',\th')$ different from $\th^*$, $r^*$,
satisfying
$$\eqalign{
\q&=\p(0,\th^*)+\p(r^*,\vth^{(k)} (0,\th^*,r^*))\cr
& =\p(0,\th')+\p(r',\vth^{(k)} (0,\th',r'))
\cr}\eqn $$
as long as $\th'$ and $\th^*$ satisfy $\abs{\th^*-\th'} > \ka$,
i.e., they are far away from each other.
This can happen because $\cB$ is a torus.
However, for every $(r^*,\th^*)$ there are at most $2\pi/\ka$ such 
$(r',\th')$'s. (Geometrically, it is obvious that there can only be a few of
them.) This complication is easily dealt with by covering the
region $\cR_b$ by sets with diameter at most $2\ga$, 
where $2\ga<\ka$:
$$
\cR_b \subset \bigcup_{k=1}^{\overline{k}} U_\ga (t_k)
.\EQN\intvls $$

\Cor{\uniquecp} Let $\ka < \ka_\zer (b) $ and fix $\q\in \cQ_\ka$.
Then the solutions of 
$$
g'(\th) = 0, \qquad \th \in \cR_b
\eqn $$
are isolated, and given by $C^1$ functions of $\q$. There is at 
most one solution in each $U_\ga (t_k)$.

\Proof Let $\ka < \ka_\zer (b) $ and $\q\in \cQ_\ka$, then 
$$
\q=\p(0,\th_\zer)+\p(r_\zer,\vth^{(k_\zer)} (0,\th,r_\zer))
.\eqn $$
Assume that 
$$
g'(\th^*) = -\nabla e (\q-\p(0,\th^*)) \cdot \del_\th \p(0,\th^*)
=0
.\eqn $$
Then by \queq{\Cthtn} and \queq{\Cfotn}, there are $r^*$, $|r^*|<\ka$, and 
$k \in \{1,2\}$ with 
$$
\q=\p(0,\th^*)+\p(r^*,\vth^{(k)} (0,\th^*,r^*))
.\eqn $$
We cannot have
$$
\q=\p(0,\th^*)+\p(r^*,\vth^{(1)} (0,\th^*,r^*))
=\p(0,\th^{**})+\p(r^{**},\vth^{(2)} (0,\th^{**},r^{**}))
$$
with $\th^*$ and $\th^{**}$ both in the same $U_\ga (t_k)$, since this
would force $\vth^{(1)} (0,\th^*,0)=\th^*$ and 
$\vth^{(2)} (0,\th^{**},0)=a(\th^{**})$ to be too close together.
By injectivity, if $k=k_\zer$ and $\abs{\th^*-\th_\zer} < \ka$ then 
$\th^*=\th_\zer$ and $r^*=r_\zer$ must hold. 
So the solutions in $\th$ are isolated, and,
by \queq{\Csith} and the inverse function theorem, $C^1$ 
in $\q$. \endproof

\sect{The easy cases}
\pni
We have already dealt with the case $\q\not\in \cQ_\ka$. By compactness,
it thus suffices to consider $\q's$ in a small ball $\cB_\ka\subset\cQ_\ka$.
By Corollary \uniquecp, we may choose the $t_k$'s of \queq{\intvls} such
that if $g'(\th)$ fails to have a zero in $U_\ga (t_k)$, then $g'(\th)$
is bounded away from zero, uniformly for $\q\in\cB_\ka$. The contribution
to $\int d\th \True{\abs{e(\q-\p(0,\th))} \le \veps}$ from these $U_\ga (t_k)$'s
is bounded as in \queq{\veasy}. This leaves $\q\in\cB_\ka$, $\th$ running
over $U_\ga (t_k)$ with $k$ such that $g'(\th)$ vanishes at precisely one
point of $U_\ga (t_k)$ and $\q\in\cB_\ka$, $\th$ running
over $U_\ka (\th^{(n)}),\ 1\le n\le N$. We now deal with the former case.

Let $\et>0$. Since $g$ is continuous, $g^{-1}(]-2\et, 2\et[)$ is open,
so
$$
g^{-1} (]-2\et,2\et[) = 
\bigcup\limits_{k \in \N} \big( a_k(\et),b_k(\et)\big)
.\EQN\Ntet $$
The compact set $g^{-1} ([-\et,\et])$ is also contained in this
union of open intervals, hence also in a finite subcovering.
Thus, choosing such a finite subcovering,  
relabelling (if necessary) the $k$'s, and defining 
$$
N_\et = \bigcup\limits_{k=1}^{k_{max}} \big( a_k(\et),b_k(\et)\big)
,\eqn $$
we have
$$
g^{-1}(]-\et,\et[) \subset g^{-1} ([-\et,\et]) \subset
N_\et \subset g^{-1}(]-2\et, 2\et[)
.\eqn $$
That is, the set of $\th$ where $|g(\th)| < \et$ is covered 
by finitely many open intervals with $|g(\th)| < 2\et$ on each
interval (this is also obvious geometrically since $g^{-1}(]-\et,\et[)$ 
is the intersection of
a translate of $-S$ with a shell around $S$). 

Let $\et < r_\zer/2$, then we can define the map 
$\quad \tilde{ }\; : N_\et \to (-2\et,2\et) \times \R/2\pi\Z$, 
$\th \mapsto (\tilde \rh,\tilde \th)$ by 
$$
\q-\p(0,\th) = \p\big(\tilde \rh(\th), \tilde \th(\th)\big)
,\EQN\Useful $$
and $\tilde \rh,\tilde\th$ are $C^{2,\hxp}$ in $\th$. 
Fix $\et$, let $\ka < \et$, and let $\veps < \ka$. 
By definition of the coordinates $\rh$ and $\th$, 
$$
\tilde\rh(\th) = e( \p(\tilde \rh, \tilde \th)) = e(\q-\p(0,\th)) = g(\th)
\EQN\geqrhtil $$
so $\tilde\rh = g \big\vert_{N_\et}$, and thus
$$
\int \limits_{\R/2\pi\Z}d\th \True{\abs{g(\th)} \le \veps} 
= 
\int\limits_{N_\et} d\th \True{\abs{\tilde \rh (\th)} \le \veps}
.\EQN\Ztet $$
Let $\th^*$ obey $\sfrac{\del \tilde \rh}{\del \th}(\th^*)=
g'(\th^*) =0$. Then, by Lemma \CriP, 
$$
\tilde \th(\th^*) = \vth^{(k)} (0,\th^*,\tilde\rh(\th^*))
\EQN\wichtli $$
Since $\vth^{(k)}$ is $C^1$, we may Taylor expand in $\tilde\rh(\th^*)$
and get
$$
\tilde \th(\th^*) - \vth^{(k)} (0,\th^*,0) =
\tilde \rh(\th^*) 
\int\limits_0^1 ds \; 
\del_\thr \vth^{(k)} (0,\th^*,s\tilde\rh(\th^*))
,\eqn $$
so that 
$$
\abs{\tilde \th(\th^*) - \vth^{(k)} (0,\th^*,0)} \le \Const 
\abs{\tilde\rh(\th^*)}
\EQN\tildiff $$
holds for $k\in \{1,2\}$. 

We have already split the integral into contributions from 
$U_\ga(t_l)$. We now only have to show that if $\th^*\in U_\ga(t_l)$,
then for all $\th \in U_\ga(t_l)$, $g''(\th) \ne 0$, 
uniformly in $\q$. 
Thus we may assume that $|\th-\th^*| \le \ka$.

Let $k=1$ in \queq{\wichtli}. Then $\vth^{(1)}(0,\th^*,0)=\th^*$
and 
$$\eqalign{
\abs{\tilde\th(\th)-\th} & \le \abs{\tilde\th(\th)-\tilde\th(\th^*)}
+ \abs{\tilde\th(\th^*)-\th^*} + \abs{\th^*-\th} \cr
& \le \Const \left(\abs{\tilde\rh(\th^*)} + \abs{\th-\th^*}\right)
.\cr}\EQN\aeuae $$
By \queq{\wpdef} and \queq{\Trick}, 
$$
g''(\th) = 2 w(\p(0,\th)) + \Ph (\th)
\eqn $$
with
$$\eqalign{
\Ph(\th) &= \Big(\del_\th \p (0,\th), \lbrack
e''(\p(\tilde \rh, \tilde \th)) - e''(\p(0,\th)\rbrack \del_\th \p (0,\th)\Big)
\cr
& - \lbrack\nabla e (\p(\tilde \rh, \tilde \th))-\nabla e(\p(0,\th))\rbrack
\cdot \del_\th^2 \p(0,\th)
\cr}\eqn $$
Since $e \in C^{2,\hxp}$, and by \queq{\tildiff},
$$
\abs{\Ph(\th)} \le \Const \left(
\abs{\tilde\rh(\th^*)} + \abs{\tilde\rh(\th)}
+ \abs{\th-\th^*} \right)^\hxp
\eqn $$
Thus, by \queq{\wz}, if $\ka > 0$ is small enough, 
$g''(\th) > 0$ holds for all $\th \in U_\ka(\th^*)$ with
$|g(\th)| = |\tilde\rh(\th)| \le \ka$, and
$|g(\th^*)| = |\tilde\rh(\th^*)| \le  2 \ka$. 
These conditions on $\tilde \rh$ are fulfilled for all 
$\veps< \ka$ because in the support
of the integrand, $|\tilde\rh(\th)| \le \veps$ and because if
$|\tilde\rh(\th^*)| > 2\ka $, $g'$ would fail to vanish on $U_\ga(t_l)$. 
Note that these conditions also hold in $U_b(\th^{(n)})$ because
$k=1$. 
 
Let $k=2$ in \queq{\wichtli}. Then $\vth^{(2)}(0,\th^*,0)=a(\th^*)$,
and therefore, as in \queq{\aeuae},
$$
\abs{\tilde\th(\th)-a(\th)} \le \Const 
\left(\abs{\tilde\rh(\th^*)} 
+\abs{\th-\th^*}+\abs{a(\th)-a(\th^*)}\right)
.\eqn $$
Moreover, by \queq{\Cnine}, \queq{\antidef}, and \queq{\deltwop}
$$
g''(\th) = w(\p(0,a(\th))) \left(1-\frac{\del a}{\del \th}(\th)\right)
\quad + \; \tilde\Ph(\th)
\eqn $$
with
$$\eqalign{
\tilde\Ph(\th) &= \Big(\del_\th \p (0,\th), \lbrack
e''(\p(\tilde \rh, \tilde \th)) - 
e''(\p(0,a(\th))\rbrack \del_\th \p (0,\th)\Big)
\cr
& - \lbrack\nabla e (\p(\tilde \rh, \tilde \th))-\nabla e(\p(0,a(\th)))\rbrack
\cdot \del_\th^2 \p(0,\th)
.\cr}\eqn $$
We have
$$
\abs{\tilde\Ph(\th)} \le \Const \left( 
\abs{\tilde\rh(\th^*)} + \abs{\tilde\rh(\th)}
+ \abs{\th-\th^*} +\abs{a(\th)-a(\th^*)} \right)^\hxp
\eqn $$
Again, H\" older continuity and \queq{\zzeri} imply that there is 
$\ka >0$ such that $g''(\th)$ is bounded away from zero, uniformly in
 $\q\in\cB_\ka$ and  $\th \in U_\ga(t_l) \cap \cR_b$. 

Taylor expansion of $g'$ gives
$$
g'(\th) = (\th-\th^*) \; \int\limits_0^1 dt \; g''(\th^*+t(\th-\th^*))
\eqn $$
If $\abs{\th-\th^*}< \ka$, the coefficient of $\th-\th^*$ is bounded away
from zero, and therefore $\abs{g'(\th)} \le \veps^{1/2}$
implies $\abs{\th-\th^*} \le \sfrac{\veps^{1/2}}{z_\zer}$.
Thus 
$$\eqalign{
\int\limits_{U_\ka(t_l)} d\th \True{\abs{g(\th)} \le \veps} & \le
\int d\th \True{\abs{g(\th)} \le \veps}\True{\abs{g'(\th)} \le \veps^{1/2}}
\cr
& + \int d\th \True{\abs{g(\th)} \le \veps}\True{\abs{g'(\th)} \ge \veps^{1/2}}
\cr
& \le \frac{2\veps^{1\over 2}}{z_\zer}+ \veps^{-{1\over 2}}
\int dx \; \True{|x| \le \veps} 
\cr
& \le 2 \veps^{1\over 2} (1+\sfrac{1}{z_\zer})
,\cr}\eqn $$
and we have proven \queq{\TOSHOW}, with $\al =\sfrac{1}{2}$,
for all contributions except those from a small neighbourhood 
of $\th^{(1)}, \ldots, \th^{(N)}$. 

\sect{The hard case}
\pni
We now turn to the case where a tangential intersection 
happens near points where $\sfrac{\del a}{\del \th}=1$. 
Let $\q$ run over $\cB_\ka$.
We have to bound the contribution from all $U_\ka (\th^{(n)})$
to the volume integral in \queq{\TOSHOW}. Let $U=U_\ka(\th^{(n)})$ and 
$\veps'=\veps^{{2\over 3}}$, then $\veps'\ge \veps$. Again, we decompose 
$$
\int\limits_{U} d\th \; \True{\abs{g(\th} \le \veps} = 
Z(\veps,\veps') + V(\veps,\veps')
\eqn $$
where 
$$\eqalign{
Z(\veps,\veps') &=\int\limits_{U\cap N_\et} 
d\th \; \True{\abs{g(\th)} \le \veps}
\True{\abs{g'(\th)} \le \veps'} \cr
V(\veps,\veps') &=\int\limits_{U\cap N_\et} 
d\th \; \True{\abs{g(\th)} \le \veps}
\True{\abs{g'(\th)} > \veps'}
.\cr}\eqn $$
Since $\et$ is fixed, we may assume that $\ka$ is chosen so small
that $U\cap N_\et$ is a single open interval.
We use that $|\tilde \rh| \le \veps$ must hold in the support 
of the integrand to replace $g'$ by a function $f$, as follows.
By \queq{\Useful} and Taylor expansion in $\tilde \rh$, 
$$\eqalign{
g'(\th) &= -\nabla e(\q-\p(0,\th)) \cdot \del_\th\p(0,\th) \cr
&= - \nabla e (\p(0,\tilde \th)) \cdot \del_\th \p(0,\th)
- \tilde \rh(\th) \ph_1(\th)\cr 
&= f(\th) - g(\th)\; \ph_\one (\th)
\cr}\eqn $$
with a  function $\ph_1$ that is uniformly bounded on $\R/2\pi\Z$,
and 
$$
f(\th) =  - \nabla e (\p(0,\tilde \th)) \cdot \del_\th \p(0,\th)
.\EQN\heresf $$
On the support of $\True{\abs{g(\th)} \le \veps}$,
$$
\abs{g'(\th)-f(\th) } \le \abs{g(\th)} \abs{\ph_\one}_\zer
\le \veps \abs{\ph_\one}_\zer
,\eqn $$
so $\abs{g'(\th)} \le \veps'$ implies 
$\abs{f(\th)}\le \veps'+\veps\abs{\ph_\one}_\zer$, thus
$$
\True{\abs{g(\th)} \le \veps}\True{\abs{g'(\th)} \le \veps'} \le 
\True{\abs{g(\th)} \le \veps}\True{\abs{f(\th)} \le
\veps' (1+\abs{\ph_\one}_\zer)}
,\eqn $$
and hence
$$
Z(\veps,\veps') \le \int\limits_{U\cap N_\et} 
d\th \; \True{\abs{g(\th} \le \veps}
\True{\abs{f(\th)} \le \veps'(1+\abs{\ph_\one}_\zer)} 
.\eqn $$
Conversely, if $\abs{g(\th)}\le \veps$ and $\abs{g'(\th)} \ge \veps'$, 
then for all 
$$
\veps \le \left(\frac{1}{2\abs{\phi_\one}_\zer}\right)^3
\EQN\fixga $$
we have
$$
\abs{f(\th)} \ge \abs{g'(\th)} - \veps \abs{\phi_\one}_\zer 
\ge \frac{1}{2}\;\veps'
.\EQN\gpbig $$
Since the constant on the right side of \queq{\fixga} is fixed,
we may assume that $\veps$ is so small that \queq{\fixga} holds.

By strict convexity, $|f(\th)|\le\Const\veps'$ implies that either
$|\tilde\th-\th|\le\Const\veps'$ or $|\tilde\th-a(\th)|\le\Const\veps'$.
In the event that $\tilde\th\approx\th$,
 $\q$ can have no representation in the form
$\p(0,\th')+\p(r',\vth^{(2)} (0,\th',r'))$ with $\th'$ in the current 
interval $U_\ka(\th^{(n)})$, since $\vth^{(2)} (0,\th',r')\approx a(\th)$.
This puts us into a ``$g'$ has isolated zeros'' setting and is handled
as in the last section.

When $\tilde \th(\th) \approx a(\th)$, we rewrite 
$$\eqalign{
f(\th) &= \nabla e (\p(0,\tilde \th)) \cdot \del_\th \p(0,a(\th))\cr
& = \nabla e (\p(0,\tilde \th)) \cdot 
\left(\del_\th \p(0,a(\th))-\del_\th \p(0,\tilde\th)\right)
\cr}\eqn $$
Taylor expansion of the second factor gives
$$
f(\th) =  (a(\th)- \tilde \th(\th)) \; \Ph (\th )
\EQN\joi $$
with 
$$
\Ph(\th) = \int\limits_0^1 ds\;
\del_\th^2 \p\Big(0,\tilde \th+s(a(\th)- \tilde \th(\th))\Big)
\cdot \nabla e (\p(0,\tilde \th(\th)))
\eqn $$
At $s=0$, the integrand is
$$
\del_\th^2 \p(0,\tilde \th(\th))\cdot \nabla e (\p(0,\tilde \th(\th)))
= - w(\p(0,\tilde \th(\th))) 
\eqn $$
which, by \queq{\wz}, is bounded below in magnitude by $w_\zer$.
By continuity, $\ka$ can be chosen so small that 
$$
\abs{\Ph(\th)} \ge \frac{w_\zer}{2} > 0
.\eqn $$
Thus
$$
Z(\veps,\veps') \le 
\int\limits_{U} d\th \; \True{\abs{g(\th)} \le \veps}
\True{\abs{\tilde\th(\th)-a(\th)} \le \Ga_\one\veps'} 
,\eqn $$
with 
$$
\Ga_\one =\sfrac{2}{w_\zer}(1+\abs{\phi_\one}_\zer),
\EQN\Gagiv $$
and it suffices to control the function $b(\th)=a(\th)- \tilde\th(\th)$
near its zeros to bound $Z(\veps,\veps')$.

First observe that
$$
\frac{b'(\th')-b'(\th)}{\th'-\th}
=\frac{a'(\th')-a'(\th)}{\th'-\th}
-\frac{1}{\th'-\th}\int_\th^{\th'}ds\ 
\frac{\partial^2\tilde\th}{\partial\th^2}(s)
$$
and that, by assumption $(H4')$, $\sfrac{a'(\th')-a'(\th)}{\th'-\th}$ is
of fixed sign and bounded below by $K_a$. We now show that 
$\sfrac{\partial^2\tilde\th}{\partial\th^2}$ is bounded in magnitude by
$\half K_a$.
Suppose that $\q= \p(0,\th^*)+\p(0,a(\th^*))$ with $\th^*$ in the current 
interval $U_\ka(\th^{(n)})$. The generalisation to
$\q=\p(0,\th^*)+\p(r^*,\vth^{(2)} (0,\th^*,r^*))$ with $r^*\ne 0$
is a small perturbation.
Differentiating \queq{\Useful} and 
recalling that $\tilde \rh (\th^*) =0$ and $\tilde \th(\th^*) = a(\th^*)$,
we get at $\th=\th^*$
$$
-\del_\th\p(0,\th^*) = 
\frac{\del \tilde \rh}{\del \th}(\th^*) \del_\rh\p(0,a(\th^*))
+
\frac{\del \tilde \th}{\del \th}(\th^*) \del_\th\p(0,a(\th^*))
.\eqn $$
Since $\del_\th\p(0,a(\th^*))=-\del_\th\p(0,\th^*))$ and since
$\del_\rh\p(0,\th) = u(\p(0,\th)) |\del_\rh\p(0,\th)|$ 
and $\del_\th\p(0,\th)$ are linearly independent, we have
$$
\frac{\del \tilde \rh}{\del \th}(\th^*)=0, \qquad
\frac{\del \tilde \th}{\del \th}(\th^*) =1
.\eqn $$
The second derivative of \queq{\Useful} at $\th=\th^*$ gives
$$ 
-\del_\th^2\p(0,\th^*) = 
\frac{\del^2 \tilde \rh}{\del \th^2}(\th^*) \del_\rh\p(0,a(\th^*))
+
\frac{\del^2 \tilde \th}{\del \th^2}(\th^*) \del_\th\p(0,a(\th^*))
+ \del_\th^2 \p(0,a(\th^*))
. $$
By \queq{\deltwop}, 
$\del_\th^2\p(0,a(\th^*))\sfrac{\del a}{\del \th}(\th^*)=
-\del^2_\th\p(0,\th^*)$
and 
$$ 
0 = \frac{\del^2 \tilde \rh}{\del \th^2}(\th^*) \del_\rh\p(0,a(\th^*))
+
\frac{\del^2 \tilde \th}{\del \th^2}(\th^*) \del_\th\p(0,a(\th^*))
+ \left(1-\frac{\del a}{\del \th}(\th^*) \right)\del_\th^2 \p(0,a(\th^*))
.\eqn $$
Recalling that $(\del_\th\p)^2 = 1$ implies that $\del_\th^2\p$ is
orthogonal to $\del_\th\p$, we dot with $\del_\th\p$ to get
$$ 
\frac{\del^2 \tilde \th}{\del \th^2}(\th^*) =  -\del_\rh\p(0,a(\th^*))\cdot
 \del_\th\p(0,a(\th^*))\frac{\del^2 \tilde \rho}{\del \th^2}(\th^*)
. $$
Substituting this in and dotting instead with $\del_\rh\p(0,a(\th^*))$ 
then yields
$$\eqalign{
\frac{\del^2 \tilde \rh}{\del \th^2}(\th^*)  &=
\left(1-\frac{\del a}{\del \th}(\th^*) \right)
X_\rh(\th,\th^*)
\cr
\frac{\del^2 \tilde \th}{\del \th^2}(\th^*)  & =
\left(1-\frac{\del a}{\del \th}(\th^*)\right)
X_\th(\th,\th^*)
\cr}\EQN\Csisi $$
where 
$$\eqalign{
X_\rh(\th,\th^*)& = -
\frac{\del_\th^2 \p(0,a(\th^*))\cdot \del_\rh\p(0,a(\th^*))}%
{\big(\del_\rh\p(0,a(\th^*))\big)^2 - 
[\del_\rh\p(0,a(\th^*))\cdot\del_\th\p(0,a(\th^*))]^2}
\cr
X_\th(\th,\th^*)& = - \del_\rh\p(0,a(\th^*)) \cdot \del_\th\p(0,a(\th^*))
X_\rh(\th,\th^*)
\cr}\eqn $$
are bounded $C^{0,\hxp}$ functions (because the change of variables
is regular and $\del_\th\p$ is a unit vector the denominator is 
bounded below by a fixed positive number). As 
$\sfrac{\del a}{\del \th}(\th^{(n)})=1$, we conclude that, if $\ka$ is small 
enough, $\sfrac{\partial^2\tilde\th}{\partial\th^2}$ is bounded above in
magnitude by $\half K_a$ and consequently $\sfrac{b'(\th')-b'(\th)}{\th'-\th}$
is of fixed sign and magnitude at least $\half K_a$.

We are now in a position to bound $Z(\veps,\veps')$.
$$\eqalign{
Z(\veps,\veps') 
&\le \int_{U} d\th \; \True{\abs{b(\th)} \le \Ga_\one\veps'} \cr
&\le \int_{U} d\th \; \True{\abs{b(\th)} \le \Ga_\one\veps'} 
\True{\abs{b'(\th)} \le \sqrt{\Ga_\one\veps'}}
+ \int_{U} d\th \; \True{\abs{b(\th)} \le \Ga_\one\veps'} 
\True{\abs{b'(\th)} \ge \sqrt{\Ga_\one\veps'}} \cr
&\le \int_{U} d\th \; \True{\abs{b'(\th)} \le \sqrt{\Ga_\one\veps'}}
+ \int_{U} d\th \; \True{\abs{b(\th)} \le \Ga_\one\veps'} 
\True{\abs{b'(\th)} \ge \sqrt{\Ga_\one\veps'}} \cr
&\le \sfrac{4}{K_a} \sqrt{\Ga_\one\veps'}
+ \int_{U} d\th \; \True{\abs{b(\th)} \le \Ga_\one\veps'} 
\True{\abs{b'(\th)} \ge \sqrt{\Ga_\one\veps'}} \cr
}$$
As $b'(\th)$ is monotone, $b(\th)$ can have at most two zeros so 
$$
\int_{U} d\th \; \True{\abs{b(\th)} \le \Ga_\one\veps'} 
\True{\abs{b'(\th)} \ge \sqrt{\Ga_\one\veps'}}
\le 4\frac{\Ga_\one\veps'}{\sqrt{\Ga_\one\veps'}}
$$
and $Z(\veps,\veps')\le 4\left(1+\sfrac{1}{K_a}\right)\sqrt{\Ga_\one\veps'}
\le 4\left(1+\sfrac{1}{K_a}\right)\sqrt{\Ga_\one}\veps^{1\over 3}$.

Finally, we turn to $V(\veps,\veps')$. By \queq{\gpbig}, 
$$
V(\veps,\veps') \le \int\limits_{U\cap N_\et} 
d\th \; \True{\abs{g(\th)} \le \veps}
\True{\abs{g'(\th)} > \veps'}
\True{\abs{f(\th)} \ge \sfrac{1}{2}\veps'}
.\eqn $$
Denoting $\abs{\Ph}_\zer = W_\one$, we have by \queq{\joi} that
$\abs{f(\th)} \ge  \sfrac{1}{2}\veps'$ implies 
$|\tilde\th(\th) - a(\th)|\ge \sfrac{1}{2W_\one}\veps'$.
We have just seen that
$f$ has only two zeros on $U \cap N_\et$ that decompose $U \cap N_\et$ 
into at most three intervals.
By \queq{\gpbig} and continuity of $g'$, 
$g$ is monotonic on each of these intervals.
Therefore, changing variables from $\th$ to $\ga = g(\th)$ on
each of these intervals, and noting that the Jacobian is bounded
by $\sfrac{2W_\one}{\veps'}$, we have 
$$
V(\veps,\veps') \le 3\; \frac{2W_\one}{\veps'}
\int d\ga \True{\abs{\ga} \le \veps} 
\le 12 W_\one \; \frac{\veps}{\veps'} =
12 W_\one \;\veps^{1\over 3}
.\eqn $$
We have thus proven \queq{\TOSHOW} with $\al = \sfrac{1}{3}$. \endproof

\appendix{D}{Properties of the Scale Zero Effective Action}

\noindent
In this appendix, we prove Lemma
\UVpart. 
The scale zero part of the propagator $C_\zer$ is a bounded function
because the function $a$ cuts off values of the denominator smaller
than $M^{-2}$. Since $C_\zer \sim \sfrac{1}{ip_\zer}$ for large
$p_\zer$, it is obvious that the integral over $|C_\zer|^n$
is finite for all $n \ge 2$. Consequently, the most delicate
part of the proof of Lemma \UVpart\ is  the proof that
$\int C_\zer$  and $\int dp \hat v(q-p) C_\zer (p)$ converge.
These integrals correspond to graphs with a loop containing
only one fermion propagator. They are, actually, the first order
graphs shown in Figure \erschte, but with $C_\zer$ as the fermion
propagator. 
As discussed, for $n=1$, the integral converges at large $p_\zer$
only because of sign cancellations that come from the 
boundary value prescription \queq{\prescri}. 
The latter implies by a contour integral argument 
that for the propagator without any cutoff,
$$
\int d^{d+1} p \; C(p_\zer, e(\p)) = \int d^d\p \; \True{e(\p) < 0}
.\eqn $$
A similar contour integral argument for 
the first order 
$\int \hat v(q-p) C(p) d^{d+1} p $ requires some analyticity properties 
of $\hat v$. The cutoff function $a$ appearing in the definition of 
$C_\zer$ cannot be analytic in any 
neighbourhood of $\R$, so the above contour argument does not apply to $C_\zer$. 
We show convergence of the integrals at large $p_\zer$  
by a different argument which does not use any analyticity properties. 
This is of course
possible since the convergence of the integral only depends on
properties of $C_\zer$ on the real line, and since the function
$a$ does not change the behaviour for large $p_\zer$. 
One advantage of this proof is that it applies to all $\hat v$ 
satisfying \AOne, which is a rather general class of potentials
(in particular, no properties of $\hat v$ for 
$p_\zer$ away from the real axis are required at all.
  
\Lem{\isii}{\Lesty Assume that  \AOne{k,\hxp} and \ATwo{k,\hxp}
hold, and let $C_\zer$ be given by \queq{\sczerC}. Then the function
$$
I(q) = \int\limits_{\R\times\cB} \hat v(q-p)\;
C_\zer (p_\zer,e(\p)) 
d^{d+1} p =
\lim\limits_{\ta\downarrow 0} \int\limits_{\R\times\cB}
\hat v(q-p)\; \frac{a(p_\zer^2+e(\p)^2)}{ip_\zer -e(\p)} 
 e^{ip_\zer \ta} d^{d+1} p 
\eqn $$
is finite, uniformly bounded, and $C^{k,\hxp}$.}

\Proof We do the case $\hat v $ independent of $q_\zer$ first.
 The integrand is bounded because of the cutoff function $a$. We do the
$p_\zer$--integration first. Fix $\p\in \cB$, and let $e=e(\p)$. 
We show that 
$$
\lim\limits_{\ta\downarrow 0}
\lim\limits_{A \to -\infty}\lim\limits_{B \to \infty}
\int\limits_A^B C_\zer(p_\zer,e) e^{i\p_\zer\ta} dp_\zer
\eqn $$
converges. It is obvious that the integral from $-1$ to $1$ converges
uniformly in $\ta$ 
since the function $a$ makes $C_\zer$ bounded. Thus we may assume
$A < -1$, $B>1$, and we have to show that the integrals over 
$[A,-1]$ and $[1,B]$ converge. Both contributions are similar, 
so we consider only the second one. Using 
$$
\frac{1}{ip_\zer-e} - \frac{1}{ip_\zer} = 
-\frac{e}{p_\zer^2+iep_\zer}
,\eqn $$
we have
$$
\int\limits_1^B dp_\zer \; \frac{a(p_\zer^2+e(\p)^2)}{ip_\zer -e(\p)}
e^{ip_\zer \ta} = 
\int\limits_1^B dp_\zer \; \frac{a(p_\zer^2+e(\p)^2)}{ip_\zer}
e^{ip_\zer \ta} - \tilde I(\tau)
\eqn $$
The integrand in $\tilde I(\tau)$ is bounded by $\sfrac{\bar
e}{p_\zer^2}\; \True{\abs{p_\zer}\ge 1}$ (here $\bar e = \max\limits_{\p
\in \cB} e(\p)$), so the integral is absolutely convergent, 
uniformly in $\ta$, to a function that is continuous in $\ta$. 
Hence we only have to show that 
$\int\limits_{1}^B \sfrac{dp_\zer }{ip_\zer} e^{ip_\zer \ta}$ converges.
The convergence of this integral at fixed $\ta$ is now  a 
consequence of 
$$
\int\limits_1^B\frac{dp_\zer}{ip_\zer} e^{ip_\zer\ta} =
\int\limits_1^B \frac{dp_\zer}{ip_\zer}\frac{1}{i\ta} 
\frac{d}{dp_\zer} e^{ip_\zer \ta} =
-\frac{e^{ip_\zer\ta}}{p_\zer \ta} \Bigg\vert^B_1 -
\int\limits_1^B \frac{dp_\zer}{\ta p_\zer^2} e^{ip_\zer\ta}
\eqn $$
Although the above bounds are not uniform in $\ta$, they show that 
at fixed $\ta$, the limit
$$
J(\ta) = \lim\limits_{A\to -\infty, B\to \infty} 
\int\limits_A^B dp_\zer \; \frac{a(p_\zer^2+e(\p)^2)}{ip_\zer -e(\p)}
e^{ip_\zer \ta} 
\eqn $$
exists. Thus we may calculate it by taking $B=L$, $A=-L$. Since 
$$\eqalign{
\int\limits_{-L}^{-1} dp_\zer \; \frac{e^{ip_\zer \ta}}{ip_\zer} &
+
\int\limits_{1}^{L} dp_\zer \; \frac{e^{ip_\zer \ta}}{ip_\zer}  =
2 \int\limits_1^L dp_\zer \; \frac{\sin(p_\zer \ta)}{p_\zer} 
= 2 \int\limits_{\ta}^{L\ta} dx \; 
\frac{\sin x}{x} \cr
& \gtoas{L\to\infty} 2 \int\limits_{\ta}^{L\ta} dx \; 
\frac{\sin x}{x} 
\gtoas{\ta \to 0} 2 \int\limits_0^\infty dx \frac{\sin x }{x} = \pi
,\cr}\eqn $$
we see that $J(\ta)$ converges as $\ta \to 0$.

If $\hat v$ is not constant, but obeys \AOne, we write the integral
as a sum of ones with the $p_\zer$--independent constant $\tilde
v(\q-\p)$ and the difference $\hat v(q_\zer-p_\zer,\q-\p) - \tilde v(\q-\p)$.
The decay assumed
in \AOne\ makes the second integral convergent and bounded uniformly
in $q_\zer$. The first one was treated above and is independent of $q_\zer$.
Convergence of the integral for
the derivatives of $I$ is even easier: apply integration by parts
to move one derivative to $C_\zer$, then the integral converges
absolutely. 

The remaining integral over $\p$ is over a compact region, so
its convergence is trivial since the integrand is bounded.
\endproof

\ssni
{\bf Proof of Lemma \UVpart:} 
The kernels $V^{(0)}_{m,r}$ of the scale zero effective actions
are sums over values of Feynman graphs, with propagator $C_\zer$
and vertex function $\hat v$. The vertex function $\hat v$ is
bounded. If we slice 
$$
C_\zer(p_\zer, e(\p)) = \sum_{n=1}^\infty C_n (p_\zer, e(\p))
\eqn $$
with $C_n(p_\zer, e(\p))$ having $p_\zer$ support contained in 
$[M^n,M^{n+2}]$ for $n \ge 2$, and $p_\zer$ support contained in 
$[0,M^3]$ for $n=1$, then 
$$\eqalign{
\sup|C_n| &\le \Const M^{-n}\cr
\hbox{ vol} \supp C_n & \le \Const M^{n}
\cr}\eqn $$
Consequently, the degree of any graph $G$ is
$$\eqalign{
& -\abs{\{\hbox{lines of }G\}} + \abs{\{\hbox{loops of }G\}} =\cr
& -\abs{\{\hbox{lines of }G\}} + \left( \abs{\{\hbox{lines of }G\}}
- \abs{\{\hbox{vertices of }G\}} + 1\right) \cr
& = 1 - \abs{\{\hbox{vertices of }G\}} 
\cr}\eqn $$
which is convergent for all graphs having strictly more than
one vertex. The graphs with only one vertex were treated in 
Lemma \isii. Lemma \UVpart\ now follows by 
the standard power counting bounds (see, e.g.\ \FST, Theorem 2.40).
\endproof

\vfill\eject
\centerline{\bfe References}
\sni
{\parskip=0.5cm
\Ref{BG}{G.\ Benfatto, G.\ Gallavotti, {\sl Perturbation Theory
of the Fermi Surface in a Quantum Liquid. A General Quasiparticle
Formalism and One-Dimensional Systems}, {\sl J.\ Stat.\ Phys.\
\bf 59} (1990) 541}
\Ref{BGPS}{G.\ Benfatto, G.\ Gallavotti, A.\ Procacci, B.\ Scoppola,
{\sl Beta Function and Schwinger functions for a Many--Fermion
System in One Dimension. Anomaly of the Fermi Surface},
{\it Comm.\ Math.\ Phys.\ \bf 160} (1994) 93 }
\Ref{BM}{F.\ Bonetto, V.\ Mastropietro, {\sl Beta Function and Anomaly
of the Fermi Surface for a $d=1$ System of Interacting Fermions
in a Periodic Potential}, {\it Comm.\ Math.\ Phys.\ \bf 172} (1995)
57}
\Ref{F}{S.\ Fujimoto, {\sl Anomalous Damping of Quasiparticles
in Two--Dimensional Fermi Systems, Journal of the Physical Society
of Japan \bf 60}(1991) 2013}
\Ref{FKLT}{J.\ Feldman, H.\ Kn\" orrer, D.\ Lehmann, E.\ Trubowitz, 
{\sl Fermi Liquids in Two Space Dimensions,} in {\sl Constructive Physics},
V. Rivasseau (ed.), Springer Lecture Notes in Physics, 1995}
\Ref{FMRT}{J.Feldman, J.Magnen, V.Rivasseau, E.Trubowitz, {\sl
An Infinite Volume Expansion for Many Fermion Green's Functions,
Helvetica Physica Acta \bf 65} (1992) 679}
\Ref{FST}{J.\ Feldman, M.\ Salmhofer and E.\ Trubowitz, 
{\sl Perturbation Theory around Non-Nested Fermi Surfaces I.
Keeping the Fermi Surface Fixed}, 
{\sl Jour.\ Stat.\ Phys.\ \bf 84} (1996) 1209}
\Ref{FT1}{J.Feldman and E.Trubowitz, {\sl Perturbation Theory for
Many--Fermion Systems, Helvetica Physica Acta \bf 63} (1990) 156}
\Ref{FT2}{J.Feldman and E.Trubowitz, {\sl The Flow of an Electron--Phonon
System to the Superconducting State, Helvetica Physica Acta \bf 64}
(1991) 213}
\Ref{GN}{G.\ Gallavotti and F.\ Nicol\` o, {\sl Renormalization Theory
in Four Dimensional Scalar Fields}, {\it Comm.\ Math.\ Phys.\ \bf 100} (1985)
545 and {\bf 101} (1985) 247}
\Ref{MW}{J.\ Mawhin and M.\ Willem, {\sl Critical Point Theory
and Hamiltonian Systems}, Springer Applied Mathematical Sciences,
vol. 74, Springer 1989}
\Ref{S}{M.\ Salmhofer, {\sl Improved Power Counting and Fermi Surface 
Renormalization}, cond-mat/9607022}
\par}

\endpage

\end